\documentclass[a4paper,reqno,11pt]{article}
\usepackage[a4paper,margin=3cm, marginparwidth=2cm]{geometry}
\usepackage[T1]{fontenc}
\usepackage{amsfonts,amsthm,amssymb,amsmath, mathdots, bbm,mathabx, bm}
\usepackage{amssymb, amscd}
\usepackage[makeroom]{cancel}
\usepackage[pdftex]{color,graphicx}
\usepackage{authblk}

\usepackage{caption}
\usepackage{subcaption}

\usepackage[all,cmtip]{xy}
\usepackage{enumitem}
\usepackage{bbold}
\usepackage{braket}

\usepackage[pagebackref, colorlinks = true, linkcolor = black, urlcolor  = black, citecolor = red]{hyperref}
\numberwithin{equation}{section}

\newtheorem{Th}{Theorem}
 
\newtheorem{Def}{Definition}[section]
\newtheorem{Prop}{Proposition}[section]
 \newtheorem{Lem}{Lemma}[section]

\newcommand{\be}{\begin{equation}}
\newcommand{\ee}{\end{equation}}

 \newcommand{\bE}{{\mathbb{E}}}
  \newcommand{\bS}{ { \bm{S} } }
 \newcommand{\bR}{{\mathbb{R}}}
 \newcommand{\bN}{{\mathbb{N}}}
 \newcommand{\cH}{{\mathcal{H}}}
 \newcommand{\cN}{{\mathcal{N}}}

\newcommand{\DeltaC}{{\Delta}}

\newcommand{\M}{{\sf M}}

\newcommand{\G}{{G}}
\newcommand{\cP}{{\mathcal{P}}}

\newcommand{\Tr}{{\mathrm{Tr}}}
\newcommand{\tr}{{\mathrm{tr}}}

\newcommand{\bsig}{ {\bm{\sigma}}}
\newcommand{\btau}{{\bm{\tau}}}

\newcommand{\pC}{{f [\bsig, \btau ]}}

\newcommand{\id}{{\mathrm{id}}}

\newcommand{\al}{{a}}
\newcommand{\bl}{{b}}

\newcommand{\TitleGamma}{{$($\protect\boldmath${\sigma}$$,\,$\protect\boldmath${\tau}$$)$ }}

\newcommand{\TitleGsig}{{\protect\boldmath${\sigma}$ }}

\newcommand{\ggam}{{\gamma_{{}_{\beta\epsilon} }}}
\newcommand{\ddel}{{\delta_{{}_{\beta\epsilon} }}}

\renewcommand\leq\leqslant

\renewcommand\geq\geqslant

\title{The tensor Harish-Chandra--Itzykson--Zuber integral II: detecting entanglement in large quantum systems}
\author[1]{Beno\^it Collins}
\author[2,3,4]{Razvan Gurau}
\author[2,5]{Luca Lionni\vspace{0.2cm}}

\affil[1]{Mathematics Department, Kyoto University, Kyoto, Japan.}
\affil[2]{Heidelberg University, Institut für Theoretische Physik, Philosophenweg 19, 69120 Heidelberg, Germany.}
\affil[3]{CPHT, CNRS, Ecole Polytechnique, Route de Saclay, 91128 Palaiseau, France.}
\affil[4]{Perimeter Institute for Theoretical Physics, 31 Caroline St. N, N2L 2Y5, Waterloo, ON, Canada.}
\affil[5]{IMAPP, Radboud University, Nijmegen, The Netherlands.}
\date{}

\begin{document}

\maketitle
\begin{abstract}
We consider the recently introduced generalization of the Harish-Chandra--Itzykson--Zuber integral to tensors and discuss its asymptotic behavior when the characteristic size $N$ of the tensors is taken to be large. This study requires us to make assumptions on the scaling with $N$ of the external tensors. We analyze a two-parameter class of 
asymptotic scaling ans\"{a}tze, uncovering several non-trivial asymptotic regimes. 

This study is relevant for analyzing the entanglement properties of multipartite quantum systems. We discuss potential applications of our results to this domain, in particular in the context of randomized local measurements. 
\end{abstract}
\vspace{0.1cm}

\tableofcontents

\vspace{0.5cm}
\section{Introduction}

For $D\ge 2$ a fixed integer, let $A$ and $B$ be self--adjoint operators on $(\mathbb{C}^N)^{\otimes D}$. The local\footnote{Local as opposed to global $U(N^D)$ transformations.} unitary transformations are denoted by:
\[
U=U^{(1)}\otimes\ldots \otimes U^{(D)}, \qquad U^{(c)} \in U(N),
\]
with $U(N)$ the group of $N\times N$ unitary matrices 
and we denote the tensor product of $D$ Haar measures
by $dU = dU^{(1)}\otimes \ldots \otimes dU^{(D)}$. The local unitary $U$ acts on $A$ and $B$ by conjugation
$A \to U A U^*$ respectively $B\to UBU^*$, where $*$ denotes the adjoint. Thus $A$ and $B$ transform as tensors with $D$ covariant and $D$ contravariant indices. The components of $A$ in the tensor canonical tensor product basis are denoted by 
$\{A_{i_1 \ldots i_D\,;\, j_1 \ldots j_D}\}_{i_c, j_c = 1, \ldots N}$.

We consider the generalization of the Harish-Chandra--Itzykson--Zuber (HCIZ) integral \cite{HarishChandra, Itzyk-Zub} introduced in \cite{CGL}:
\[
t\to I_{D,N}(t, A, B) =  \left \langle \exp \bigl(t \, \Tr (AUBU^* )\bigr)\right \rangle_U =  \int dU \; e^{t \Tr( A U BU^*) } \; ,
\]
which we refer to as the \emph{tensor HCIZ integral}. 
We study the expansion of its logarithm:
\[
C_{D,N}(t, A, B)  = \log I_{D,N}(t, A, B) =\sum_{n\ge 1} \frac{t^n }{n!}   \;  C_n\bigl(\Tr (AUBU^* )\bigr)  \; ,
\]
as a power series in $t$ and a Laurent series in $N$ and, in particular, the behavior of this object in the large $N$ limit.  

The coefficients 
$C_n\bigl(\Tr (AUBU^* )\bigr)$ are the \emph{cumulants} of the tensor HCIZ integral.
In \cite{CGL}, we expanded them on trace-invariants (a class of polynomials invariant under local unitary transformations) of the external tensors $A$ and $B$ multiplied by   connected Weingarten functions.
We studied the $1/N$-expansion of the Weingarten functions and showed that the coefficients of these expansions are generalizations of monotone double Hurwitz numbers.

However, we did not address the $1/N$-expansion of the tensor HCIZ integral per se. This expansion is subtle because it depends on how the trace-invariants of $A$ and $B$ scale with $N$.
In this paper, we study a class of asymptotic scaling ans\"{a}tze and show that there exists a $1/N$ expansion for each of them. We classify the various large $N$ limits. 

\paragraph{Motivations from quantum physics.} Apart from the motivations detailed in \cite{CGL} (\emph{i.e.} random tensor models),  the tensor HCIZ integral is relevant to the study of entanglement in \emph{multipartite quantum systems}, that is, quantum systems in which several sub-systems interact non-locally \cite{LU-1, LU-2, LU-3, LU-4, DNL}.

We consider a closed quantum system composed of $D$ subsystems represented by complex vector spaces with equal dimension $\{\cH_c=\mathbb{C}^N\}_{1\le c \le D}$, that is, the multipartite system is ``balanced''.
The indices $c = 1,\ldots, D$ labeling the subsystems are called {\it colors}. The density matrix $\rho$ of a mixed state is a  Hermitian, positive and normalized (of trace $1$) linear
operator on the tensor product space $\cH=\cH_1\otimes \ldots \otimes \cH_D$ which in general does not factor as a tensor product. 

In the tensor HCIZ integral, we interpret one of the external tensors, say $B$, as a density matrix up to normalization and the other tensor as an observable.  The expectation of $A$ in the state  $B$ is $\langle A \rangle_B = \Tr(AB)$.   

If $\Ket{\Psi}$ is a pure state of the full system, the set of states equivalently entangled to $\Ket{\Psi}$ is obtained by acting with local unitary transformations $U^{(1)} \otimes \cdots \otimes U^{(D)} \Ket{\Psi}$ \cite{LU-1, LU-2, LU-3, LU-4}. For mixed states, the set of density matrices equivalently entangled to $B$ is: 
\[
\bigl\{B_U=UBU^* \ \vert \ U=U^{(1)}\otimes \cdots \otimes U^{(D)}, \; U^{(c)}\in U(N) \;\; \forall c \bigr\} \;.
\]

The integral $\int dU \, \Tr(AUBU^*)=\bigl \langle \langle A\rangle_{B_U} \bigr\rangle_{U}$ is the average of the expectation of the observable $A$ over the states $B_U$ equivalently entangled to $B$. It is also the expectation of $A$ in a random quantum state $B_U$  equivalently entangled to $B$. This quantity is expected to provide information on the entanglement properties of $B$ and not on the local degrees of freedom due to the averaging. More information on the entanglement is obtained from the higher moments:
\be 
\label{eq:moments-tensor-hciz-intro}
\int dU \left(\Tr(AUBU^*)\right)^n=\bigl\langle \left(\langle A\rangle_{B_U} \right)^n\bigr\rangle_{U}. 
\ee

Random measurements based on random local unitary transformations have received increased attention recently as a method for characterizing correlations between subsystems of a multipartite system  \cite{random-meas-1, random-meas-2, random-meas-3, random-meas-4, random-meas-5}.  In this context, the observable $A$ is fixed only up to random unitary transformations (distributed on the tensor product of $D$ Haar measures).  A substantial advantage of this is that, contrary to many other entanglement criteria, it does not require the alignment of local reference frames with respect to a global shared reference frame. This alignment is very challenging to implement experimentally.  It has been shown for certain systems of a few qudits, that is, certain small values of $D, N$, 
 that
the first 
moments \eqref{eq:moments-tensor-hciz-intro} allow one to detect weak forms of entanglement that are not detected by other standard methods such as the positive-partial-transpose (PPT) criterion \cite{random-meas-2, random-meas-4, random-meas-5}.

\emph{The tensor HCIZ integral is the exponential generating function of the moments} \eqref{eq:moments-tensor-hciz-intro}. Our results on its asymptotic at large $N$ (which is the common dimension of the $\cH_c$s) provide new criteria for detecting entanglement in large multipartite quantum systems. 

\paragraph{Main result.}
We consider ans\"{a}tze for the asymptotic scaling with $N$ of the trace-invariants of $B$ and $A$ depending on two parameters (see Sections \ref{sec:notation} and \ref{sec:Asymptotic-regimes} for the detailed notation):
\be\label{eq:introscale}
 \Tr_{\btau}(B) \sim N^{\beta \sum_{1\le c\le D} \#(\tau_c) + \epsilon \sum_{1\le c_1<c_2 \le D}\# ( \tau_{c_1} \tau_{c_2}^{-1})}  \; \tr_{\btau}(\bl)\;,
\ee
where $\btau$ is a $D$-uple of permutations 
$\btau = (\tau_1, \dots, \tau_D)$ which indexes the trace-invariants, $\#(\tau)$ denotes the number of cycles of $\tau$, $\tr_{\btau}(\bl)$ denotes the {\it rescaled trace-invariant} which stays finite at large $N$ and:
\begin{itemize}
 \item the parameter $\beta$ multiplies a contribution factored over the subsystems $c\in\{1, \dots, D\}$ and is expected to be large if $B$ is a tensor product state or a convex combination thereof. 
 \item the parameter $\epsilon$  multiples a contribution which does not factor over the subsystems and is expected to be large for entangled states. 
\end{itemize}

We consider both the case where the trace-invariants of $A$ scale precisely as those of $B$, and the case when all the trace-invariants of $A$ are of order $1$.

\

 In the context of random local measurements \cite{random-meas-1, random-meas-2, random-meas-3, random-meas-4, random-meas-5} in multipartite quantum systems,  $A= \otimes_c A_c$ is a tensor product of local observables $A_c$. A relevant example consists in choosing  $\mathrm{Rank}(A_c)\sim O(1)$ (corresponding for instance to projectors on local pure quantum states),  in which case the trace-invariants of $A$ are of order $1$.  We now focus on this case. 
 
 The conclusions of this paper are as follows. The large $N$ moments in \eqref{eq:moments-tensor-hciz-intro} do not discriminate between the different scaling ans\"{a}tze for the trace-invariants: in Theorem~\ref{th:moments} we show that they are universal and hold no information on $\beta$ and $\epsilon$. 
On the contrary, the cumulants $C_n$ do discriminate between them (Theorem \ref{th:other-ent-microscopic-regimes}):
\begin{itemize}
 \item for any $(\beta,\epsilon)$ there exist $\ddel$ and $\ggam$ independent on $n$, $A$ and $B$ (but dependent $\beta,\epsilon$) such that:
\[
\lim_{N\to \infty}  \frac{1}{N^{\ddel}} \; C_n\left( N^{\ggam} \Tr(AUBU^\star)\right) = c_n(a,b) <\infty \;, \qquad c_n(a,b) \neq 0
\;.
\]
\item for any $(\beta,\epsilon)$ the $N$ independent coefficient $c_n(a,b)$ is a sum over a subset of the rescaled trace-invariants.
\item the $(\beta,\epsilon)$ plane splits into regions corresponding to different large $N$ \emph{regimes}. In each regime the subset of trace-invariants contributing to $c_n(a,b)$ is fixed. This subset changes drastically between regimes.
\item one regime (\ref{micAentitemV} in Theorem \ref{th:other-ent-microscopic-regimes}) is 
called ``entangled''. For $0\le \beta < \min \{1/D , \epsilon \}$ we find:
\[
\lim_{N\rightarrow +\infty} 
\frac 1 {N^{\beta D} } 
C_n\Bigl( N^{D- \epsilon \frac{D(D-1)}2 }\Tr (AUBU^* )\Bigr)  =   (n-1)! \    \tr(\al^n) \,  \tr (\bl^n) \; ,
\]
with $\tr(a^n) = \Tr(A^n) \sim O(1)$ and $\tr(b^n)
 = N^{-\beta D -n \epsilon \binom{D}{2} } \Tr(B^n)\sim O(1)
$ where $\Tr(\dots)$ is the usual operator trace.
\end{itemize}

\paragraph{Applications to random local measurements.}
Our task is to derive information about $\epsilon$ and $\beta$. In an actual experiment, it should be possible to make a numerical fit for
\[
C_n\left( \Tr(AUBU^\star)\right) = N^{\ddel -  \ggam n} c_n(a,b) \;.
\]
and to identify the corresponding large $N$ regime precisely. At the very least, one can check whether or not the system is in the entangled regime \ref{micAentitemV}, that is, whether or not:
\[
c_n \approx   (n-1)! \    \tr(\al^n) \,  \tr (\bl^n) \;.
\]

Assuming that we find that a system is in the regime \ref{micAentitemV}, several conclusion can be drawn:
\begin{itemize}
 \item $B$ cannot be a separable state as this asymptotic regime arises only at $\epsilon>0$. Thus $B$ is entangled (hence the name of the regime).
 \item if $\ddel>0$ then $\beta>0$ and $B$ cannot be a pure state, see Section \ref{subsec:Gaussian}.
\item if $\ddel=0$ then $\beta = 0$. We conjecture that in this case, $B$ \emph{is asymptotically pure}. Furthermore, we argue in Section \ref{sub:maxrank} that tensors with $\beta + \epsilon(D-1)=1$ saturate the number of degrees of freedom. If this is the case, then the maximal possible value of $ \epsilon $ is $1/(D-1)$ and a state with $\beta=0$ and $\epsilon=1/(D-1)$ is an entangled state which maximizes the number of degrees of freedom. We conjecture that any such state is \emph{maximally entangled} (note that under certain assumptions, maximally entangled states for balanced systems are expected to be pure \cite{maximally-entangled-pure}, hence our first conjecture). We give a confirmatory example of this in Section \ref{subsec:Gaussian}.

\end{itemize}

The $\epsilon = \beta <1/D$ line is a threshold of detection of entanglement. Any scaling for which the strength $\epsilon$ of the entangled part exceeds the strength $\beta$ of the separated part will be correctly detected as entangled by the large $N$ cumulants. In particular, if the strength $\epsilon$ of the entangled part exceeds $1/D$, entanglement will always be detected.

\

The paper is organized as follows. In Section \ref{sec:notation} we introduce the notation and recall some facts on the tensor HCIZ integral. In Section \ref{sec:Asymptotic-regimes} we give an overview and detailed discussion of our results. Section \ref{sec:Col-graphs} is technical and introduces some combinatorial tools used in the rest of the paper. Section \ref{sec:D>1} contains our main theorems. Some technical points are relegated to the Appendices \ref{sec:D1}, \ref{sec:microregproofs} and \ref{sec:symregproofs}.

\section*{Acknowledgements}

B.C. was partially supported by JSPS Kakenhi 17H04823, 20K20882, 21H00987,  and by the Japan-France Integrated action Program
(SAKURA), Grant number JPJSBP120203202.
R.G. and L.L. are supported by the European Research Council (ERC) under the European Union’s Horizon 2020 research and innovation program (grant agreement No818066) and by Deutsche Forschungsgemeinschaft (DFG, German Research Foundation) under Germany's Excellence Strategy  EXC-2181/1 - 390900948 (the Heidelberg STRUCTURES Cluster of Excellence). During most of this project, L.L.~was at Radboud University, supported by the START-UP 2018 programme with project number 740.018.017, financed by the Dutch Research Council (NWO). L.L.~thanks JSPS and Kyoto University, where the discussions at the origin of this project took place.

\newpage

\section{Notations and previous results}
\label{sec:notation}

We follow the notations of \cite{CGL}.
We denote by $A^*$ the adjoint of $A$.  $S_n$ stands for the group of permutations of $n$ elements and 
the number of disjoint cycles of $\sigma\in S_n$ is denoted by $\#(\sigma)$. $D$-uples of permutations are denoted in bold, and 
$\bS_n=\{ \bsig = (\sigma_1,\ldots ,\sigma_D) |  \,\sigma_c \in S_n, \,\forall c \}$.

\

We denote by $\cP(n)$ the set of partitions of the set $\{1,\ldots , n\}$ and $\pi$ its elements; $|\pi|$ is the number of blocks of $\pi$, $B\in \pi$ denotes the blocks of $\pi$ and $|B|$ the cardinal of $B$. The refinement partial order is denoted by $\le$, that is, $\pi'\le \pi$ if all the blocks of $\pi'$ are subsets of the blocks of $\pi$. In this case, $\pi'$ is said to be finer than $\pi$ and $\pi$ coarser than $\pi'$.  Furthermore,  $\vee$ denotes the joining of partitions; that is, $\pi\vee\pi'$ is the finest partition coarser than both $\pi$ and $\pi'$. 

\

A partition $\pi$ of an ordered set $C$ is said to be \emph{non-crossing} if there are no four elements $p_1<q_1<p_2<q_2 \in C$ such that $p_1, p_2\in B$ and $q_1, q_2\in B'$ for $B\neq B'$ two different blocks of $\pi$.

\ 

The partition induced by the transitivity classes of the permutation $\sigma$ (i.e. the disjoint cycles of $\sigma$) is denoted by $\Pi(\sigma)$, hence $ |\Pi(\sigma)| = \#(\sigma)$; $d_p(\sigma)$ denotes the number of cycles of $\sigma$ with $p$ elements, thus $d_1(\sigma)$ is the number of fixed points of $\sigma$ and $ \sum_{p\ge 1} d_p(\sigma) = \#(\sigma)$.

\ 

We denote by $\Pi(\bsig) = \bigvee_{c=1}^D 
\Pi(\sigma_c) $ the partition induced by the transitivity classes of the group generated by $(\sigma_1, \dots, \sigma_D) =\bsig\in \bS_n$ and $| \Pi(\bsig) |$ its number of blocks; $\Pi(\bsig,\btau) \equiv \Pi(\bsig) \vee \Pi(\btau)$.

\
 
We say that a permutation $\tau:C\to C$ of an ordered set $C$ is \emph{non-crossing} if the partition $\Pi(\tau)$ of $C$ is non-crossing and the ordering of the cycles of $\tau$ agrees with the order on $C$, that is, any cycle 
$(a \, \tau(a) \,\tau^2(a) \dots)$ of $\tau$ can be ordered such that $a < \tau(a) < \tau^2(a) <\dots $.

\paragraph{Trace-invariants and graphical representation.}
For $A$ an operator on $(\mathbb{C}^N)^{\otimes D}$ and $ \bsig = (\sigma_1,\ldots ,\sigma_D)\in \bS_n$, the \emph{trace-invariant}:
\[
\Tr_{\bsig}(A)=\sum_{\rm{all\ indices}} 
 \left( \prod_{s=1}^n A_{j^1_s\ldots j^D_s, i^1_s\ldots i^D_s } \right) 
\prod_{   c\,\in\{1,\ldots, D\}  } \left( \prod_{s=1}^n \delta_{i^c_s,j^c_{\sigma_c(s)}}  \right) \; ,
\]
is a polynomial invariant under conjugation $A\to UAU^*$ by local unitary transformations $U = U^{(1)} \otimes \ldots \otimes U^{(D)}$, $U^{(c)} \in U(N)$. The superscript $c$ of the index $i_s^c$ is the \emph{color} of $i_s^c$. 

\begin{figure}[!ht]
\centering
\includegraphics[scale=0.7]{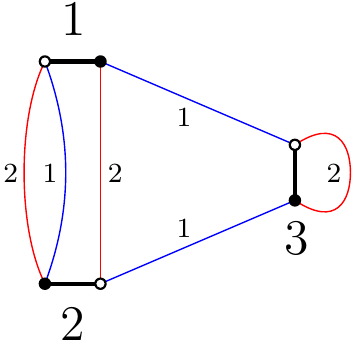}
\caption{Graphical representation of $\bsig = \big( (123) , (12)(3) \big)$. 
}
\label{fig:ExInvariant}
\end{figure}

The trace-invariants can be canonically represented as graphs; see Fig.~\ref{fig:ExInvariant} for an example.
For each $A_{j^1_s\ldots j^D_s, i^1_s\ldots i^D_s }$, we draw a black and a white vertex connected by a thick edge and labelled $s\in\{1,\dots, n\}$. We attach an outgoing half-edge of color $c$ for each $i_s^c$ to the white vertex and an incoming half-edge of color $c$ for each $j^c_s$ to the black one.
The outgoing half-edge $i^c_s$ is joined to the incoming half-edge $j^c_{\sigma_c(s)}$ for all $c$ and $s$. 
The resulting graph is denoted by $\bsig$, as it is a canonical representation of the $D$-uple of permutations. The graph $\bsig$ has $| \Pi(\bsig) |$
connected components. 

 For $D=1$, $\Tr_{\sigma}(A)=\Tr_{\sigma^{-1}}(A)$. For $D>1$, $\Tr_{\bsig}(A) = \Tr_{\bsig^{-1}}(A)$ if and only if the graphs $\bsig$ and $\bsig^{-1}$ are automorphic, which is not always the case.

\paragraph{Cumulants.}
For a random variable $X$, the cumulant (connected moment) $C_n(X)$ is defined by 
$ C(t) = \log \bE (\exp tX)= \sum_{n\ge 1}t^n \, C_n(X)/ n!$. In our case:
 \be
 \label{eq:general-cumulants}
 C_{D,N}(t,A,B) = \log\big\langle \exp(t\, \Tr (AUBU^* ))\big\rangle_U = \sum_{n\ge 1} \frac{t^n}{n!} \;  C_n\bigl(\Tr (AUBU^* )\bigr)  \;.
 \ee
 Of course $ C_n\bigl(\Tr (AUBU^* )\bigr)$ depends on $N$, but we keep this implicit to simplify the notation. 
The starting point of the present paper is Thm.~4.1 of \cite{CGL} which gives a formula for the cumulants of the tensor HCIZ integral. In the present paper, we use only the result at leading order in $N$, which we report here:
\begin{Th}[\cite{CGL}] 
\label{th:previous-results}
The cumulants of the tensor HCIZ integral have the asymptotic expansions:
\begin{align}
\label{eq:WC-asym2}
C_n\bigl(\Tr (AUBU^* )\bigr)= 
\frac 1 {N^{2nD }}\sum_{  \bsig,\btau \, \in \bS_n }
   \Tr_{\bsig}(A) \, \Tr_{\btau^{-1}}(B) \,N^{s(\bsig, \btau) } \pC \; \big(1+O(1/ {N^2}) \big)\;,
\end{align}
where the scaling exponent with $N$ of a term is: 
\be
\label{eq:scaling-of-weingarten}
s(\bsig, \btau) = \sum_{c=1}^D \lvert\Pi(\sigma_c\tau_c^{-1}) \rvert- 2 \lvert \Pi(\bsig, \btau)\rvert + 2 \;,
\ee
and the coefficient $\pC$ is given by:
\begin{itemize}
 \item denoting by $\nu_c=\sigma_c\tau_c^{-1}$ and by $\nu_{c|B}$ the restriction of $\nu_c$ to the block $B$ of a partition $\pi_c \ge \Pi(\nu_c)$ we have:
 \be
\label{eq:comb-expr-leading-cum-Weing}
\begin{split}
 \pC =&  \sum_{\substack{{\pi_1\ge \Pi(\nu_1) ,\ \ldots\ ,\  \pi_D\ge \Pi(\nu_D)}\\{| \Pi(\bsig, \btau)\vee\pi_1\vee\ldots\vee\pi_D | = 1 }\\{\sum_c \Pi(\nu_c)  - \sum_c \lvert \pi_c \rvert - \lvert \Pi(\bsig, \btau)\rvert + 1=0 } } }  
 (-1)^{nD - \sum_{c=1}^D |\Pi (\sigma_c\tau_c^{-1}) | }
 \crcr
& \hspace{2.1cm} \times 
\prod_{p=1}^n \Biggl(\frac {(2p)!}{p!(p-1)!}\Biggr)^{ \sum_c d_p(\nu_c) } 
\   \prod_{c=1}^D \prod_{B\in \pi_c} \frac{(2 |B| + |\Pi ({{\nu_c}_{|_B}})| - 3)!}{(2| B | ) !} \;.
\end{split}
\ee
  
 \item if $(\bsig, \btau)$ act transitively on $\{1,\ldots , n\}$, that is, $\lvert \Pi(\bsig, \btau)\rvert=1$, then $\pC$ reduces to a product of $D$
 Moebius functions on the lattice of non-crossing partitions \cite{NicaSpeicher}:
\[
\pC = \prod_{c=1}^D  \M(\sigma_c\tau_c^{-1}) \;,\qquad 
\M(\nu) = \prod_{p\ge 1 }  \left[  \frac{ (-1)^{p-1} }{p} 
    \binom{ 2 p-2 }{ p-1  }  \right]^{d_p(\nu)} \; .
\]

\item if $(\bsig, \btau)$ act transitively on $\{1,\ldots , n\}$ and moreover $\bsig=\btau$, then  $\pC  =  1  $.
\end{itemize}

\end{Th}

\ 

The scaling exponent $s(\bsig, \btau)$ and the leading order coefficient $\pC$ can be understood graphically.


\paragraph{The graph $(\bsig,\btau)$.}

\begin{figure}[h!]
\centering
\includegraphics[scale=0.7]{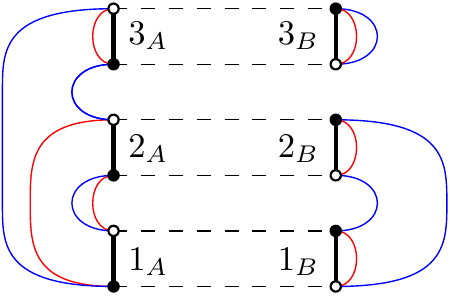}
\caption{The graph $(\bsig,\btau)$ for 
$\bsig = \bigl((123),(12)(3)\bigr)$ and 
$\btau = \bigl((12)(3), (1)(2)(3)\bigr)$.}
\label{fig:Excolgraph}
\end{figure}

We connect the vertices of the graph $\bsig$ associated  to $\Tr_{\bsig}(A) $ with the ones of the graph $\btau^{-1}$ associated  to $\Tr_{\btau^{-1}}(B) $ by joining the white (resp.~black) vertex $s$ of $\bsig$ with the black (resp.~white) vertex $s$ of $\btau^{-1}$ via a dashed edge. We denote this graph 
by $(\bsig, \btau)$, as it is a canonical representation
of the pair of $D$-uples of permutations.
An example is displayed in  Fig.~\ref{fig:Excolgraph}, where $1_A$, $2_A$ and $3_A$ denote the thick edges of $\bsig$ and $1_B$, $2_B$ and $3_B$ those of 
$\btau^{-1}$.

For each $s\in\{1,\ldots,n\}$, the four vertices labeled $s$ are connected into a quadrangle by dashed and thick edges. 
These vertices represent the four instances of $s$: black and white vertices for the inputs (incoming indices $j_s$), respectively the outputs (outgoing indices $i_s$), and two copies for $A$ and $B$.

\paragraph{The scaling exponent.}
The scaling exponent $s(\bsig, \btau) = \sum_{c=1}^D \lvert\Pi (\sigma_c\tau_c^{-1}) \rvert- 2 \lvert \Pi(\bsig, \btau)\rvert + 2 $ in \eqref{eq:scaling-of-weingarten}
has a simple interpretation in terms of the graph $(\bsig , \btau)$:
\begin{itemize}
 \item  $| \Pi(\bsig,\btau) |$, the number of transitivity classes of  the group generated by $(\bsig,\btau)$, 
is the number of connected components of $(\bsig , \btau)$.
 \item $ |\Pi(\sigma_c\tau_c^{-1})| = \#(\sigma_c\tau_c^{-1}) $ counts the cycles of the permutation $\sigma_c\tau_c^{-1}$, that is, the cycles made of alternating dashed edges and edges of color $c$ in the graph $(\bsig , \btau)$.
\end{itemize}

\paragraph{The graph $(\Pi,\{ \pi_c\}_c ; \{\Pi_c\}_c)$.} \label{subs:GrPi}
Concerning $\pC$ defined  \eqref{eq:comb-expr-leading-cum-Weing}, an abstract graph \cite{CGL} can be used to encode the constraints over the partitions $\{\pi_c\}_c$ indexing the sum. 

\begin{Def}\label{eq:defgraphpart}
Consider $2D+1$ partitions $\Pi,\{\pi_c\}_{1\le c\le D} ,\{\Pi_c\}_{1\le c\le D} $ on $\{1,\dots, n\}$ such that
 $ \Pi\ge \Pi_c $ and $ \pi_c \ge \Pi_c$ for all $c$. We build the abstract graph $(\Pi,\{ \pi_c\}_c ; \{\Pi_c\}_c)$ as follows:
 \begin{itemize}
 \item for every block $B$ of the partition $\Pi$, we draw a square vertex.
 \item for all $c$, for every block $B_c$ of $\pi_c$, we draw a triangular
 $c$--colored vertex.
 \item for all $c$, every block $b_c$ of $\Pi_c$ is at the same time contained in a block $B(b_c)$ of $\Pi$ and a block $B_c(b_c)$ of $\pi_c$. We join the vertices corresponding to $B(b_c)$ and $B_c(b_c)$ by a $c$--colored edge corresponding to $b_c$.  
\end{itemize}
\end{Def}
 
The graph $(\Pi, \{ \pi_c\}_c; \{ \Pi_c\}_c )$ has $ | \Pi\vee \pi_1 \dots \vee \pi_D|$ connected components\footnote{Two blocks $B\in \Pi$ and $B_c \in \pi_c$ are connected by an edge $b_c\in \Pi_c$ if and only if $b_c\subset B, B_c$, hence both $B$ and $B_c$ belong to the block of $  \Pi\vee \pi_1 \dots \vee \pi_D$ which contains $b_c$.}, 
$ \lvert \Pi \rvert +  \sum_c |\pi_c| $ vertices and $\sum_c \lvert \Pi_c \rvert$ edges, hence:
\[
\sum_c \lvert \Pi_c \rvert -
\sum_c |\pi_c| -  \lvert \Pi \rvert  + 
| \Pi\vee \pi_1 \dots \vee \pi_D| \ge 0 \;,
\]
as this is the number of \emph{excess edges} in the graph, that is, the number of edges in the complement of a spanning forest\footnote{A spanning forest in a graph is a set of edges which is a spanning tree in each connected component of the graph.}. The sum in  \eqref{eq:comb-expr-leading-cum-Weing} runs over the $\{\pi_c\}_c$ for which $\big( \Pi(\bsig, \btau) , \{\pi_c\}_c ; \Pi(\nu_c) \big) $ is a tree.

  \newpage
  
\section{Overview of the results}
\label{sec:Asymptotic-regimes} 
  
We consider sequences of tensors $\{A_N,B_N\}_{N\in\bN}$ such that $\Tr_{\btau}(B_N)$ (and similarly for $A_N$) obeys:
\[
\lim_{N\rightarrow  \infty}\frac 1 {N^{s_B (\btau)}}
\, \Tr_{\btau}(B_N) = \tr_{\btau}(\bl) < \infty \;, \qquad
 \text{that is,}\qquad \Tr_{\btau}(B_N)  \sim N^{s_B(\btau)}  \tr_{\btau}(\bl) \; , 
\]
where $\tr_{\btau}(\bl)\neq 0$ denotes the {\it rescaled trace-invariant}, which stays finite at large $N$. The rescaled trace-invariants can be interpreted \cite{Spei03} as the trace-invariants of a formal variables $\bl$. To simplify the notation, we suppress the subscript $N$ on $B_N$ (and $A_N$).

The interesting asymptotic ans\"{a}tze are those $s_{A}(\bsig), s_B(\btau)$ for which the cumulants in  \eqref{eq:WC-asym2} have the same asymptotic behavior in $N$ irrespective of $n$. Allowing for a rescaling of $t$:
\[
 C_D(t,a,b)  = \lim_{N\rightarrow \infty} \frac 1 {N^{\delta} }\log I_{D,N}(tN^{\gamma}, A, B) \; ,
\]
is then an infinite series in $t$ with coefficients of order $O(1)$ for some constants $\delta $ and $\gamma$. Such an ansatz leads to an exponential approximation
$
I_{D,N}(tN^{{\gamma}}, A, B) \sim \exp ( N^{ {\delta } } C_D(t,a,b) ) $ for the tensor HCIZ integral itself.
 
\paragraph{General scaling ansatz.} In this paper we consider the general scaling ansatz for $\btau \in \bS_n$:
\be
 \label{eq:general-scaling-assumption}
 s_B(\btau)= \alpha_B n + 
\beta_B \sum_{c=1}^D  \# (\tau_c) + \epsilon_B \sum_{ 1\le c_1<c_2  \le D} \# ( \tau_{c_1} \tau_{c_2}^{-1}) \;,
\ee
with $\alpha_B,\beta_B$ and $\epsilon_B$ some parameters independent\footnote{The subscript only indicates that they are associated to the trace-invariants of $B$, as opposed to the ones of $A$.} on $B$. We focus on this ansatz because it is one of the simplest ones, which leads to a change in the large $N$ behavior of the cumulants when varying the strength $\beta_B$ of the ``separable'' part (the one which factors over the $D$-subsystems) versus the strength $\epsilon_B$ of the ``entangled'' part (the one which does not). 
We say that the asymptotic scaling is \emph{asymptotically separable} if $\epsilon_B=0$ and \emph{asymptotically entangled} if not.

The trace-invariant $\Tr_{(\id,\dots, \id)}(B)\equiv \Tr(B)$ with $\id\in S_1$ is the trace of $A$ in the sense of operators on $(\mathbb{C}^N)^{\otimes D}$. It is associated to the graph consisting of a pair of vertices connected by all the $D+1$ edges, and: 
\[
 \Tr(B) \sim N^{\alpha_B + \beta_B D + \epsilon_B \binom
 {D}{2} } \; \tr(\bl)\;,
\]
which can always be set to $O(1)$ by choosing $\alpha_{B} = -\beta_{B} D -\epsilon_{B} \binom{D}{2}$. This normalization is essential if $B$ represents a density matrix of a multipartite quantum system: in this case we have furthermore $\Tr(B) = \tr(\bl) = 1$.

\paragraph{$A$-microscopic and symmetric scaling ans\"{a}tze.} We will focus on the following two cases for $D\ge 2$:
\begin{itemize}
\item The  $A$-\emph{microscopic} scaling ans\"{a}tze consist in taking the trace-invariants of $A$ of order~$1$, that is, $\Tr_{\bsig}(A)\sim \tr_{\bsig}(a)=O(1)$. We relabel $\epsilon_B\equiv\epsilon$ and  $\beta_B \equiv \beta$.
This case is relevant for random measurements
where $A=A_1\otimes \cdots \otimes A_D$ is a tensor product of local observables of small rank $\mathrm{Rank}(A_c) \sim O(1)$. The simplest examples consist in taking the $A_c$s as projectors on pure states in $\cH_c$.

\item The \emph{symmetric} scaling ans\"{a}tze consist in choosing the same scaling parameters for $A$ and $B$, namely 
$\epsilon_A = \epsilon_B\equiv\epsilon$ and $\beta_A = \beta_B \equiv \beta$. This generalizes the scaling of the original HCIZ integral \cite{Itzyk-Zub}.

\end{itemize}

In both cases, the asymptotic scaling is indexed by the two independent parameters 
$\beta,\epsilon$.

\paragraph{Cumulants at large $N$.} The cumulants  \eqref{eq:WC-asym2} write in terms of rescaled trace-invariants, and of the scalings \eqref{eq:scaling-of-weingarten} and \eqref{eq:general-scaling-assumption}, including an overall scaling factor $N^{\ggam}$, as:
\begin{align}\label{eq:WC-norm}
& C_n\bigl( {N^{\ggam}}\Tr (AUBU^* )\bigr)  
\\
&\qquad = \sum_{ \bsig,\btau\, \in\, \bS_n}
  N^{n \ggam  -2nD  + s(\bsig, \btau)+s_A(\bsig) + s_B(\btau)}      \; \tr_{\bsig}(\al) \,\tr_{\btau^{-1}}(\bl) \; \pC  \; \big(1+O(1/N^2) \big) \; . \nonumber
\end{align}
We will show below that for any $\beta, \epsilon \ge 0$, 
there exist $\ggam$ and $\ddel$ such that the cumulant $C_n$ has a $1/N$ expansion, that is, the scaling exponent of any term above reads: 
\[
  n ( \ggam - 2 D )    + s(\bsig, \btau) +s_A(\bsig) + s_B(\btau) = \ddel - h_{_{\beta\epsilon}}(\bsig,\btau)\;,
\]
with $h_{_{\beta\epsilon}}(\bsig,\btau) \ge 0$ for any $(\bsig,\btau)$. 
The \emph{leading order} graphs are the set $\{(\bsig, \btau) | \;  h_{_{\beta\epsilon}}(\bsig ,\btau)=0 \}$, and:
\[
\lim_{ N\to \infty} \frac{1}{N^{\ddel} } \;  C_n\bigl( {N^{\ggam}}\Tr (AUBU^* )\bigr)   = 
  c_n(a,b) = \sum_{ \substack{ {\bsig,\btau\, \in\, \bS_n} \\ { h_{_{\beta\epsilon}}(\bsig,\btau)=0} } } \; \tr_{\bsig}(\al) \,\tr_{\btau^{-1}}(\bl) \; \pC \;.
\]

Note that as we allow for a rescaling $\ggam$, 
the linear terms in $n$ in $s_{A}(\bsig)$ and $s_{B}(\btau)$ can always be reabsorbed in a shift of $\ggam$. While this comes to working with non-normalized $A$ and $B$, in the following, we will sometimes use $\alpha_A = \alpha_B=0$ to simplify the discussion.

\paragraph{Asymptotic regimes.} We call \emph{asymptotic regime}, or regime for short, a set of values of the parameters $(\beta, \epsilon)$  that lead to the same set of leading order graphs $(\bsig, \btau)$. We say that a regime is  \emph{(combinatorially) richer} than another one if the set of leading order graphs of the latter is strictly included in that of the former, and we call \emph{combinatorially prolific} a regime such that for a given $\bsig$, there exists more than one $\btau$ such that $(\bsig, \btau)$ is a leading order graph. 

Our choices of names for the regimes below are motivated by the applications to the study of entanglement in multipartite quantum systems. In this context, each color corresponds to a subsystem, and $B$ is a density matrix. 

\subsection{Tensors satisfying the asymptotic scaling ans\"{a}tze}
\label{subsec:concrete-examples}

Before presenting the various asymptotic regimes, let us first give some examples of tensors satisfying the scaling ans\"{a}tze and discuss their entanglement. 

The Hilbert space of a multipartite quantum system is a tensor product $\cH = \bigotimes_c \cH_c$.
We denote by $\Ket{i_c}$ the standard basis in $\cH_c$, and  $\Psi_{i_1\dots i_D} = \Braket{ i_1 \dots i_D| \Psi } $ the components of the vector $\Ket{\Psi}\in \cH$ in the tensor product basis $\Ket{i_1\dots i_D} = \bigotimes_c \Ket{i_c}$. A state of the system is a density matrix, that is, a positive semi-definite Hermitian operator $\rho$ on $\cH$ with trace $1$. The state of the system is:
\begin{itemize}
 \item \emph{pure} if its density matrix is a one dimensional projector $\rho = \Ket{\Psi}\Bra{\Psi}$ for some $\Ket{\Psi}\in \cH$.
 \item \emph{separable} if its density matrix can be written as:
\[
\rho = \sum_{k=1}^K p_k \, \rho_1^{(k)} \otimes \ldots \otimes \rho_D^{(k)} \,, 
\]
with  $K\ge 1$, $p_k>0$ such that $\sum_k p_k = 1$ and $\rho_c^{(k)}$ are density matrices. A state is \emph{entangled} if it is not separable. 
\end{itemize}

\subsubsection{States with microscopic scaling}
\label{subsub:example-micro}

If a state is both pure and separable then its density matrix is a tensor product of rank $1$ projectors $\rho = \bigotimes_c \Ket{ \Psi_c } \Bra{ \Psi_c}$. \emph{Pure separable states display the microscopic scaling} $\epsilon=\beta=0$:
\[
\Tr_{\bsig}( \Ket{ \Psi_1 } \Bra{\Psi_1 }\otimes \cdots \otimes \Ket{ \Psi_D} \Bra{ \Psi_D }) =1 \;.
\]
The microscopic scaling is also obtained if 
$\rho = \bigotimes_c \rho_c$ with ${\rm Rank}(\rho_c)$ a constant independent of $N$, and more generally for any family of states whose trace-invariants stay finite at large $N$, regardless of whether they are pure, separable, etc.

\subsubsection{States with asymptotically separable scalings}
\label{subsub:example-separable}

\paragraph{Tensor product state.}
Consider a tensor product state $\rho = \bigotimes_c \rho_c$ such that
$\textrm{Rank}(\rho_c)= N^{ \beta}$ with $0\le \beta\le 1$. We assume that the eigenvalues of $\rho_c$ are all of the same order  $N^{-\beta}$ (for instance $\rho_c$ could be proportional to the identity on its image). The trace-invariants decouple:
\[
\Tr_{\bsig}( \rho_1\otimes \cdots \otimes \rho_D) = \prod_{c=1}^D \Tr_{\sigma_c} (\rho_c) = \prod_{c=1}^D \prod_{p\ge 1} 
\left[ \Tr(\rho_c^p) \right]^{d_p(\sigma_c)} \;,
\]
and a cycle of $\sigma_c$ of length $p$ contributes $N^{\beta (1 - p)}$, leading to:
\[
\Tr_{\bsig}( \rho_1\otimes \cdots \otimes \rho_D)\,\sim  N^{\beta \sum_c \big[ \#(\sigma_c) -n \big]  } \;.
\]

\paragraph{Maximally mixed state.} The maximally mixed state is the state with density matrix $\frac 1 {N^D} \mathbb{1}^{\otimes D} $, where $\mathbb{1}$ the identity. It is asymptotically separable with $\epsilon=0$ and $\beta=1$:
\[
\Tr_{\bsig}\Bigl(\frac 1 {N^D} \mathbb{1}^{\otimes D} \Bigr) = N^{ \sum_c \big[ \#(\sigma_c) -n \big]  } \;,\qquad \Tr\Bigl(\frac 1 {N^D} \mathbb{1}^{\otimes D} \Bigr) = 1 \;.
\]

\paragraph{General separable states.}
By multilinearity, a trace-invariant evaluated over a general separable state is:
\be
\label{eq:n-linear-sep}
\Tr_\bsig(\rho) = \sum_{k_1, \ldots, k_n=1}^K p_{k_1} \cdots p_{k_n}\prod_{c=1}^D\, \prod_{{\eta \textrm{ cycle }}{\textrm{of $\sigma_c$}}} \Tr\biggl( \overrightarrow{\prod}_{i\in\eta}\rho_c^{(k_i)} \biggr),
\ee
where the product along the cycle $\eta$ is ordered.
We note that \emph{a separable state must have $\epsilon=0$}: as the traces factor over the cycles of $\sigma_c$, there is no way to obtain an asymptotic scaling with $\epsilon>0$. A simple upper bound on the trace-invariant is:
\[
 \left|  \Tr_\bsig(\rho) \right| \le 
 \sum_{k_1, \ldots, k_n=1}^K p_{k_1} \cdots p_{k_n}
 \; \left( \max_{c,i} \big\{ {\rm Rank}(\rho_c^{(k_i)}) \big\}\right)^{\sum_c \#( \sigma_c)} 
 \prod_{c=1}^D \prod_{i=1}^n \lvert| \rho_c^{(k_i)} \rvert|
 \;,
\]
where $\lvert| \rho_c^{(k_i)}\rvert|$ denotes the operator norm of $\rho_c^{(k_i)}$. Assuming that all the individual density matrices have ${\rm Rank}(\rho_c^{(k_i)})= N^{ \beta}$ with $0\le \beta\le 1$ and that all their eigenvalues are of the same order, we have $\lvert| \rho_c^{(k_i)} \rvert| \sim N^{-\beta}$, hence:
\[
 \left|  \Tr_\bsig(\rho) \right|  \le \;  N^{ \beta \sum_c \big[ \#( \sigma_c) - n \big] }   \sum_{k_1, \ldots, k_n=1}^K p_{k_1} \cdots p_{k_n}  = N^{ \beta \sum_c \big[ \#( \sigma_c) - n \big] } 
 \; .
\]

However, this is an upper bound and not the asymptotic behavior. Additional assumptions are needed in order to obtain a lower bound of the same order. The contribution to \eqref{eq:n-linear-sep}  that gathers the terms with all $k$s equal is:
\begin{equation}
\label{eq:lower-bound-separable}
\sum_{k=1}^K p_{k}^n \; \prod_{c} \, \prod_{{\eta \textrm{ cycle }}{\textrm{of $\sigma_c$}}} \Tr\biggl( \overrightarrow{\prod}_{i\in\eta}\rho_c^{(k)} \biggr) \sim 
\left(  \sum_k p_k^n  \right)  N^{ \beta \sum_c \big[ \#( \sigma_c) - n \big] }  \;,
\end{equation}
but the sum \eqref{eq:n-linear-sep}  over all the attributions of $k$ can be significantly smaller, as the terms with distinct $k$s can be of the same order of magnitude and do not have a definite sign. 
Such terms can be rendered inoffensive under various assumptions:
\begin{itemize}
 \item if all the $\rho_c^{(k)}$s are commuting, then all the terms in  \eqref{eq:n-linear-sep} are positive.
 \item if  $ {\rm Rank} \big( \rho_c^{(k_i)}\rho_c^{(k_j)} \big)  =  N^{\beta'}$ with $\beta'<\beta$, then all the terms with distinct $k$s are smaller in scaling than the ones with equal $k$s.
\end{itemize}
In these cases, \eqref{eq:lower-bound-separable} is a lower bound on $\Tr_\bsig(\rho)$, and we can conclude that the asymptotic dependency in $N$ is precisely $N^{ \beta \sum_c \big[ \#( \sigma_c) - n \big] } $. 

\subsubsection{States with asymptotically entangled scalings}
\label{subsec:Gaussian}

\paragraph{Pure states.} 
The pure density matrix $\rho=  \Ket{\Psi}\Bra{\Psi}$ with $\Ket{\Psi}\in \cH$
reads in the standard basis:
\[
 \Braket{j_1\dots j_D| \rho| i_1 \dots i_D} 
  = \Psi_{j_1\dots j_D} \Psi^*_{i_1\dots i_D}  \;,
\]
where $*$ denotes complex conjugation. We note that $\rho$ does not connect the  outgoing indices $i^1_s, \ldots, i^D_s$ with the incoming ones $j^1_s, \ldots, j^D_s$ of the same $s$:
\[
 \Tr_{\bsig}(\rho) =\sum_{\rm{all\ indices}} 
 \left( \prod_{s=1}^n \Psi^*_{i^1_s\ldots i^D_s } \right) 
\left( \prod_{c=1}^D    \prod_{s=1}^n \delta_{i^c_s,j^c_{\sigma_c(s)}}  \right) 
 \left( \prod_{s=1}^n \Psi_{j^1_s\ldots j^D_s } \right) 
\; ,
\]
hence \emph{pure states must have} $\beta=0$ as there is no way to obtain an asymptotic scaling depending on the number of cycles of $\sigma_c$.

\paragraph{A 1-uniform state.} A pure density matrix $\rho=  \Ket{\Psi}\Bra{\Psi}$ with 
$\Ket{\Psi} \in \cH = \otimes_c \cH_c$  is said to be \emph{1-uniform} if, after  partially tracing all the subsystems except $\cH_c$, one gets the maximally mixed state:  
\be
\label{eq:1-uniform}
 \sum_{i_2,\dots, i_D} 
  \Braket{j_1 \,  i_2 \dots i_D| \rho| i_1 \, i_2 \dots i_D}  = \frac 1 N \delta_{i_1  j_1}.
\ee
A pure 1-uniform density matrix is considered to be \emph{maximally entangled} \cite{uniform1, uniform2, uniform3}. 

We decompose each $\cH_c$ as a tensor product of $D-1$ Hilbert spaces, each of them with dimension $N^{1/(D-1)}$. We denote this by $\cH_{c}=\cH_{c(1)}\otimes \ldots \widehat{\cH_{c(c)} } \ldots \otimes  \cH_{c(D)}$, where the hat signifies an element missing in a list. The canonical basis of $\cH_c$ is a tensor product basis:
\[ 
\Ket{i_c} = \bigotimes_{ c', \; c'\neq c } \Ket{i_{c(c')}} \equiv
 \Ket{  i_{c(1)} \ldots \widehat{ i_{c(c)} } \dots i_{c(D)} }  \;,
\]
that is, every index is split into a multi-index $i_c \equiv (i_{c(1)},  \ldots  \widehat{ i_{c(c)} },  \dots, i_{c(D)})$, and each sub-index $i_{c(c')}$ ranges from 1 to $N^{1/ {(D-1)}}$. 
We consider the pure density matrix:
\[
\begin{split}
& \rho=  \Ket{\Psi} \Bra{\Psi}  \;,  
\qquad
  \Ket{\Psi} = \Psi_{i_1\dots i_D} \bigotimes_c \Ket{i_c}
   =   {N^{-\frac{D}4}} \big(  \prod_{c_1<c_2}  \delta_{i_{c_1(c_2)} i_{c_2(c_1)}}   \big)   \; \bigotimes_{\substack{ { c,c'} \\ {c' \neq c} } } 
   \Ket{i_{c(c')}} \;.
\end{split}
\]

To compute the value of a trace-invariant evaluated on $\rho$, we note that:
\[
 \Tr_{\bsig} (\rho) =  N^{-n\frac{D}{2} }
 \sum_{\rm{all\ indices}} 
 \left( \prod_{s=1}^n 
   \prod_{c_1<c_2} \delta_{ i^s_{c_1(c_2)} i^s_{c_2(c_1)} }  
\delta_{ j^s_{c_1(c_2)} j^s_{c_2(c_1)} }  
\right) 
\prod_{   c\,\in\{1,\ldots, D\}  } \left( \prod_{s=1}^n \prod_{ c'\neq c} \delta_{ i^s_{c(c')} j^{\sigma_c(s)}_{c(c')} }   \right) \; ,
\]
and an index $i^s_{c_1(c_2)}$ is identified with $i^s_{c_2(c_1)}$ which is identified with $j^{\sigma_{c_2}(s)}_{c_2(c_1)} $ which is identified with $j^{\sigma_{c_2}(s)}_{c_1(c_2)} $ which is identified to  $i^{\sigma_{c_1}^{-1}\sigma_{c_2}(s)}_{c_1(c_2)} $, and so on. Consequently, we get a free sum over an index $i_{c_1(c_2)}$ for every cycle of $\sigma_{c_1}^{-1} \sigma_{c_2}$. As the range of $i_{c(c')}$ is $N^{1/(D-1)}$, we obtain:
\[
\Tr_{\bsig} (\rho) = N^{\frac 1 {D-1} \sum_{c_1<c_2} \big[ \#(\sigma_{c_1}\sigma_{c_2}^{-1}) - n \big] } \;.
\]
Moreover, the partial trace of $\rho$ over all the subsystem except $1$ is:
\[
 N^{-\frac{D}{2}} \sum_{\{ i_{c(c')},j_{c(c')} \}_{c>1} } \big( \prod_{c_1<c_2} \delta_{ i_{c_1(c_2)} i_{c_2(c_1)} } \delta_{ j_{c_1(c_2)} j_{c_2(c_1)} }  \big) \big( \prod_{c>1,c'\neq c} \delta_{i_{c(c')}j_{c(c')}}  \big) = \frac{1}{N} \prod_{c\neq 1} \delta_{i_{1(c)} j_{1(c)}} \;,
\]
as there are $\binom{D-1}{2}$ free sums over indices of size $N^{1/(D-1)}$. Thus $\rho$ satisfies \eqref{eq:1-uniform} and is 1-uniform.

\subsubsection{Interpolation: states with $\beta + \epsilon(D-1) \le 1$}
\label{subsub:interpolation}

Writing $N =N^{1-\beta - \epsilon(D-1) } N^{\beta} (N^\epsilon)^{D-1}$, we proceed similarly to the previous example and split each Hilbert space  $\cH_c$ as a tensor product
$\cH_c = \cH_c^{1} \otimes \cH_c^s \otimes \cH_c^{e} $, and furthermore we split $\cH_c^e$ as 
$\cH^e = \bigotimes_{c', c\neq c'}\cH^e_{c(c')}$. We fix the dimensions of the various Hilbert spaces to:
\[
\dim(\cH_c^1)=N^{1-\beta - \epsilon(D-1) } \;,\qquad
\dim(\cH_c^s) = N^\beta \;, \qquad 
\dim(\cH_{c(c')} ) = N^{\epsilon} \; .
\]
Such a splitting is possible only if $ \beta + \epsilon(D-1) \le 1 $. The  standard basis in $\cH_c$ is a tensor product basis:
\[
\Ket{i_c} =\Ket{p_c} \otimes \Ket{X_c} \bigotimes_{c',\, c'\neq c} \Ket{i_{c(c')}}  \equiv \Ket{ p_c \, X_c \,  i_{c(1)}  \ldots \widehat{i_{c(c)}} \ldots  i_{c(D)}} \;,
\]
that is, we split
$i_c \equiv ( p_c \, X_c \, i_{c(1)} \ldots \widehat{i_{c(c)}} \ldots  i_{c(D)} ) $ where $p_c$ ranges from 1 to $N^{1-\beta - \epsilon(D-1)}$, $X_c$ from 1 to $N^{\beta}$ and each $i_{c(c')}$ from 1 to $N^{{\epsilon}}$.

The idea is to chose a density matrix which is a rank 1 projector on $\bigotimes_c \cH_c^1$, is maximally mixed on $\bigotimes_c \cH_c^s$,  and is the 1-uniform state of Sec. \ref{subsec:Gaussian} on $\bigotimes_c \cH_c^e$, that is, $ \rho = \; N^{-\beta D-\epsilon \binom{D}{2} }  \; \rho^1 \otimes \rho^s \otimes \rho^e $:
\[
\begin{split}
& \rho^1 = \bigotimes_c \Ket{\Psi^1_c} \Bra{\Psi^1_c} 
 \;, \qquad \Ket{\Psi^1_c} \in \cH_c^1 \;,\crcr
 & \rho^s = \bigotimes_{c} \mathbb{1}^s_c \;,  \qquad 
 \mathbb{1}^s_c = \sum_{X_c} \Ket{X_c} \Bra{X_c} \;, \quad \Ket{X_c} \; \text{basis in} \; \cH^s_c \crcr
 & \rho^e = \Ket{\Psi^e} \Bra{\Psi^e} \;,\qquad \Ket{\Psi^e} = 
 \big( \prod_{c_1<c_2}  \delta_{i_{c_1(c_2)} i_{c_2(c_1)}} \big) \; \bigotimes_{\substack{ { c,c'} \\ {c\neq c'} } } 
   \Ket{i_{c(c')}} \;  .
\end{split}
\]
Using the sections~\ref{subsub:example-micro},~\ref{subsub:example-separable} and~\ref{subsec:Gaussian}, we get:
\[
\Tr_{\bsig} (\rho) =
N^{-n\beta D - n\epsilon \binom{D}{2}}  \;
 \Tr^1_{\bsig}(\rho^1) \Tr^s_{\bsig}(\rho^s) 
 \Tr^{e}_{\bsig}(\rho^e)
 =
N^{ \beta\sum_{c=1}^D \big[ \#(\sigma_c)  -n \big]+ \epsilon\sum_{c_1<c_2} \big[  \#(\sigma_{c_1}\sigma_{c_2}^{-1}) -n\big]  } \;  ,
\]
where $\Tr^1$, $\Tr^s$ and $\Tr^e$ denote the traces over $\bigotimes_c \cH_c^1$, $\bigotimes_c \cH_c^s$, and 
 $\bigotimes_c \cH_c^e$ respectively.
 
 \

The (macroscopic) family of states with $1= \beta +\epsilon(D-1)$ should be compared to other one-parameter families of states interpolating between a maximally entangled state and the maximally mixed state, such as the isotropic state or the Werner state for $D=2$ (see \cite{LU-1}, Sec.~VI.9), which are known to be separable up to some threshold value of the parameter.

\subsection{Asymptotic moments of the tensor HCIZ integral}
\label{sub:moments}

One can try to detect the entanglement using the moments of the tensor HCIZ integral, as done in \cite{random-meas-1, random-meas-2, random-meas-3, random-meas-4, random-meas-5} for systems with finite $D,N$. From Prop.~2.5 and Cor.~2.4 of \cite{CGL}, we get:
\[
\begin{split}
\left\langle [ \Tr (AUBU^* ) ]^n\right\rangle_U& =N^{-2nD }\sum_{\bsig, \btau \in \bS_n}     \tr_{\bsig}(\al) \, \tr_{\btau^{-1}}(\bl) \crcr & 
 \qquad \times N^{\sum_{c=1}^D\#(\sigma_c\tau_c^{-1}) + s_A(\bsig) +   s_B(\btau)}  \prod_{c=1}^D\M(\sigma_c\tau_c^{-1})
  \bigl(1+O(N^{-2}))  \; .
 \end{split}
\]
\emph{The large $N$ asymptotics of the moments is insensitive to the scaling hypothesis} because:
\[
 \sum_c | \Pi(\sigma_c \tau_c^{-1}) | + \beta_A \sum_c \Pi(\sigma_c) + \epsilon_A\sum_{c_1<c_2} \Pi(\sigma_{c_1} \sigma_{c_2}^{-1} ) 
  + \beta_B \sum_c \Pi(\tau_c) + \epsilon_B\sum_{c_1<c_2} \Pi(\tau_{c_1} \tau_{c_2}^{-1} ) \;,
\]
is maximized trivially for $\sigma_c =\tau_c = \id$, $\forall c$, as each term is maximized individually. 

 \begin{Th}
 \label{th:moments}
For $\epsilon_A,\epsilon_B,\beta_A,\beta_B \ge 0$, taking the normalized scaling ansatz:
 \[
 s_A(\bsig)= 
 \beta_A \sum_{c=1}^D \left[  \# (\sigma_c) -n \right] + \epsilon_A \sum_{ 1\le c_1<c_2  \le D} \left[ \# ( \sigma_{c_1} \sigma_{c_2}^{-1}) -n \right] \;,
 \]
and similarly for $B$, the moments of the tensor HCIZ integral obey:
 \[
 \lim_{N\rightarrow + \infty} \left \langle  \left[  \Tr(AUBU^*) \right]^n\right\rangle_U = N^{-nD} \left(\tr(a)\tr(b)\right)^n \;.
 \]
\end{Th}

\subsection{Asymptotic regimes for the cumulants in $D=1$}
\label{subsec:D1}

In $D=1$, there are no $\epsilon$ terms and no entanglement to speak of. We include it here, as it provides a good introduction to the $D\ge 2$ case. We take the opportunity to gather and generalize some results scattered throughout the literature.

The asymptotic scaling depends only on $\beta_{A}$ and $\beta_B$, and we can assume $\beta_A\le \beta_B$.
We include non-realistic  scaling ans\"{a}tze for the traces: $\beta < 0$ corresponds to 
matrices of rank smaller than 1, $\beta>1$ to rank larger than $N$, see Sec.~\ref{subsub:example-separable}. Although there is no sequence of matrices with such behavior, one can still derive a  formal large $N$ limit for the cumulants.
Note that in $D=1$, we have $\Tr_{\tau^{-1}} (B)= \Tr_{\tau} (B) $.

\begin{figure}[!ht]
\centering
\includegraphics[scale=0.6]{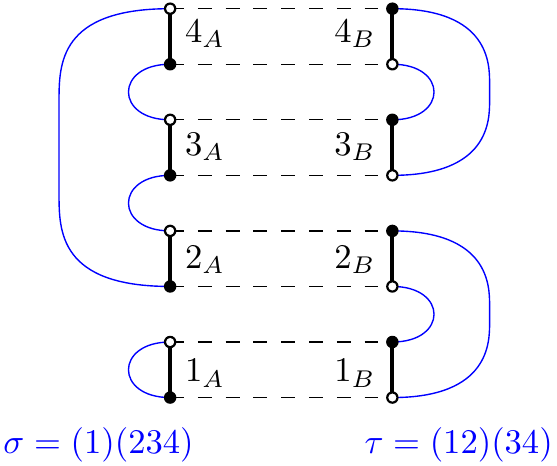}
\qquad
\includegraphics[scale=0.6]{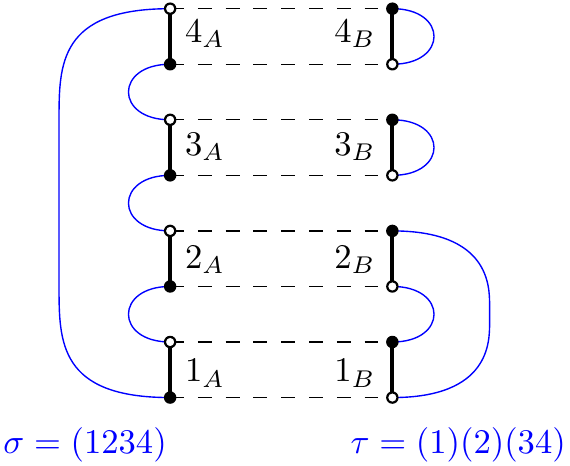}
\qquad
\includegraphics[scale=0.6]{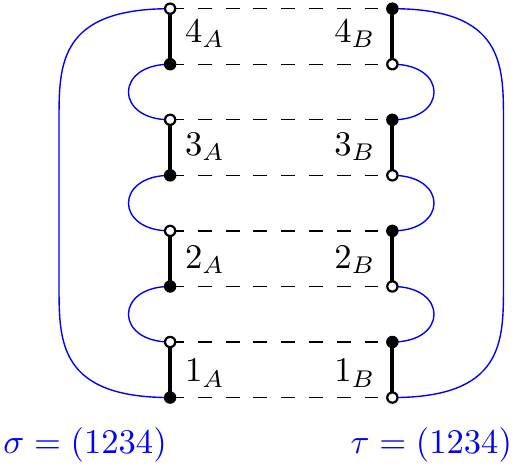}
\\[+3ex]
\includegraphics[scale=0.6]{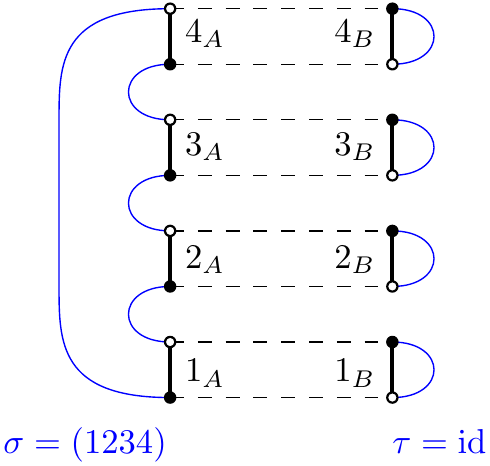}
\hspace{1.1cm}
\includegraphics[scale=0.6]{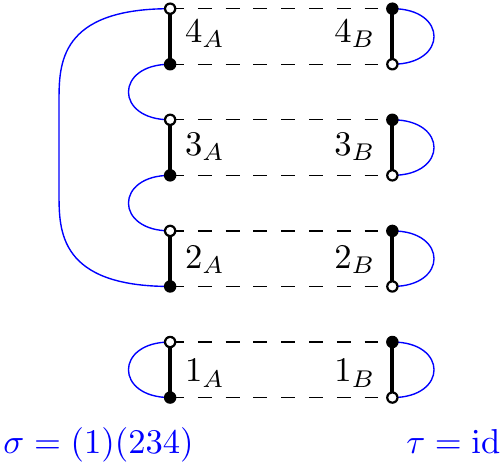}
\hspace{1.5cm}
\includegraphics[scale=0.6]{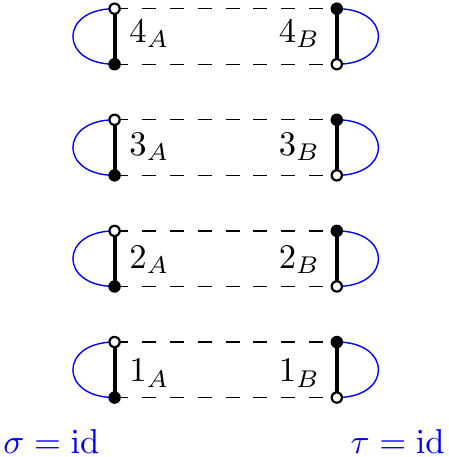}
\hspace{0.5cm}
\caption{From left to right and top to bottom examples for the regimes \ref{D=1item1}, \ref{D=1item2},
\ref{D=1item3}, \ref{D=1item4}, \ref{D=1item5}, \ref{D=1item6} in $D=1$.}
\label{fig:D1-ex}
\end{figure}

\begin{Th} 
\label{th:RegimesD1}
For any $\beta_A, \beta_B \in \bR$, if $\Tr_{\sigma} (A)  \sim N^{\beta_A \#(\sigma)} \tr_\sigma(\al)$ and similarly for $B$, then the limit:
\be
\label{eq:limitD1}
 \lim_{N\rightarrow +\infty} \frac 1 {N^\ddel} C_n\bigl(N^\ggam \Tr (AUBU^* )\bigr)  \;,
\ee
of the cumulants in  \eqref{eq:WC-norm} exists and is non-trivial  if and only if: 
\be
\nonumber
\ggam = 3 - \max(\beta_A, 1) - \max( \beta_B, 1) \quad \textrm{and}\quad \ddel = \beta_A+\beta_B + 2 - \max(\beta_A, 1) - \max( \beta_B, 1).
\ee
Furthermore, the graphs $(\sigma,\tau)$ that contribute to 
 \eqref{eq:limitD1} are (see Fig.~\ref{fig:D1-ex} for examples): 
\begin{enumerate}
\item\label{D=1item1} for $\beta_A=\beta_B=1$, the non-necessarily connected planar graphs:
\[
\lim_{N\rightarrow +\infty} \frac 1 {N^2} C_n\bigl(N \Tr (AUBU^* )\bigr)  =  \sum_{  \substack{{\sigma,\tau\,\in S_n\textrm{ s.t.}}\\{(\sigma , \tau) \textrm{ planar} }}  }
    \tr_{\sigma}(\al) \,  \tr_{\tau} (\bl) \,  f \bigl[\sigma, \tau\bigr]\;.
\]
\item\label{D=1item2} for $\beta_A<\beta_B=1$, the planar graphs such that $\sigma$ is a cycle of length $n$. In this case,  $\tau$ is a  non-crossing permutation on $\sigma$, and denoting by $\sigma_0 = (12\ldots n)$, we have:
\[
\lim_{N\rightarrow +\infty} \frac 1 {N^{\beta_A +1}} C_n\bigl(N \Tr (AUBU^* )\bigr)  =  (n-1)!\; \tr(
\al^n)\hspace{-0.1cm}\sum_{  \substack{{\tau\textrm{ non-cross.}}\\{\textrm{on }\sigma_0}}  }
     \tr_{\tau} (\bl) \, \M \bigl(\sigma_0\tau^{-1}\bigr).
\]
\item\label{D=1item3} for $\beta_A\le \beta_B<1$, the graphs such that $\sigma$ and $\tau$ are the same cyclic permutation:
\[
\lim_{N\rightarrow +\infty} \frac 1 {N^{\beta_A + \beta_B}} C_n\bigl(N \Tr (AUBU^* )\bigr)  =  (n-1)!\,   \tr(\al^n) \,  \tr (\bl^n)\;.
\]
\item\label{D=1item4} for $\beta_A < 1 <\beta_B$, the graphs such that $\sigma$ is a cycle of length $n$ and $\tau$ is the identity:
\[
\lim_{N\rightarrow +\infty} \frac 1 {N^{\beta_A + 1}} C_n\bigl(N^{2-\beta_B} \Tr (AUBU^* )\bigr)  = (-1)^{n-1} \frac {(2n-2)!}{n!}\,   \tr(\al^n) \,  \tr (\bl)^n   .
\]
 \item\label{D=1item5} for $\beta_A = 1 <\beta_B$, the graphs such that $\tau$ is the identity and $\sigma$ is any permutation:
\[
\lim_{N\rightarrow +\infty} \frac 1 {N^{2}} C_n\bigl(N^{2-\beta_B} \Tr (AUBU^* )\bigr)  =  \tr(b)^n \sum_{\sigma \in S_n} \tr_\sigma(\al) f[\sigma, \id].
\]
 \item\label{D=1item6} for $1 <\beta_A\le \beta_B$, the graphs such that  $\sigma=\tau=\id$:
\[
\lim_{N\rightarrow +\infty} \frac 1 {N^{2}} C_n\bigl(N^{3-(\beta_A + \beta_B)} \Tr (AUBU^* )\bigr)  =  
(-1)^{n-1} \frac{(3n-3)!}{(2n)!} \bigl[ 2  \tr(\al) \,  \tr (\bl)\bigr]^n \, .
\]
\end{enumerate}
\end{Th}
\proof
See Appendix \ref{sec:D1}.

\qed

The following pattern emerges: whenever $\beta<1$ for a matrix, the corresponding permutation is a cycle; whenever $\beta >1$, the corresponding permutation is the identity. We obtain combinatorially prolific regimes only for $\beta_A \le \beta_B \le 1$.

For $A$ microscopic  ($\beta_A=0$) and $B$ macroscopic  ($\beta_B=1$), the leading order graphs are non-crossing permutations (in the universality class of plane trees). For the symmetric macroscopic scaling $\beta_A = \beta_B=1$, the leading order graphs are planar maps, which form a richer universality class.

This theorem generalizes several results in the literature. 
Item \ref{D=1item1}, $\beta_A = \beta_B=1$ leading to planar graphs corresponds to the scaling of the original HCIZ integral \cite{HarishChandra, Itzyk-Zub}. The coefficients $f[\sigma,\tau]$ are related to double Hurwitz numbers in genus zero (see \cite{CGL}  and references therein). 

Item \ref{D=1item2}, $ \beta_A<\beta_B=1$  is the scaling considered by Zinn-Justin in \cite{ZJ} and subsequently by the first author in \cite{Collins03} to obtain alternative proofs and generalize to weaker hypothesis Voicolescu's result that independent random matrices are asymptotically free. With this assumption, the coefficients of the cumulants are that of the primitive of the R-transform. 

A particular case of item \ref{D=1item3}, $\beta_A \le \beta_B<1$ is considered in \cite{CHS} in the context of the study of sum and product operations of randomly rotated eigenvalues of fixed rank.
This is related to a notion of cyclic monotone free independence, that generalizes
the more well-known notion of monotone independence.
Note that the present manuscript is more general than the point of view of
\cite{CHS} because we also treat
matrices of small but non necessarily fixed rank, and have an additional continuum of parameters for the behaviour of the limiting moments. 

In Item \ref{D=1item5}, the HCIZ integral is formally (since the scaling ansatz involves matrices of rank larger than $N$) a generating function for the genus zero monotone single Hurwitz numbers $\vec H_0(\sigma)$ \cite{CGL}:
\[
f[\sigma, \id] =  (-1)^{\#(\sigma) - 1} \frac{\prod_{p\ge 1} d_p(\sigma) ! }{n!} \vec H_0(\sigma). 
\]
This reproduces the similar result obtained in \cite{HCIZ-Hurwitz-1} by assuming $\lim_{N \rightarrow \infty} \frac 1 N \Tr(B^k) = \delta_{1,k}$, which fixes $\tau=\id$ at leading order in $1/N$. 

 \subsection{Asymptotic regimes for the cumulants for $D\ge 2$}

 \begin{figure}[!ht]
\centering
\begin{subfigure}[]{0.4\textwidth}
\includegraphics[scale=0.4]{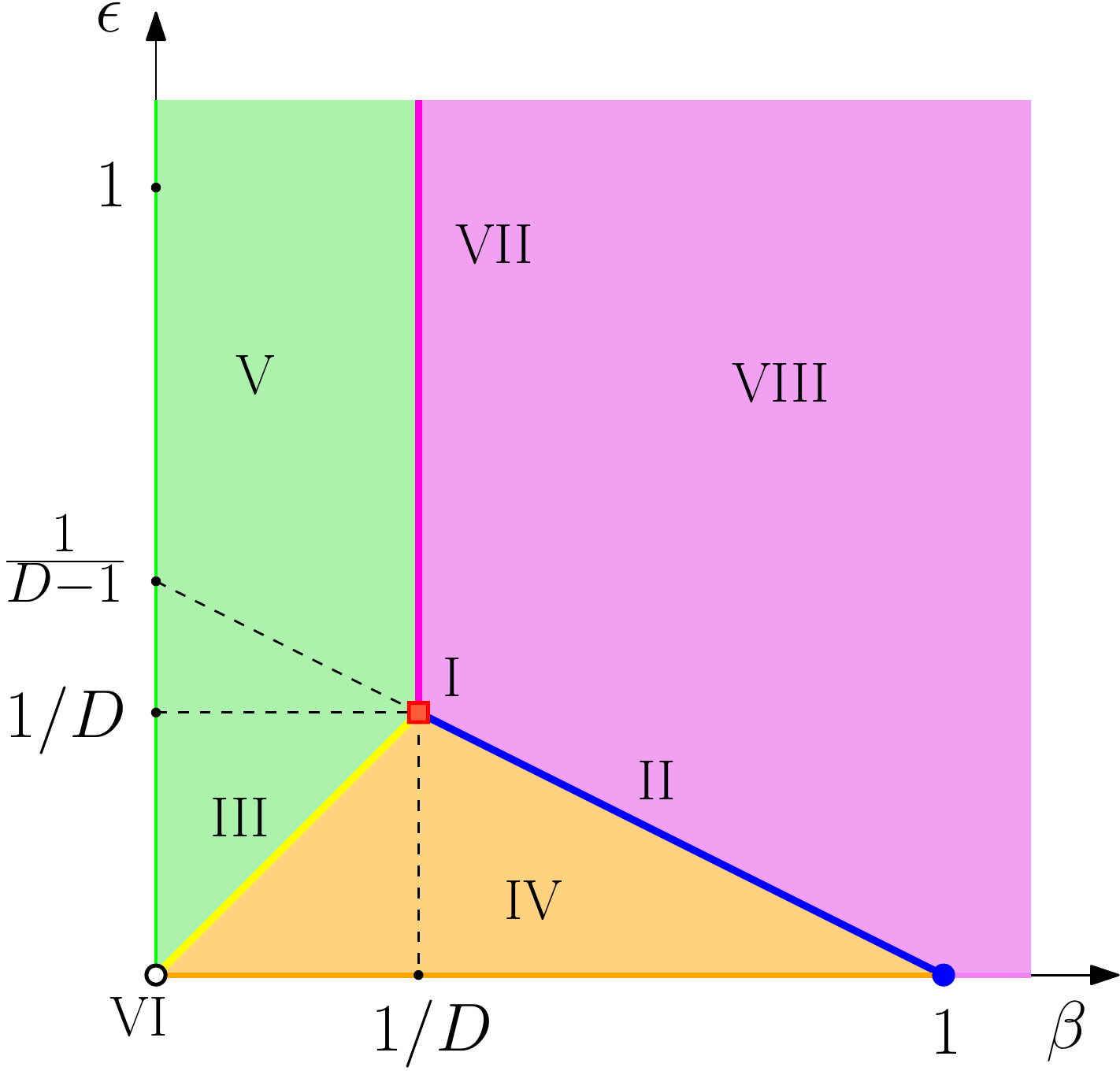} 
\caption{Microscopic A}
\label{fig:summary-results-0-micro}
\end{subfigure}
\hspace{0.15\textwidth}
\begin{subfigure}[]{0.4\textwidth}
\includegraphics[scale=0.4]{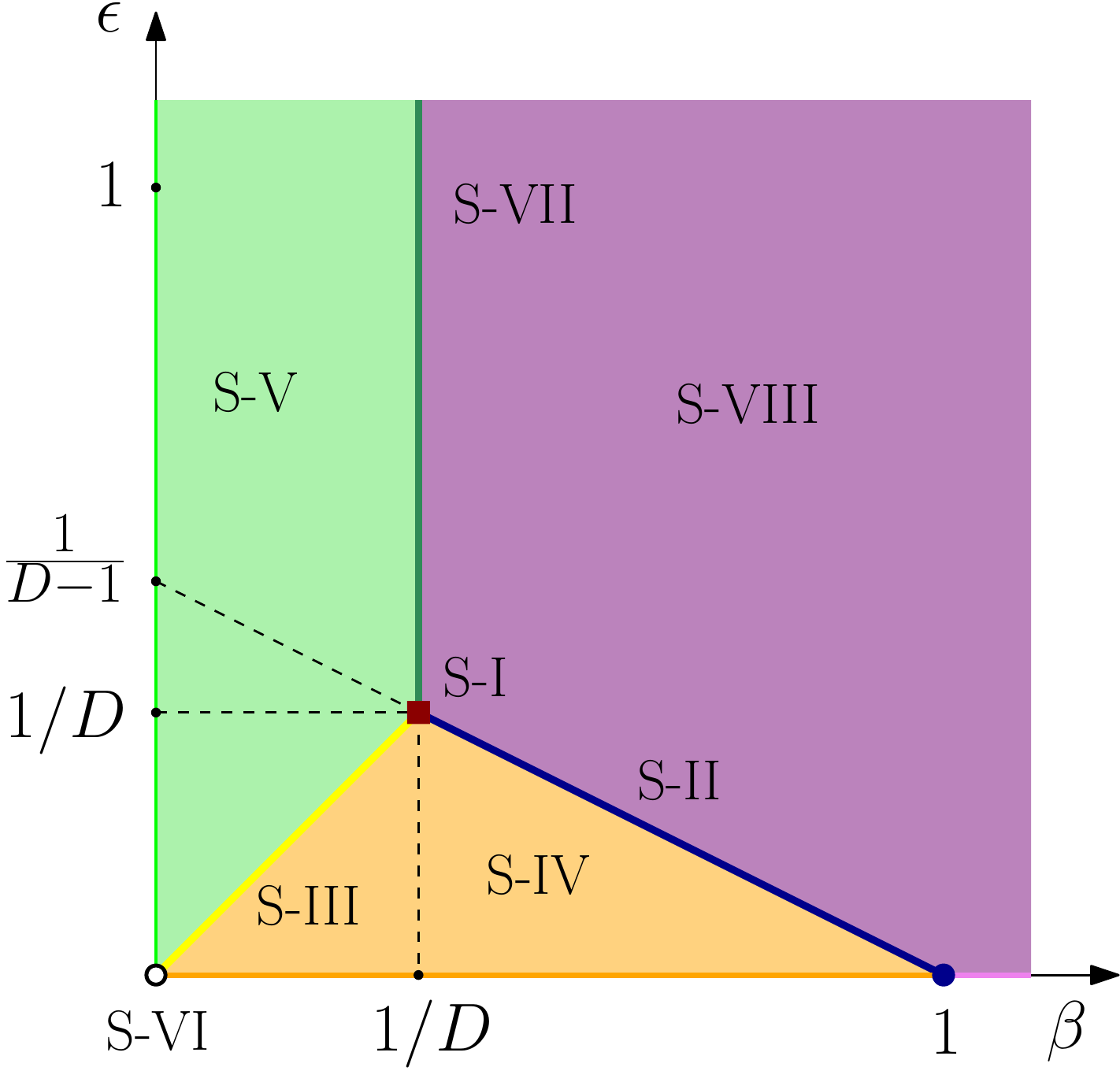}
\caption{Symmetric scalings}
\label{fig:summary-results-0-sym}
\end{subfigure}
\caption{The different regions of the $\beta-\epsilon$  diagram correspond to the different asymptotic regimes for $A$ microscopic  (left) and for symmetric scalings (right). 
}
\label{fig:summary-results-0}
\end{figure}

The asymptotic behavior of the cumulants turns out to be more interesting than that of the moments. Fig.~\ref{fig:summary-results-0} presents the 
asymptotic regimes we find for $A$ microscopic   and for symmetric scalings. 

 \subsubsection{Asymptotic regimes for $A$ microscopic  }
 \label{sec:result-micro}
For $A$ microscopic,  the general scaling ansatz is:
\[
\Tr_{\bsig}(A) \sim O(1) \;,\qquad \Tr_{\btau}(B) \sim N^{ \beta \sum_c \#(\tau_c) + \epsilon \sum_{ c_1<c_2  }  \# ( \tau_{c_1} \tau_{c_2})^{-1}  }\; \tr_{\btau}(b) \; .
\]
The resulting regimes are listed in our first main theorem, Theorem \ref{th:other-ent-microscopic-regimes}, and displayed in Fig.~\ref{fig:summary-results-0-micro}.
For all the regimes except \ref{micAentitemVII} and \ref{micAentitemVIII}, we can build states with the right asymptotic behavior using Section \ref{subsub:interpolation}. 
 
The combinatorially prolific regimes are \ref{micAentitemI}, \ref{micAentitemII} and \ref{micAentitemVII}:
\begin{itemize}
\item \underline{The macroscopic separable regime:} (\ref{micAentitemII} in Fig.~\ref{fig:summary-results-0-micro}) is obtained along the line  $\beta=1-\epsilon(D-1)>1/D $ and $\epsilon \ge 0$. It takes its name from its endpoint $\epsilon=0$ and $\beta=1$ which has asymptotically separable scaling and is called \emph{macroscopic} because (at $\epsilon=0$) it is obtained for a tensor product of $D$ matrices of full rank.  
 \item  \underline{The macroscopic boundary regime:} (\ref{micAentitemI} in Fig.~\ref{fig:summary-results-0-micro}) is the point $\beta=\epsilon= 1/D$. We call it \emph{boundary} as it is found on the boundary of  the   entangled regime \ref{micAentitemV} discussed below. 
\emph{ This regime \ref{micAentitemI} is 
richer than all regimes but \ref{micAentitemVI}}.
  \item  \underline{The hyper-macroscopic boundary regime:}
  (\ref{micAentitemVII} in Fig.~\ref{fig:summary-results-0-micro}) is the vertical line $\epsilon>\beta=1/D$. We have not been able to exhibit a sequence of tensors  $\{B\}_{N\to \infty}$ displaying  asymptotic scalings in this region, see the discussion in Sec.~\ref{sub:maxrank}. 
 \end{itemize}

The following regimes are not combinatorially prolific: 
  \begin{itemize}
\item \underline{The microscopic regime:}
(\ref{micAentitemVI} in Fig.~\ref{fig:summary-results-0-micro}) is the point $\epsilon=\beta=0$ and is realized by tensors whose trace-invariants do not scale in $N$, such as for instance 
tensor products of one dimensional projectors, Sec. \ref{subsub:example-micro}.

\item  \underline{The mesoscopic separable regime:}  
(\ref{micAentitemIV} in Fig.~\ref{fig:summary-results-0-micro}) gathers $0 \le \epsilon<\beta<1-\epsilon(D-1)$. We call it \emph{mesoscopic}, as at $\epsilon=0$ it is obtained for a tensor product of matrices that are neither rank one (finite) nor full rank\footnote{In $D=1$ the mesoscopic and microscopic regimes collapse (see Sec.~\ref{subsec:D1}).}. The regime \ref{micAentitemII} is richer than this regime \ref{micAentitemIV}.

\item  \underline{The mesoscopic boundary regime:} 
(\ref{micAentitemIII} in Fig.~\ref{fig:summary-results-0-micro}) is obtained for $0<\epsilon=\beta<1/D$. 
The regime  \ref{micAentitemI} is richer than this regime \ref{micAentitemIII}, which in turn  is richer than  \ref{micAentitemIV}.

\item  \underline{The entangled regime:}
(\ref{micAentitemV} in Fig.~\ref{fig:summary-results-0-micro}) is obtained for $\epsilon>0$, 
$\beta < \min ( 1/D , \epsilon)$ and:
\[
\lim_{N\rightarrow +\infty} 
\frac 1 {N^{\beta D} } 
C_n\Bigl( N^{D- \epsilon \frac{D(D-1)}2 }\Tr (AUBU^* )\Bigr)  =   (n-1)! \    \tr(\al^n) \,  \tr (\bl^n) \; .
\]
\emph{This regime is combinatorially trivial} and obtained for instance for a $1$-uniform state, see Sec.~\ref{subsec:Gaussian}.

\item  \underline{The hyper-macroscopic regime:}
(\ref{micAentitemVIII} in Fig.~\ref{fig:summary-results-0-micro}) is obtained for $\beta>\max(1/D , 1-\epsilon(D-1))$, and we have not been able to exhibit states with such asymptotic scalings.
\end{itemize}
 
\paragraph{Lower bound on the number of leading order graphs.} The graphs $(\bsig,
\btau)$ have quadrangles
of dashed and thick edges labeled from 1 to $n$. Due to relabeling, their number grows like $n!$, that is, super-exponentially, but in the expansion \eqref{eq:general-cumulants} the cumulant $C_n$ is divided by $1/n!$, which takes out this super-exponential growth.

All the regimes of the $\beta-\epsilon$ diagram are richer than the entangled regime \ref{micAentitemV}, 
which is combinatorially trivial, as it has only one leading order graph up to relabelling for each $n\ge 1$. Moreover:

\begin{Th} 
\label{th:lower-bound-number-LO}
For $r= $ \normalfont{I} to \normalfont{VIII}, let $\cN_{r}(D,n)$ be the number of leading order graphs for the cumulant $C_n$ in the regime $r$ of the $\beta-\epsilon$ diagram. Then  $\cN_{\mathrm{\ref{micAentitemV}}}(D,n) =  (n-1)!$, and for $r\neq \mathrm{\ref{micAentitemV}}$:
\[
\frac{\cN_r(D,n)}{\cN_{\mathrm{\ref{micAentitemV}} }(D,n)} \,\ge \frac 1 {nD+1} \binom{nD+1}{n} \sim_n c_D\, n^{-\frac 3 2} (d_D)^n.
\]
\end{Th}

\proof From Theorem \ref{th:other-ent-microscopic-regimes} we have that (see Sec. \ref{sec:Col-graphs} for the relevant definitions):
\begin{itemize}
 \item $\cN_{\mathrm{\ref{micAentitemV}}}(D,n)$ is the number of cycles over $n$ vertices, that in  $(n-1)!$
 \item for any other regime, the leading order graphs include at least one graph $(\bsig, \cdot)$ for any  connected $(D+1)$-melonic graph $\bsig$. There are $(n-1)! \frac{1}{nD+1} \binom{nD+1}{n}$ such $\bsig$ over $n$ edges of color $D+1$ labelled by $s\in\{1, \dots, n\}$, see e.g.~\cite{FLT}.
\end{itemize}  
  \qed

 \subsubsection{Asymptotic regimes for symmetric scalings}
  
For symmetric scalings:
\[
\begin{split}
 \Tr_{\bsig}(A) \sim & N^{ \beta \sum_c \#(\sigma_c) + \epsilon \sum_{ c_1<c_2  }  \# ( \sigma_{c_1} \sigma_{c_2} )^{-1}  }\; \tr_{\bsig}(a) 
\;, \crcr
 \Tr_{\btau}(B) \sim &  N^{ \beta \sum_c \#(\tau_c) + \epsilon \sum_{ c_1<c_2  } \# ( \tau_{c_1} \tau_{c_2})^{-1}  }\; \tr_{\btau}(b) \; ,
\end{split}
\] 
the regimes are listed in our second main theorem, Theorem \ref{th:other-sym-regimes}, and displayed in Fig.~\ref{fig:summary-results-0-sym}. 
The large $N$ regimes lie \emph{in the same regions} of the $\beta-\epsilon$ plane as the ones for  $A$ microscopic. 
  
The combinatorially prolific symmetric regimes are
\ref{symAentitemII} and \ref{symAentitemI}:
 \begin{itemize}
 \item \underline{The symmetric  macroscopic separable regime:} (\ref{symAentitemII} in Fig.~\ref{fig:summary-results-0-sym}) is obtained for the symmetric scalings $\beta=1-\epsilon(D-1)>1/D$ and $\epsilon \ge 0$ and is richer than the regime \ref{micAentitemII}.
 
 \item  \underline{The symmetric macroscopic boundary regime:} (\ref{symAentitemI} in Fig.~\ref{fig:summary-results-0-sym}) is obtained for $\beta=\epsilon= 1/D$ and is richer than the regime \ref{symAentitemII}. 
 We expect this regime \ref{symAentitemI} to be combinatorially richer than the regime \ref{micAentitemI}, however we have not been able to prove it. 
 \end{itemize}

The symmetric regimes  \ref{symAentitemIII}, \ref{symAentitemIV},  \ref{symAentitemV}, and \ref{symAentitemVI}  are \emph{identical} (have the same leading order graphs) to the corresponding regimes  \ref{micAentitemIII}, \ref{micAentitemIV},  \ref{micAentitemV}, and \ref{micAentitemVI} .
The regimes \ref{symAentitemI}, \ref{symAentitemII}, \ref{symAentitemVII} and \ref{symAentitemVIII} are new, but the last two are not combinatorially prolific and expected not to be realizable.

\subsection{Hyper-macroscopic regimes and degrees of freedom} 
 \label{sub:maxrank}

  
In Sec.~\ref{subsub:interpolation}, we have exhibited states for all $\beta+ \epsilon(D-1) \le 1$. The construction breaks down for  $\beta+ \epsilon(D-1) > 1$, because the Hilbert spaces $\cH_c$ do not have enough dimensions to be split\footnote{This is particularly transparent for $\epsilon =0$, in which case a scaling $\beta>1$ would  be realized by a tensor product states $\otimes_c\rho_c$ with $\rho_c$ of rank $N^{\beta}$.} as $N =N^{1-\beta - \epsilon(D-1)} N^{\beta} N^{\epsilon(D-1)}$.
This suggests that the line $\beta+\epsilon(D-1) = 1$ with $\epsilon, \beta \ge 0$ is the frontier at which the tensors $\rho$ maximize the number of degrees of freedom in the subsystems.

As will be shown in Theorem \ref{th:other-ent-microscopic-regimes} and Theorem \ref{th:other-sym-regimes}, this is supported by the overall scaling of the cumulants: in the regimes \ref{micAentitemIII} and 
\ref{micAentitemIV}, the overall scaling exponent is $\ddel = \beta + \epsilon(D-1)$, that is, $N^{\ddel}$ is exactly the expected number of degrees of freedom in a subsystem (in the entangled regime \ref{micAentitemV}, the scaling is less than the expected number of degrees of freedom). In the 
macroscopic regimes \ref{micAentitemI} and \ref{micAentitemII}, $\ddel$ reaches $1$ and the scaling factor is $N$, the dimension of the Hilbert space $\cH_c$. Beyond, in the hyper-macroscopic regimes \ref{micAentitemVII} and \ref{micAentitemVIII}, the overall scaling gets stuck to  its maximal value $N$. The same holds mutatis mutandis for symmetric scalings.

\newpage
 
\section{Combinatorial facts}
\label{sec:Col-graphs}

Our analysis of the large $N$ regimes of the cumulants 
relies on several combinatorial results, which we gather in this section.

\begin{figure}[h!]
\centering
\includegraphics[scale=0.6]{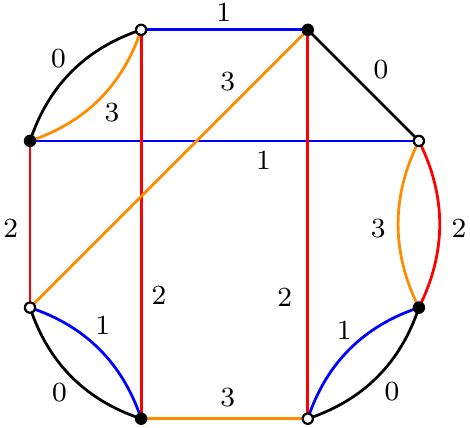}
\caption{A connected 4-colored graph.
}
\label{fig:Non-melo}
\end{figure}

\subsection{Colored graphs}
\label{subsec:GeneralitiesColGr}
\addtocontents{toc}{\protect\setcounter{tocdepth}{2}}

We will use several classical results on colored graphs (see \cite{Gurau:2011xp, Gurau:2019qag} and references therein), which we now review.
Let us fix $q$ an integer larger or equal to $3$.
A bipartite $q$-edge-colored graph $\G$, or $q$-colored graph for short, is a graph such that (see Fig.~\ref{fig:Non-melo} for an example):
\begin{itemize}
    \item the vertices are either black or white and all the vertices are $q$--valent,
    \item an edge connects a black and a white vertex and has a color $c\in\{1,\dots, q\}$, such that all the edges incident to a vertex have different colors. 
\end{itemize}

We denote by $V(\G)$ and $ K(\G)$ the numbers of vertices and connected components of $\G$. The number of edges of $\G$ is $q V(\G)/2 $.  The \emph{faces} of $\G$ are the bi-colored cycles of edges in the graph. We denote by $F^{(c,c')} (\G)$ the number of faces of color $(c,c')$, that is, the number of subgraphs obtained by keeping only the edges of color $c$ and $c'$ and we furthermore denote by:
\[
 F(\G)=\sum_{c<c'}F^{(c,c')} (\G) \qquad \textrm{ and }\qquad F_c(\G)=\sum_{c'\neq c}F^{(c,c')} (\G) \;,
\]
the total number of faces of $\G$, respectively the number of faces of $G$ containing the color $c$.

\paragraph{Degrees.}  The {\it degree}\footnote{Sometimes called the {\it reduced}-degree.} of a $q$-colored graph $\G$ is the non-negative half integer 
\cite{Gurau:2011xp,Gurau:2019qag}: 
\be\label{eq:degree}
\omega(\G)= (q-1) K(G) + \frac{(q-1)(q-2) }{4} V(\G) -   F(\G) \ge 0 \;.
\ee
The fact that the degree is non-negative is not trivial, and it is the basis of the $1/N$ expansion in random tensors 
\cite{Gurau:2011xp}. Besides being positive, the degree has another useful property  \cite{Gurau:2011xp,Gurau:2019qag}: denoting by $G^{\hat c}$ the graph obtained from $G$ by deleting the edges of color $c$, we have:
\be\label{eq:tare}
\omega(G)  \ge \omega(G) - \omega(G^{\hat c}) \ge \frac{1}{q-1} \omega(G)  \;.
\ee

The {\it $c$-degree} of a $q$-colored graph $\G$ is the non-negative half integer \cite{Gurau:2019qag, FLT}:
\be\label{eq:Ccdegree} 
\Omega_c(\G)= K(\G) + \frac{ q-2 }{2}  V(\G)  - F_c(\G) \ge 0 \;.
\ee

The degree and the $c$-degree are related by observing that the number of faces containing the color $c$ in $G$ can be written as the total number of faces minus the number of faces that do not contain it, that is, $F_c (G) =  F(G) - F(G^{\hat c})$ and simple algebra leads to: 
\be\label{eq:truc}
 \Omega_c(G) = (q-2) \big[ K(G^{\hat c}) - K(G) \big] 
  + \omega(G) - \omega(G^{\hat c}) \; .
\ee
As $G^{\hat c}$ must have at least a connected component for each connected component of $G$, this together with 
 \eqref{eq:tare} provides the proof that the $c$-degree is non-negative.

\subsubsection{The $q=3$ case}
\label{subsec:VanishingDeg3}

A 3-colored graph $\G$ can be canonically embedded in a surface by cyclically ordering counterclockwise the edges in the order $(0,1,2)$ around every white vertex 
and in the order $(0,2,1)$ around every black one. This promotes $G$ to a combinatorial map. The bi-colored cycles of $G$ are exactly the faces of the map, and the Euler relation reads $
F(\G) - V(\G)/2 = 2K(\G) - 2g(\G) $.
Thus for $q=3$, the degree is twice the genus.

\begin{figure}[h!]\centering
\includegraphics[scale=0.6]{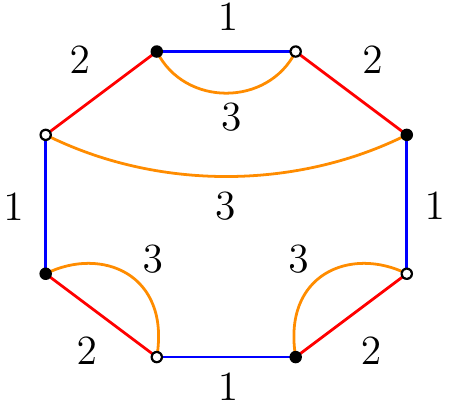}
\caption{A connected 3-colored graph with $\Omega_3(G)=0$.}
\label{fig:NCpairing}
\end{figure}

\paragraph{Non-crossing pairings.} For $q=3$, the graph $G^{\hat c}$ reduces to the faces of $G$ which do not contain $c$, that is, $K(G^{\hat c}) = F(G^{\hat c})$. In particular $\omega(G^{\hat c}) = 0$ always, and using  \eqref{eq:truc},
the $c$-degree of $G$ reads:
\[
\Omega_c(\G) = \big[ F(\G^{\widehat{c} }) - K(G) \big] + 2g(\G) \; .
\]
It follows that the $c$-degree vanishes if and only if each connected component of $G$ (see Fig.~\ref{fig:NCpairing}):
\begin{itemize}
 \item has exactly one face of color $(c_1,c_2)$ with $c_1,c_2\neq c$ for each connected component;
 \item in each connected component, the edges of color $c$ form a planar non-crossing pairing (a non-intersecting chord diagram).
\end{itemize}

We summarize this subsection in the following Proposition.

\begin{Prop}\label{thm:coloredgr3}
The degree of a $3$-colored graph $\G$ is twice its genus. The $c$-degree of a $3$-colored graph $\G$ vanishes if and only if each connected component of $\G$ is a non-crossing pairing, that is, $\G$ is planar and has a single face with colors $c_1,c_2\neq c$ per connected component.
\end{Prop}

\subsubsection{The $q>3$ case}
\label{subsub:vanishing-degrees}

\paragraph{Melonic graphs.} 

\begin{figure}[h!]\centering
\raisebox{3.5ex}{\includegraphics[scale=0.8]{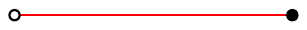}}
\hspace{2.5cm}
\raisebox{3.5ex}{\begin{tabular}{@{}c@{}}\small insertion\\$\rightleftarrows$\\ \small removal\end{tabular} }
\hspace{1.5cm}
\includegraphics[scale=0.8]{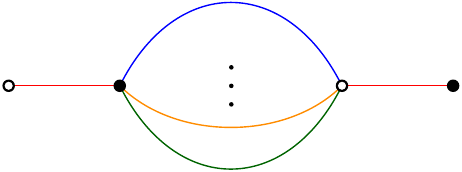}
\caption{Melonic insertion/removal.}
\label{fig:def-mel}
\end{figure}

\begin{figure}[h!]\centering
\raisebox{4ex}{\includegraphics[scale=0.7]{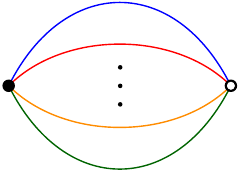}}
\hspace{2cm}
\includegraphics[scale=0.6]{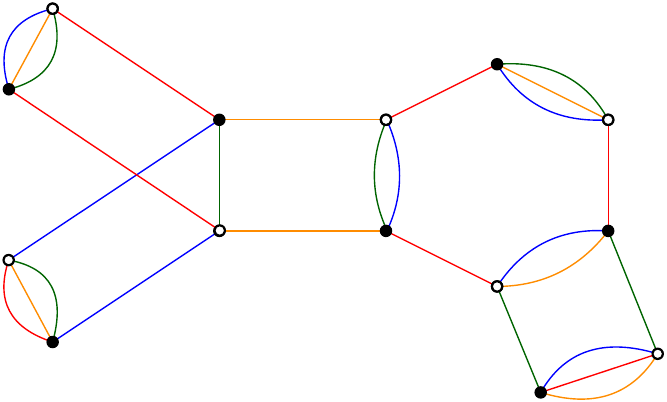}
\caption{The unique $q$-colored graph with two vertices (left), and a melonic graph (right). 
}
\label{fig:ex-mel}
\end{figure}

A \emph{melonic insertion} on an edge in a graph (see Fig.~\ref{fig:def-mel}) consists in splitting the edge and inserting two vertices connected by $q-1$ edges respecting the colors. A connected $q$-colored graph is called \emph{melonic} (right in Fig.~\ref{fig:ex-mel}) if it can be obtained by melonic insertions starting from the unique $q$-colored graph with two vertices (left in Fig.~\ref{fig:ex-mel}). 

\begin{Prop}\label{thm:coloredgr}
\cite{Gurau:2011xp,Gurau:2019qag} For $q>3$, the connected colored graphs of vanishing degree are the melonic graphs. From  \eqref{eq:tare} and \eqref{eq:truc}, the connected colored graphs of vanishing $c$-degree are the melonic graphs that stay connected when removing all the edges of color $c$. We will call such graphs \emph{$c$-melonic}.
\end{Prop}

\subsection{The graphs \TitleGsig}

The graph $\bsig$ associated to the trace-invariant $\Tr_{\bsig}(A)$  is a $(D+1)$-colored graph: the edges $(i^c_s, j^c_{\sigma_c(s)} ) $ inherit the color $c\in\{1, \dots, D\}$ of the index they represent, and we assign the color $D+1$ to the thick edges labeled $s$. The graph $\bsig$ has:
\begin{description}
\item[\it{Vertices--}] $2n$ vertices, $n$ white and $n$ black, labeled $s\in\{1,\ldots,n\}$. The white and the black vertices with label $s$ are linked by a thick edge of color $D+1$. 
\item[\it{Edges--}] $n (D+1)$ edges colored $\{1,\dots, D+1\}$.  The edges of color $1$ to $D$ track the indices of $A$ and encode the $\sigma_c$'s; the edges of color $D+1$ connect the vertices with the same label $s$.
\item[\it{Faces--}] the faces of  $\bsig$ fall in two categories:
     \begin{itemize}
         \item[-] $F^{(c_1,c_2)} (\bsig) =  \#(\sigma_{c_1}\sigma_{c_2}^{-1})$ faces with colors $(c_1,c_2)$ with $1\le c_1 <c_2 \le D$. 
         \item[-] $ F^{(c,D+1)}(\bsig) = \# (\sigma_c)$ faces with colors $(c,D+1)$ with $1\le c\le D$. 
     \end{itemize}    
     
\item[\it{Connect components--}] the graph $\bsig$ has $\lvert \Pi(\bsig)\rvert$ connected components.
     
\item[\it{Degree--}] the degree \eqref{eq:degree}  of $\bsig$ reads:
\be
\label{eq:degree-bsig}
 \omega(\bsig) = D\lvert \Pi(\bsig)\rvert +n\frac{D(D-1)}2 - \bigl(\sum_{c} \#(\sigma_c) + \sum_{c_1<c_2}\#(\sigma_{c_1}\sigma_{c_2}^{-1})\bigr) \ge 0 \;.
 \ee
\item[\it{$(D+1)$-degree--}]  the $c$-degree \eqref{eq:Ccdegree} of $\bsig$ for $c=D+1$ is:
\be
\label{eq:c-degree-bsig}
 \Omega_{D+1}(\bsig) = \lvert \Pi(\bsig)\rvert+n(D-1) - \sum_{c} \#(\sigma_c) \ge 0 \; .
\ee
\end{description}

\subsection{The graphs \TitleGamma}

The graph $(\bsig , \btau)$ of Section \ref{sec:notation} is a $(D+2)$-colored graph with:
\begin{description}
\item[\it{Vertices--}] $4n$ vertices: $n$ white and $n$ black coming from $A$, and $n$ white and $n$ black coming from $B$. They are labeled $s\in\{1, \dots, n\}$, and the four vertices labeled $s$ are connected into a quadrangle by thick edges of color $D+1$ and dashed edges of color $0$.

\item[\it{Edges--}] $2n (D+2) $ edges colored by $\{0,1,\dots, D+1\}$:
      \begin{itemize}
\item[-] the edges of color $0$ are dashed
and connect the $A$ vertices with the $B$ vertices with the same label $s$. Deleting the color-0  edges, we are left with the graphs $\bsig$ and $\btau^{-1}$.
\item[-] the edges with colors $c\in\{1,\dots, D+1\}$ are the ones of the graphs $\bsig$ and $\btau^{-1}$.
      \end{itemize}

\item[\it{Faces--}] the faces of  $(\bsig , \btau) $ fall in four categories:
     \begin{itemize}
\item[-] $ F^{(c_1,c_2)} (\bsig, \btau) =  \#(\sigma_{c_1}\sigma_{c_2}^{-1}) +    \#(\tau_{c_1}\tau_{c_2}^{-1})$ faces with colors $(c_1,c_2)$ with $1\le c_1 <c_2 \le D$. 
\item[-] $F^{(0,c)}(\bsig, \btau)=\#(\sigma_c\tau_c^{-1})$  faces with colors $(0,c)$ with $1\le c\le D$.
\item[-] $F^{(c,D+1)}(\bsig, \btau) = \# (\sigma_c) + \# (\tau_c)$ faces with colors $(c,D+1)$ with $1\le c\le D$.         \item[-] $F^{(0,D+1)}(\bsig, \btau) = n$ faces with colors $(0,D+1)$. 
     \end{itemize}    
\item[{\it Connect components--}] the graph $(\bsig , \btau)$ has $\lvert \Pi(\bsig, \btau)\rvert$ connected components. 
\end{description}

\subsection{The graphs $(\sigma_c , \tau_c)$ }
\label{subsec:Gammac}

\begin{figure}[!ht]
\centering
\includegraphics[scale=0.7]{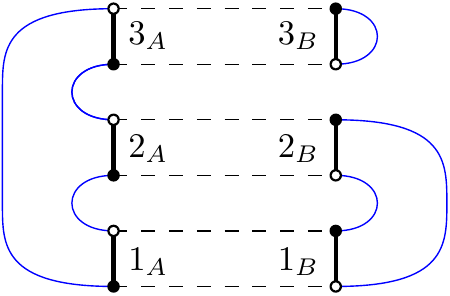}
\hspace{2cm}
\includegraphics[scale=0.7]{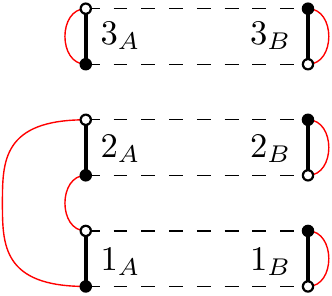}
\caption{Graphs $(\sigma_1 , \tau_1), (\sigma_2 , \tau_2)$ for the graph $(\bsig,\btau)$ in  Fig.~\ref{fig:Excolgraph}.  }
\label{fig:monocol-gr}
\end{figure}

For a given pair $(\bsig, \btau)$ and a color $c\in\{1,\ldots,D\}$, we consider the 3-colored graph $(\sigma_c , \tau_c)$ obtained from $(\bsig , \btau)$ by keeping only the edges of color 0, $c$, and $D+1$, as shown in  Fig.~\ref{fig:monocol-gr}.

This graph has $4n$ vertices, $6n$ edges, $F(\sigma_c, \tau_c)=\#(\sigma_c) + \#(\tau_c) +\#(\sigma_c\tau_c^{-1}) +n$ faces and $|\Pi(\sigma_c,\tau_c)|$ connected components (not to be confused with 
$|\Pi(\sigma_c\tau_c^{-1})| = \#(\sigma_c\tau_c^{-1}) $).
The Euler relation for $(\sigma_c,\tau_c)$ reads:
\be
\label{eq:EulerCharGammac}
\#(\sigma_c\tau_c^{-1}) = n+2\lvert \Pi(\sigma_c, \tau_c)\rvert - 2 g(\sigma_c, \tau_c)  - \#(\sigma_c) - \#(\tau_c) \;.
\ee

\begin{figure}[!ht]
\centering
\raisebox{2ex}{\includegraphics[scale=0.7]{mono1.pdf}}
\hspace{1cm}\raisebox{6ex}{\huge$\rightarrow$}\hspace{1cm}
\includegraphics[scale=0.7]{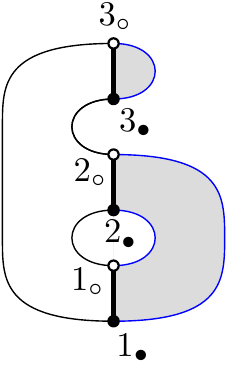}\hspace{0.5cm}
\caption{Obtaining a non-crossing pairing from the 3-colored graph $(\sigma_1 , \tau_1)$ in Fig.~\ref{fig:monocol-gr}.  }
\label{fig:mono-NCP}
\end{figure}

\paragraph{$(\sigma_c , \tau_c)$ and non-crossing permutations.} We will encounter below the case when 
the $3$-colored graph $(\sigma_c,\tau_c)$ is planar and moreover $\sigma_c$ is a cycle of length $n$. 
An example is the graph $(\sigma_1,\tau_1)$ in Fig.~\ref{fig:monocol-gr}, which we reproduce in 
Fig.~\ref{fig:mono-NCP} on the left. Without loss of generality, we take $\sigma_c=(12\ldots n)$.

The quadrangles made of thick edges of color $D+1$ and dashed edges of color $0$ are labeled $s\in \{1, \dots, n\}$. As $\sigma_c$ is a cycle, the graph $\sigma_c$ (which is a particular case of graph $\bsig$ for only one color) is a cycle of alternating thick edges and edges of color $c$ that visits the thick edges in the order $\{1, \dots, n\}$, see again  Fig.~\ref{fig:mono-NCP} on the left.

Collapsing the quadrangles labeled $s$ to thick edges by contracting the dashed edges, we conclude that if  $\sigma_c$ is a cycle of length $n$, $(\sigma_c,\tau_c)$ is planar if and only if the edges of $\tau_c^{-1}$ (or equivalently $\tau_c$) give a non-intersecting cord diagram on the cycle $\sigma_c$. 
In this case, $\Pi(\tau_c)$ is a non-crossing partition of 
the set $\{1,\ldots,n\}$ ordered according to $\sigma_c$. The cycles of $\tau_c$ correspond to 
the shaded regions on the right of Fig.~\ref{fig:mono-NCP}, and the non-crossing condition means that the shaded regions do not intersect.

Thus $(\sigma_c, \tau_c)$ is planar and $\#(\sigma_c)=1$ if and only if $\tau_c$ is a non-crossing permutation on the image of $\sigma_c$. Note that it is indeed $\tau_c$ which agrees with the ordering induced by $\sigma_c$, and not $\tau_c^{-1}$. 
This is summarized in the proposition below.

\begin{Prop}
\label{prop:non-cross-cc}
If $\Pi(\tau_c) \le \Pi(\sigma_c)$ and $(\sigma_c , \tau_c)$ is planar, $g(\sigma_c , \tau_c)=0$, then the permutation $\tau_c$ restricted to the blocks of $\Pi(\sigma_c)$ is non-crossing. We will denote this by $\tau_c \preceq \sigma_c$. $\preceq$ is a partial order relation between permutations and we say that $\tau_c$ is non-crossing on $\sigma_c$. We use the notation  $\btau \preceq\bsig$ if $\tau_c \preceq \sigma_c$ for all $c$.
Note that $\bsig \preceq \bsig$ for any 
$\bsig\in\bS_n$.
\end{Prop}

For example, on the right of Fig.~\ref{fig:monocol-gr}, we have $\tau_2\preceq \sigma_2$. The following simple properties of the graph $(\sigma_c,\tau_c)$ will be useful.

\begin{Prop} 
\label{lem:supports-included}
We have 
$\sum_c g(\sigma_c, \tau_c)\ge 0$ and $\sum_c \bigl(|\Pi(\sigma_c)| - \lvert \Pi(\sigma_c, \tau_c)\rvert \bigr) \ge 0$ with equality if and only if $\btau \preceq \bsig$, in which case $|\Pi(\bsig,\btau)| = \Pi(\bsig)$. If moreover $\Pi(\tau_c) = \Pi(\sigma_c)$ for all $c$, then $\btau = \bsig$. In detail:
\begin{itemize}
 \item $|\Pi(\sigma_c)| = \#(\sigma_c)\ge \lvert \Pi(\sigma_c, \tau_c)\rvert$
with equality if and only if $\Pi(\tau_c) \le \Pi(\sigma_c) $.
\item If for all $c\in\{1,\ldots,D\}$ we have $|\Pi(\sigma_c)| =  \lvert \Pi(\sigma_c, \tau_c)\rvert$, then $  \Pi(\bsig, \btau) =  \Pi(\bsig)$.
\item  If $|\Pi(\sigma_c)| =|\Pi(\tau_c)|  =\lvert \Pi(\sigma_c, \tau_c)\rvert$ and $g(\sigma_c, \tau_c)=0$, then $\sigma_c=\tau_c$.
\end{itemize}

\end{Prop}
\proof
The first item is trivial, as $ \Pi(\sigma_c, \tau_c) 
= \Pi(\sigma_c)\vee \Pi(\tau_c)$. The second one follows by observing that if $\Pi(\tau_c) \le \Pi(\sigma_c)$, then $\Pi(\sigma_c,\tau_c) =\Pi(\sigma_c) $ and 
$ \Pi(\bsig, \btau)= \bigvee_{c=1}^D \Pi(\sigma_c,\tau_c)  = \Pi(\bsig)$.

Finally, if $|\Pi(\sigma_c)| =|\Pi(\tau_c)|  =\lvert \Pi(\sigma_c, \tau_c)\rvert$, then $\Pi(\sigma_c) = \Pi(\tau_c)$ and therefore any component of $(\sigma_c , \tau_c)$ consists in exactly one cycle of $\sigma_c$ and one cycle of $\tau_c^{-1}$. If furthrmore $g(\sigma_c,\tau_c)=0$, then by Prop.~\ref{prop:non-cross-cc} the permutation $\tau_c$ is non-crossing on $\sigma_c$, which is only possible if $\sigma_c=\tau_c$.

\qed

\subsection{Auxiliary non-negative numbers}
\label{subsec:min-conn-gr}
For any graph $(\bsig,\btau)$, let us define:
\be
\label{eq:def-DeltaK}
\DeltaC(\bsig, \btau) = n(D-1)  + \lvert \Pi(\bsig, \btau)\rvert - \sum_{c=1}^D \, \lvert \Pi(\sigma_c, \tau_c)\rvert \; .
\ee
Observe that if $\btau \preceq \bsig$, then $\DeltaC(\bsig, \btau) = \Omega_{D+1}(\bsig)$.

\begin{Prop}
\label{lem:DeltaK-pos}
For any $\bsig, \btau \in \bS_n$, we have $
\DeltaC(\bsig, \btau) \ge 0$.
\end{Prop}
\proof 

\begin{figure}[!ht]
\centering
\includegraphics[scale=0.5]{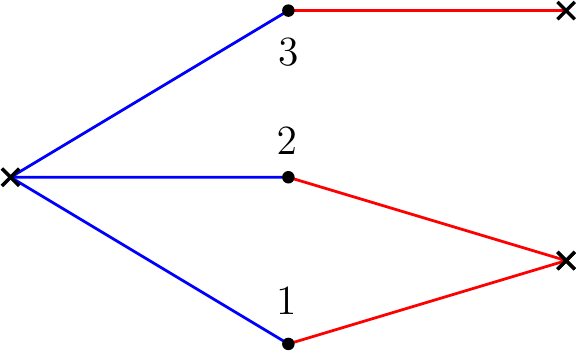}
\caption{The graph $G_{_\Delta}(\bsig, \btau)$ for the example of Fig.~\ref{fig:Excolgraph}.
}
\label{fig:Gconn}
\end{figure}

We define the abstract bipartite graph $G_{_\Delta}(\bsig, \btau)$ (see Fig.~\ref{fig:Gconn} for an example) having:
\begin{itemize}
 \item[-] $\sum_{c=1}^D |\Pi(\sigma_c,\tau_c)|$ cross vertices, one for each connected component of each $(\sigma_c , \tau_c)$. The cross vertices inherit the color $c$. 
 \item[-] $n$ round vertices, one for each quadrangular face with colors $(0, D+1)$ of $(\bsig , \btau)$. These vertices inherit the label $s\in\{1,\dots, n\}$ of the quadrangles.
 \item[-] each round vertex $s$ is connected, for all $c$, by an edge of color $c$ to the  cross vertex of color~$c$ corresponding to the connected component of 
 $(\sigma_c,\tau_c)$ to which it belongs. 
\end{itemize}

The important remark is that $(\bsig , \btau)$ and $G_{_\Delta}(\bsig, \btau)$ have the same number of connected components: indeed,
two round vertices in 
 $G_{_\Delta}(\bsig, \btau)$ are incident to the same cross vertex with color $c$ if and only if the corresponding quadrangles in $(\bsig,\btau)$ are connected by a path of edges of color $c$ (be it edges of $\bsig$ or of $\btau^{-1}$). Consequently, two  
 round vertices are connected by a simple path in $G_{_\Delta}(\bsig, \btau)$ if and only if the corresponding quadrangles are connected by a simple path in $(\bsig,\btau)$\footnote{A simple path is a sequence of edges such that two consecutive edges share a vertex, and all the vertices in the sequence are distinct}.

The number of excess edges of $G_{_\Delta}(\bsig, \btau)$ is:
$$ nD - \bigg( n + \sum_{c=1}^D\lvert \Pi(\sigma_c, \tau_c)\rvert \bigg) + \lvert \Pi(\bsig, \btau)\rvert \ge 0 \;.$$
which proves the proposition.
The bound is saturated when $G_{_\Delta}(\bsig, \btau)$ is a forest.

\qed

For any $(\bsig,\btau)$, the graph  $G_{_\Delta}(\bsig, \btau)$ is bipartite: every edge connects a cross vertex with a round vertex. For every color $c$, any round vertex is connected by one edge of color $c$ to a cross vertex of color $c$. It follows that the round vertices are $D$ valent, while the valency of the cross vertices is not constrained. A univalent cross vertex of color $c$ hooked to a round vertex $s$ signifies a fixed point of both permutations $\sigma_c,\tau_c$, that is, $\sigma_c(s) = \tau_c(s) = s$.

\paragraph{Minimally connected graphs.}

\begin{figure}[!ht]
\centering
\includegraphics[scale=0.6]{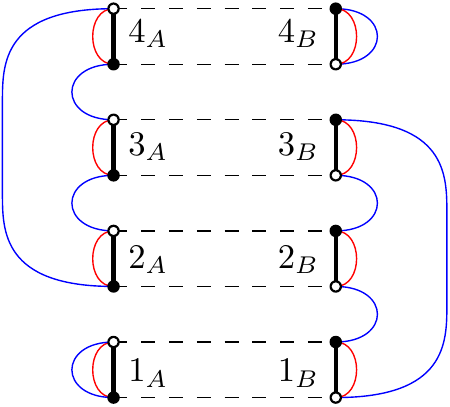}
\hspace{1.5cm}
\includegraphics[scale=0.6]{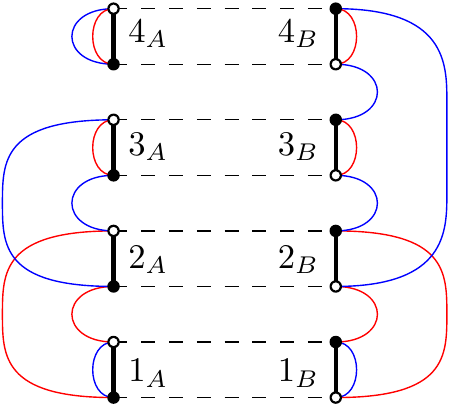}
\caption{Minimally connected graphs in $D=2$. For later reference, note that in these examples $g(\sigma_c , \tau_c)=0$ for every $c$, $\bsig$ is not connected, and $\btau\npreceq\bsig$.
}
\label{fig:D2-min-conn}
\end{figure}

We call \emph{$\Delta$-arborescent} the graphs $(\bsig,\btau)$ for which $G_{_\Delta}(\bsig, \btau)$ is a forest, which are the graphs for which $\DeltaC(\bsig,\btau)=0$. Some examples are displayed in Fig.~\ref{fig:D2-min-conn}.
Reminiscent of the melons, $\Delta$-arborescent graphs can be constructed recursively, in a way that corresponds to building the $G_{_\Delta}(\bsig, \btau)$ by recursive insertions of leaves. 
\begin{figure}[!ht]
\centering
\includegraphics[scale=0.9]{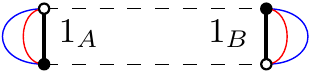}
\caption{The graph $(\mathbf{id},\mathbf{id})\in  \bS_1$ for $D=2$. }
\label{fig:ididS1}
\end{figure}

Denoting by $\mathbf{id}= (\id, \dots, \id)\in \bS_1$, we observe 
that $(\mathbf{id},\mathbf{id})$ is the unique graph $(\bsig,\btau)$  with exactly one  quadrangle with colors $(0,D+1)$: the two $A$ vertices and the two $B$ vertices are connected by all the color-$c$ edges (Fig.~\ref{fig:ididS1}). 
In particular, $(\mathbf{id},\mathbf{id})$ is melonic as a $D+2$ color graph.

Since $G_{_\Delta}(\bsig, \btau)$ is a forest, there must exist a round vertex $s$  connected to $D-1$ univalent cross vertices (leaves).
  
\begin{figure}[!ht]
\centering
\includegraphics[scale=0.9]{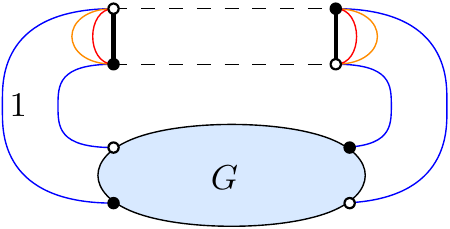}
\hspace{2cm}
\includegraphics[scale=0.9]{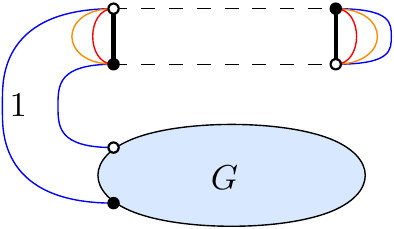}
\caption{Two possibilities for a chain-quadrangle with external color 1 for $D=3$. $G$ is a non-empty, non-necessarily connected portion of the graph.  }
\label{fig:melo-quadr}
\end{figure}

As explained above, for all $c'\neq c$, we have 
$\sigma_{c'}(s) = \tau_{c'}(s) = s $:  for both $\bsig$ and $\btau^{-1}$ the two vertices $s$ are connected by $D$ edges: one for each color different from $c$, and one thick edge of color $D+1$. We call this a \emph{chain-quadrangle} with external color $c$ (Fig.~\ref{fig:melo-quadr}). 

\begin{figure}[!ht]
\centering
\includegraphics[scale=0.9]{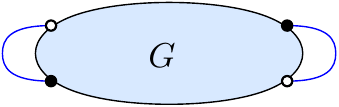}
\hspace{2cm}
\includegraphics[scale=0.9]{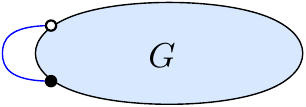}
\caption{The graphs obtained from Fig.~\ref{fig:melo-quadr} by deleting the chain-quadrangle. }
\label{fig:melo-quadr-DEL}
\end{figure}

Given a chain-quadrangle, if we delete its four vertices and all the edges linking them, there is a unique way to reconnect the remaining edge(s) of color $c$, as illustrated in Fig.~\ref{fig:melo-quadr-DEL}. We call this operation the \emph{deletion of a chain-quadrangle}.

\begin{Prop}\label{prop:mincon}
 A graph $(\bsig,\btau)$ is $\Delta$-arborescent if and only if it reduces to a collection of graphs  $(\mathbf{id},\mathbf{id})$ 
 by 
iterative deletions of chain-quadrangles. 
\end{Prop}

\proof Consider $(\bsig,\btau)$ with a chain-quadrangle with external color $c$, and let $(\bsig',\btau')$ be obtained from $(\bsig,\btau)$ by deleting it. Then $(\bsig',\btau')$ has $n-1$ quadrangles $(0, D+1)$ and has lost one connected component for each $(\sigma_{c'}, \tau_{c'})$, $c'\neq c$, and these exactly compensate, so that:  
$$\DeltaC(\bsig', \btau') = \DeltaC(\bsig, \btau) + \bigl(\lvert\Pi(\bsig',\btau')\rvert - \lvert\Pi(\bsig,\btau)\rvert\bigr) - \bigl(\lvert\Pi(\sigma'_c,\tau'_c)\rvert - \lvert\Pi(\sigma_c,\tau_c)\rvert\bigr).$$
There are three cases:
\begin{itemize} 
\item The number of connected components of $(\bsig, \btau)$ is raised by one, then so is $\lvert\Pi(\sigma_c,\tau_c)\rvert$, and $\DeltaC(\bsig', \btau')=\DeltaC(\bsig, \btau)$. 
\item Both the number of connected components of $(\bsig, \btau)$ and $(\sigma_c,\tau_c)$ remain the same, in which case again $\DeltaC(\bsig', \btau')=\DeltaC(\bsig, \btau)$.
\item The number of connected components of $(\bsig, \btau)$ remains the same, but that of $(\sigma_c,\tau_c)$ is raised by one. In that case, $\DeltaC(\bsig', \btau')=\DeltaC(\bsig, \btau) - 1$. 
\end{itemize}
Therefore, if $(\bsig,\btau)$ is $\Delta$-arborescent, then the third case cannot occur, and the connected components of $(\bsig',\btau')$ are also $\Delta$-arborescent. As already discussed,  $(\bsig,\btau)$ being $\Delta$-arborescent, there necessarily exists in $G_{_\Delta}(\bsig, \btau)$ a round vertex connected to $D-1$ cross leaves, so there necessarily exists a chain-quadrangle in $(\bsig,\btau)$. If $n>1$, we may delete it, obtaining one or two smaller $\Delta$-arborescent graphs. For each connected component, either it is $(\mathbf{id},\mathbf{id})$, or it has more than one quadrangle $(0, D+1)$ and it contains a chain-quadrangle which we may delete, and so on, until we are left with a union of graphs  $(\mathbf{id},\mathbf{id})$.

\qed

\paragraph{Bounds on $\DeltaC$.} In the following, we will need certain bounds on $\DeltaC$. In order to establish them, we note that $\DeltaC$ is related to the 
$(D+1)$-degrees of $\bsig$ and $\btau$, see  \eqref{eq:c-degree-bsig}. Let us denote by:
\be\label{eq:boxdef}
\Box_{\bsig}(\bsig, \btau) =\DeltaC(\bsig, \btau) - \Omega_{D+1}(\bsig) =  \lvert \Pi(\bsig, \btau)\rvert - \lvert \Pi(\bsig)\rvert + \sum_{c=1}^D \big(\Pi(\sigma_c) - \lvert \Pi(\sigma_c, \tau_c)\rvert\big)   \; ,
\ee
and we stress that $\DeltaC(\bsig, \btau)$ is symmetric in $\bsig$ and $\btau$, while $\Box_{\bsig}(\bsig, \btau)$ is not. 
Observe that if $\btau \preceq \bsig$, then $\Box_{\bsig}(\bsig,\btau) = 0 $. In particular, for any $\bsig$ we have $\Box_{\bsig}(\bsig,\bsig) = 0 $.

\begin{figure}[!ht]
\centering
\includegraphics[scale=0.8]{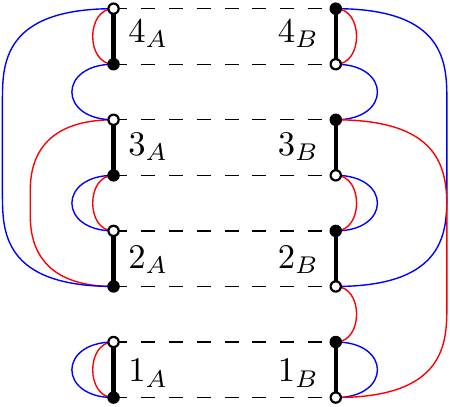}
\hspace{1cm}
\includegraphics[scale=0.8]{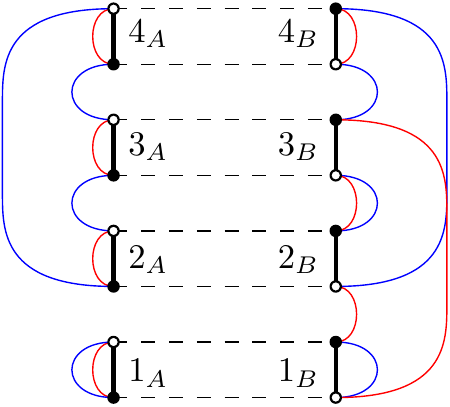}
\hspace{1cm}
\includegraphics[scale=0.8]{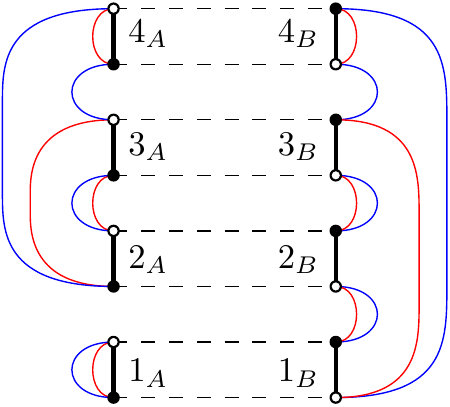}
\caption{Examples of graphs with vanishing $\Box_{\btau}$. The graph on the left has $\Box_{\bsig} = \Box_{\btau}=0$ but $\DeltaC>0$, while the other two have $\Box_{\btau}=0$ but $\Box_{\bsig}>0$.
}
\label{fig:vanish-box}
\end{figure}

\begin{Prop} 
\label{lem:positive-box}
For any $(\bsig , \btau)$, we have 
$\Box_{\bsig}(\bsig, \btau) \ge 0$
and $\Box_{\btau}(\bsig, \btau) \ge 0 $.
\end{Prop}

\begin{figure}[!ht]
\centering
\includegraphics[scale=0.8]{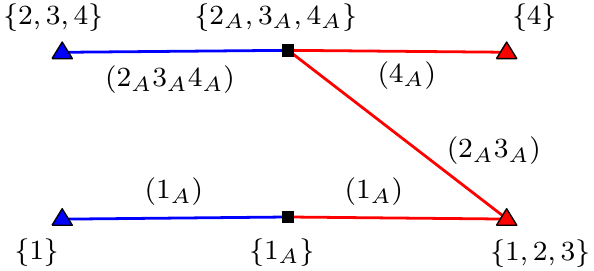}
\includegraphics[scale=0.8]{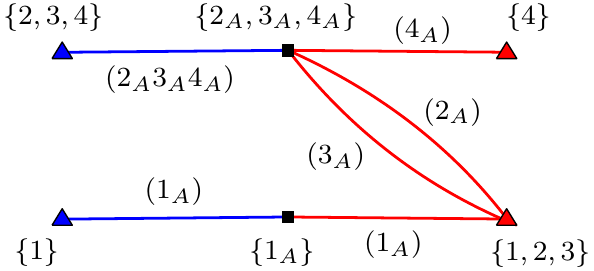}
\includegraphics[scale=0.8]{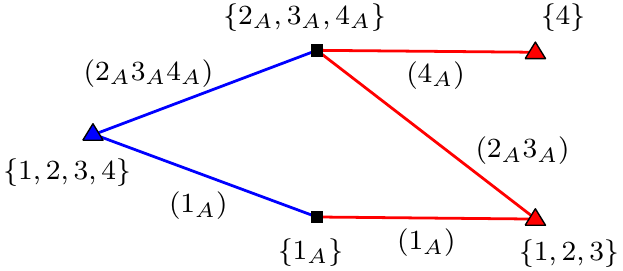}
\caption{Graphs $G_{_{\Box_\bsig}}$ for the examples of Fig.~\ref{fig:vanish-box}. The triangular vertices on the left (resp.~right) of each figure correspond to connected components of $(\sigma_1 , \tau_1)$ (resp.~$(\sigma_2 , \tau_2)$). 
}
\label{fig:graph-box}
\end{figure}

\proof This is a direct consequence of the construction in Definition~\ref{eq:defgraphpart}.
The partitions $\Pi(\bsig)$, $\{\Pi(\sigma_c, \tau_c)\}_c$ and $ \{ \Pi(\sigma_c)\}_c$ are such that
 $\Pi(\bsig) \ge \Pi(\sigma_c)$ and $\Pi(\sigma_c, \tau_c) \ge \Pi(\sigma_c)$.
Building the graph $G_{_{\Box_\bsig}} = \big(\Pi(\bsig), 
\{\Pi(\sigma_c, \tau_c)\}_c,  \{ \Pi(\sigma_c)\}_c \big)$ as in Def.~\ref{eq:defgraphpart}, the number of excess edges of $G_{_{\Box_\bsig}}$ is:
\[
 \sum_c |  \Pi(\sigma_c)|  -\sum_c
 \lvert \Pi(\sigma_c, \tau_c)\rvert
 - |\Pi(\bsig)| + 
 |\Pi(\bsig)\bigvee_{c=1}^D \Pi(\sigma_c,\tau_c)|\ge 0 \; ,
\]
and we conclude by observing that 
$ \Pi(\bsig)\bigvee_{c=1}^D \Pi(\sigma_c,\tau_c) = \Pi(\bsig) \vee \Pi(\btau) = \Pi(\bsig,\btau)$. The graphs $G_{_{\Box_\bsig}} $ for the examples of Fig.~\ref{fig:vanish-box} are shown in Fig.~\ref{fig:graph-box}.

\qed

We stress that the relation between 
$G_{_{\Box_\bsig}}$ and $G_{_\Delta}(\bsig, \btau)$ is \emph{not} straightforward. While the triangular vertices of $G_{_{\Box_\bsig}}$ are the cross vertices of $G_{_\Delta}(\bsig, \btau)$ and the square vertices of $G_{_{\Box_\bsig}}$ are obtained by collapsing together the round vertices of $G_{_\Delta}(\bsig, \btau)$ associated to the labels $s$ belonging to the same connected component of $\bsig$, the edges encode very different things in the two cases.

\begin{Prop} [Lower bounds on $\Delta$]
\label{lem:positive-box-2}
As $ \Box_{\bsig}(\bsig, \btau)=\DeltaC(\bsig, \btau) - \Omega_{D+1}(\bsig) \ge 0$, we have:
 \[
 \begin{split}
& \DeltaC(\bsig, \btau)\ge \Omega_{D+1}(\bsig) \;, \qquad \DeltaC(\bsig, \btau)\ge \Omega_{D+1}(\btau)
 \;, \crcr
& \DeltaC(\bsig, \btau)\ge  \Box(\bsig, \btau) = \frac{\Box_{\btau}(\bsig, \btau) + \Box_{\bsig}(\bsig, \btau)} 2  \;.
 \end{split}
\]
In particular, if $(\bsig,\btau)$ is $\Delta$-arborescent then both $\bsig$ and $\btau^{-1}$ are (not necessarily connected) $(D+1)$-melonic graphs. The converse of this statement is \emph{not true}. 

If $\Pi(\tau_c) \le \Pi(\sigma_c)$ for all $c$, then $ \Pi(\bsig, \btau) = \Pi(\bsig) $ and $\Pi(\sigma_c,\tau_c)=\Pi(\sigma_c)$. In this case, $ \DeltaC(\bsig, \btau)=  \Omega_{D+1}(\bsig)$ and $(\bsig , \btau)$ is  $\Delta$-arborescent if and only if $\bsig$ is $(D+1)$-melonic. 

Finally, if $\bsig$ is $(D+1)$-melonic then $\DeltaC(\bsig,\bsig) =0 $ (as can be seen by deleting iteratively the chain-quadrangles associated to the melonic insertions of $\bsig$).
\end{Prop}

\paragraph{Graphs with $\Box_{\btau}=0$.}
We will encounter below the family of all the graphs with $\Box_{\tau}=0$. While this family includes the $\Delta$-arborescent graphs, it is strictly larger (see Fig.~\ref{fig:vanish-box}): graphs with $\Box_{\btau}=0$ but $\DeltaC = \Omega_{D+1}(\btau)>0$ are among the leading order graphs in regime \ref{micAentitemI} of Theorem \ref{th:other-ent-microscopic-regimes}. We have not yet found a satisfactory recursive construction for the family of graphs with $\Box_{\btau}=0$.

\newpage

\section{Asymptotic regimes for $D\ge 2$}
\label{sec:D>1}
The asymptotic expansion of the cumulants of the tensor HCIZ integral \eqref{eq:WC-norm} for $D$ arbitrary is: 
\vspace{+1ex}
\begin{equation*}
C_n\bigl( {N^{\ggam}}\Tr (AUBU^* )\bigr)= N^{n( \ggam - 2D ) }\sum_{  \bsig,\btau }
  N^{s(\bsig, \btau) +s_A(\bsig) + s_B(\btau)}   \tr_{\bsig}(\al) \, \tr_{\btau^{-1}}(\bl) \pC(1+O(1)) \;, 
\end{equation*}
where \eqref{eq:scaling-of-weingarten} and 
the asymptotic scaling ansatz~\eqref{eq:general-scaling-assumption} with $\alpha_A=0$ read:
\[
s(\bsig, \btau) = \sum_{c=1}^D\#(\sigma_c\tau_c^{-1})- 2 \big[ \lvert \Pi(\bsig, \btau)\rvert-1 \big] \;, \qquad s_A(\bsig)=\beta_A \sum_{c=1}^D \#(\sigma_c) + \epsilon_A \sum_{c_1<c_2} \#(\sigma_{c_1}\sigma_{c_2}^{-1}) \; ,
\]
and similarly for $B$. Note that summing \eqref{eq:EulerCharGammac} over $c$, the total scaling exponent with $N$ of a graph reads:
 \be\label{eq:anfang}
 \begin{split}
& n(\ggam - D) - 2 \big[ \lvert \Pi(\bsig, \btau)\rvert-1 \big]  + 2 \sum_{c=1}^D\lvert \Pi(\sigma_c, \tau_c)\rvert - 2 \sum_{c=1}^D g(\sigma_c, \tau_c) \crcr
&\;\; -(1-\beta_A) \sum_{c=1}^D\#(\sigma_c) -(1-\beta_B) \sum_{c=1}^D \#(\tau_c)
+ \epsilon_A \sum_{c_1<c_2} \#(\sigma_{c_1}\sigma_{c_2}^{-1})
+ \epsilon_B \sum_{c_1<c_2} \#(\tau_{c_1}\tau_{c_2}^{-1}) \;,
 \end{split}
 \ee
where $\lvert \Pi(\sigma_c, \tau_c)\rvert$ and $g(\sigma_c, \tau_c)$ are respectively the number of connected components and the total genus of $(\sigma_c , \tau_c)$.

In the rest of this paper, we will identify for each $\beta_A,\beta_B,\epsilon_A,\epsilon_B$, the $\ggam$ and $\ddel$ such that: 
\[
\begin{split}
& \ddel \ge n(\ggam - D) - 2 \big[ \lvert \Pi(\bsig, \btau)\rvert-1 \big]  + 2 \sum_{c=1}^D\lvert \Pi(\sigma_c, \tau_c)\rvert - 2 \sum_{c=1}^D g(\sigma_c, \tau_c) \crcr
&\;\; -(1-\beta_A) \sum_{c=1}^D\#(\sigma_c) -(1-\beta_B) \sum_{c=1}^D \#(\tau_c)
+ \epsilon_A \sum_{c_1<c_2} \#(\sigma_{c_1}\sigma_{c_2}^{-1})
+ \epsilon_B \sum_{c_1<c_2} \#(\tau_{c_1}\tau_{c_2}^{-1}) \;,
 \end{split}
\]
and the equality is saturated for an infinite family of $(\bsig, \btau)$, $n\ge 1$. We furthermore classify
the leading order graphs.

 The general strategy consists in writing for each asymptotic ansatz the scaling in \eqref{eq:anfang} as a sum of non-positive terms and then identifying the terms which maximize it.
  To get acquainted with the typical structures, we will first analyze in detail some of the regimes.
 
\subsection{The microscopic regime}
\label{subsec:MicroSep}
\begin{figure}[!ht]
\centering
\includegraphics[scale=0.6]{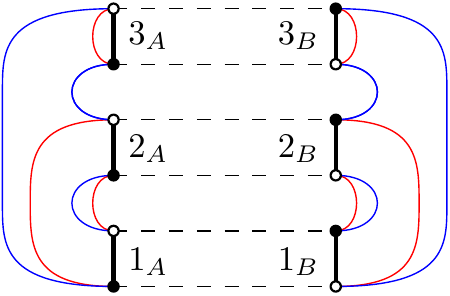}
\caption{Leading order graph in the microscopic regime.}
\label{fig:Ex-mic-sep}
\end{figure}

We first consider that both $A$ and $B$ are microscopic, that is,
$\epsilon_A = \epsilon_B = \beta_A=\beta_B=0$:
\[
\Tr_{\bsig}(A)\sim \tr_{\bsig}(\al)=O(1) \qquad \textrm{ and } \qquad \Tr_{\btau}(B)\sim \tr_{\btau}(\bl)=O(1).
\]
\begin{Lem}
\label{lem:micromicro}
For $\epsilon_A = \epsilon_B = \beta_A=\beta_B=0$, we have (see Fig.~\ref{fig:Ex-mic-sep}):
\[ 
\lim_{N\rightarrow +\infty} 
C_n\bigl( {N^{D}}\Tr (AUBU^* )\bigr)  =   \sum_{  \bsig\,\in\, \bS_n \ \mathrm{ connected}}
    \tr_{\bsig}(\al) \,  \tr_{\bsig^{-1}} (\bl)  \; .
\]
\end{Lem}

\proof 
The scaling exponent in~\eqref{eq:anfang} becomes:
\[
 n( \ggam - D ) - 2 \big(\lvert \Pi(\bsig, \btau)\rvert-1 \big)
   - \sum_c \big(2g(\sigma_c, \tau_c) + |\Pi(\sigma_c) | - \lvert \Pi(\sigma_c, \tau_c) | + |\Pi(\tau_c)| - \lvert \Pi(\sigma_c, \tau_c)\rvert \big) \;.
\]
Setting $\ggam=D$ eliminates the first term, and the others are non-positive, hence the scaling is maximal when they vanish. From Prop. \ref{lem:supports-included}, it follows that $\sigma_c = \tau_c$, thus $\Pi(\bsig,\btau) = \Pi(\bsig)=1$. Finally, from Thm.~\ref{th:previous-results}, since $\bsig$ is connected and $\bsig=\btau$, $f[\bsig, \btau]=1$. 

\qed

\subsection{The purely entangled regime}
\label{subsec:purely-entangled}

\begin{figure}[!ht]
\centering
\includegraphics[scale=0.7]{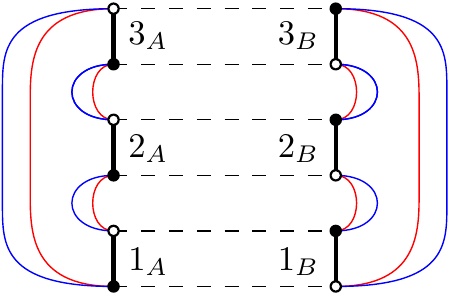}
\caption{Leading order graph for the entangled regime: all the colors are the same cycle.}
\label{fig:first-regime-ent}
\end{figure}

The purely entangled (non-necessarily symmetric) scaling ansatz is $\beta_A=\beta_B=0$ and $\epsilon_A,\epsilon_B>0$:
\[
\Tr_{\bsig}(A)\sim N^{\epsilon_A \sum_{c_1<c_2} \#(\sigma_{c_1}\sigma_{c_2}^{-1})} \tr_{\bsig}(\al) \qquad \textrm{ and } \qquad \Tr_{\btau}(B)\sim N^{\epsilon_B\sum_{c_1<c_2} \#(\tau_{c_1}\tau_{c_2}^{-1})}\tr_{\btau}(\bl).
\]
\begin{Lem}
\label{thm:all-purely-entangled}
For  $\beta_A=\beta_B=0$ and $\epsilon_A,\epsilon_B>0$, the leading order graphs $(\bsig,\btau)$ are such that  
all the $\sigma_c$ and $\tau_c$ are the same cycle of length $n$ (see Fig.~\ref{fig:first-regime-ent}):
\[ 
\lim_{N\rightarrow +\infty} 
C_n\Biggl( \frac{N^D}{N^{(\epsilon_A+\epsilon_B)\frac{D(D-1)}2}}\Tr (AUBU^* )\Biggr)  =  (n-1)!\, \tr(\al^n) \,  \tr (\bl^n) .
\]

The same regime is obtained if $\epsilon_A=0,\epsilon_B>0$.
\end{Lem}
\proof We rewrite the scaling exponent in  \eqref{eq:anfang} in terms of the degrees \eqref{eq:degree-bsig} of $\bsig$ and $\btau$ as:
\[
\begin{split}
&   n\bigl[ \ggam -D + (\epsilon_A+\epsilon_B) \frac{D(D-1)}2 \bigr]  - 2 (\lvert \Pi(\bsig, \btau)\rvert - 1) \crcr &\qquad -\sum_c\Bigl( 2g(\sigma_c, \tau_c) +|\Pi(\sigma_c)| - \lvert \Pi(\sigma_c, \tau_c)\rvert +  |\Pi(\tau_c)| - \lvert \Pi(\sigma_c, \tau_c)\rvert \Bigr)
\label{eq:exp-N-mic-ent-1}\crcr
& \qquad- \epsilon_A \omega(\bsig) - \epsilon_B \omega(\btau) -\epsilon_A\bigg[\sum_c  \#(\sigma_c) - D\lvert \Pi(\bsig)\rvert\bigg]-\epsilon_B\bigg[\sum_c  \#(\tau_c) - D\lvert \Pi(\btau)\rvert\bigg] \;, 
\end{split}
\]
which is a sum of non-positive terms, because for every connected component of $\btau$, each $\tau_c$ must have at least one cycle, so that $\sum_c\#(\tau_c)\ge D\lvert \Pi(\btau)\rvert $. 

For this scaling to be maximal, $(\bsig, \btau)$ must be connected. From Prop.~\ref{lem:supports-included}, we have $\bsig = \btau$, hence $|\Pi(\btau)| = 1$. It follows that
$\sum_c\#(\tau_c) -D\lvert \Pi(\btau)\rvert= \sum_c\#(\tau_c) -D$, which is minimal if and only if every $\tau_c$ is a cycle of length $n$. But in this case, from \eqref{eq:degree-bsig}, $\omega(\btau)=0$ if and only if 
$\tau_c$ is the {\it same} cycle
for all colors $c$.

Finally, from Thm.~\ref{th:previous-results}  we have $f[\bsig, \btau]=1$ since $(\bsig, \btau)$ is connected and $\bsig=\btau$. The factor $(n-1)!$ is the number of cycles of length $n$. 

\qed

\subsection{Regimes with $A$  microscopic: First main theorem }

We now assume that $A$ is microscopic: $\epsilon_A=0,\beta_A=0$.

\paragraph{The $B$-macroscopic separable regime.}
\label{subsec:MacroSep}

\begin{figure}[!ht]
\centering
\includegraphics[scale=0.6]{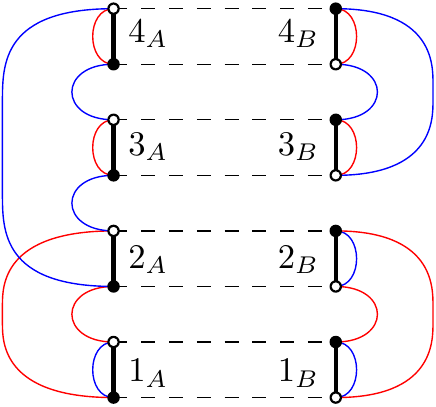}
\hspace{1.2cm}
\includegraphics[scale=0.6]{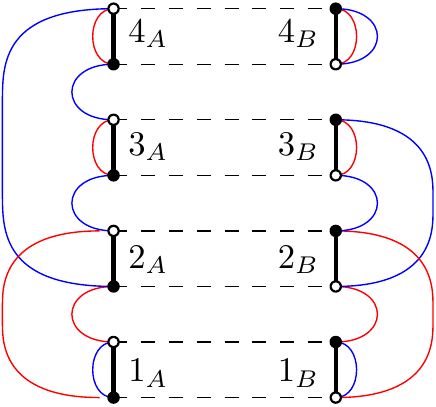}
\hspace{1.2cm}
\includegraphics[scale=0.6]{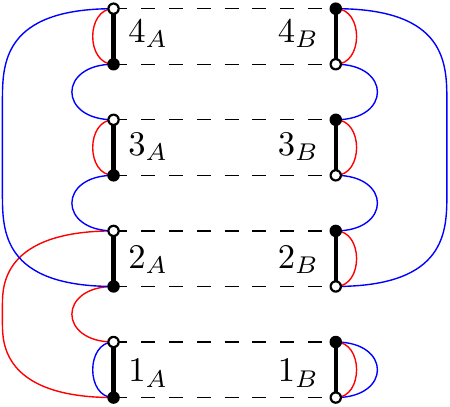}
\caption{Examples of graphs
$(\bsig, \btau)$ with $\bsig$
$(D+1)$-melonic and $\btau \preceq \bsig$.}
\label{fig:D2-gen-NC}
\end{figure}

Here, $\epsilon_A=\beta_A=0$ and $\epsilon_B=0$, $\beta_B=1$, that is:
\[
\Tr_{\bsig}(A)\sim \tr_{\bsig}(\al)=O(1) \qquad \textrm{ and } \qquad \Tr_{\btau}(B)\sim N^{\sum_{c=1}^D \#(\tau_c)}\tr_{\btau}(\bl) \, .
\]
\begin{Lem}
\label{thm:MacroSep}
For $\epsilon_A=\beta_A=0$ and $\epsilon_B=0$, $\beta_B=1$, we have (see Fig.~\ref{fig:D2-gen-NC}):
\[
\lim_{N\rightarrow +\infty} 
\frac 1 {N } 
C_n\bigl( N\Tr (AUBU^* )\bigr)  =  \sum_{  \substack{{\bsig \, \in \bS_n \text{ connected}}\\{ (D+1)-\textrm{melonic}}}} \, \, 
\sum_{\btau , \,  \btau \,\preceq\, \bsig  }\, 
    \tr_{\bsig}(\al) \,  \tr_{\btau^{-1}} (\bl)\,  \prod_{c=1}^D \M(\sigma_c\tau_c^{-1}) \; . 
\]
\end{Lem}

For $D=1$, the conditions $\bsig$
$(D+1)$-melonic and $\btau \preceq \bsig$ reduce to $\sigma$ a cycle and $\tau$ non-crossing on $\sigma$, that is, the regime \ref{D=1item2} of Thm.~\ref{th:RegimesD1}. 

\proof The scaling exponent in \eqref{eq:anfang} can be written as:
\[
1 +  n( \ggam -1) -  (\lvert \Pi(\bsig, \btau)\rvert - 1)  -2 \sum_{c=1}^D g(\sigma_c, \tau_c) - \sum_{c=1}^D\,\bigl[\#(\sigma_c) - \lvert \Pi(\sigma_c, \tau_c)\rvert\bigr] -\DeltaC(\bsig, \btau) \; , 
\]
with $\DeltaC(\bsig, \btau)$ defined in  \eqref{eq:def-DeltaK}. Setting $\ggam=1$ takes out the linear term in $n$, and the remaining ones are non-positive. The exponent of $N$ is maximal for graphs satisfying:
\[
\forall c,\; g(\sigma_c, \tau_c)=0, \qquad\#(\sigma_c) = \lvert \Pi(\sigma_c, \tau_c)\rvert, \qquad \lvert \Pi(\bsig, \btau)\rvert = 1, \quad\textrm{ and }\quad \DeltaC(\bsig, \btau)=0 \;.
\]

The leading order graphs $(\bsig, \btau)$ are connected. From Prop.~\ref{lem:supports-included}, $\bsig$ is also connected 
and $\btau\preceq\bsig$. As $\btau\preceq \bsig$, from Prop.~\ref{lem:positive-box-2}, 
$ \DeltaC(\bsig, \btau) =0$ if and only if 
$\bsig$ is $(D+1)$-melonic.

\qed

\paragraph{All the $B$-separable regimes.}
\label{subsec:other-sep-microscopic}
The general $A$-microscopic  and $B$-separable scaling ansatz is: 
\[
\Tr_{\bsig}(A)\sim \tr_{\bsig}(\al)=O(1) \qquad \textrm{ and } \qquad \Tr_{\btau}(B)\sim N^{ \beta \sum_{c=1}^D\#(\tau_c)}\tr_{\btau}(\bl) \;,
\]
which covers the entire $\epsilon=0$ axis of Fig.~\ref{fig:summary-results-0-micro}.

\begin{figure}[!ht]
\centering
\includegraphics[scale=0.7]{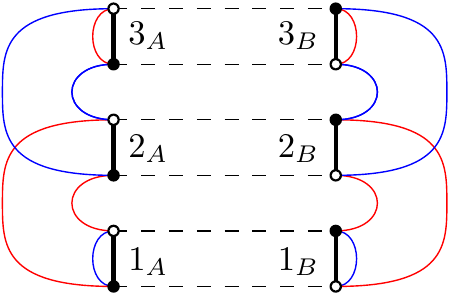}
\hspace{2cm}
\includegraphics[scale=0.7]{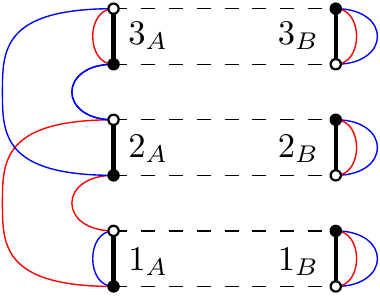}
\caption{Examples of leading order graphs for the regimes 
\normalfont{IV} ($0<\beta<1$) and 
\normalfont{VIII} ($1<\beta$).  }
\label{fig:other-sep-microscopic}
\end{figure}

\begin{Th}
\label{th:other-sep-microscopic-regimes}
For $A$ microscopic  ($\epsilon_A=\beta_A=0$) and $B$ separable   ($\epsilon_B=0,\beta_B=\beta$), the leading order graphs $(\bsig,\btau)$ are (see Fig.~\ref{fig:other-sep-microscopic}):
\begin{itemize}
\item[\normalfont{VI}]\label{sepitemVI} -  For $\beta = 0$,  $\bsig=\btau$ and $\bsig$  is connected.
\item[\normalfont{IV}]\label{sepitemIV} - For $0<\beta<1$, $\bsig=\btau$ and $\bsig$  connected $(D+1)$-melonic.
\item[\normalfont{II}]\label{sepitemII} - For $\beta=1$, $\bsig$ is connected  $(D+1)$-melonic and $\btau\preceq\bsig$.
\item[\normalfont{VIII}]\label{sepitemVIII} - For $1<\beta$, $\tau_c = \mathrm{id}$ for all $c$ and  $\bsig$ is connected $(D+1)$-melonic. This generalizes the regime \ref{D=1item4} of Thm.~\ref{th:RegimesD1}.
\end{itemize}
\end{Th}

\proof
Items \normalfont{VI} ($\beta=0$) and \normalfont{II} ($\beta=1$) have already been addressed in Lemmata~\ref{lem:micromicro} and \ref{thm:MacroSep}. The rest of the items are particular cases of Thm.~\ref{th:other-ent-microscopic-regimes}. 

\qed 

We get four different regimes, one each for $B$ microscopic ($\beta=0$), mesoscopic ($0< \beta< 1$), macroscopic ($\beta=1$) and ``hyper-microscopic'' ($1 < \beta$). In all cases $\bsig$ is connected, and for $\beta > 0$ it is $(D+1)$-melonic; $\btau = \bsig$ for
$\beta<1$, $\btau \preceq \bsig$ for $\beta=1$ and $\btau = \id$ for $\beta>1$.

\paragraph{The $B$-macroscopic boundary regime.}
\label{subsec:rich-entangled}
The $B$-macroscopic boundary regime is obtained for $\beta_B = \epsilon_B = 1/D$:
\[
\Tr_{\bsig}(A)\sim  \tr_{\bsig}(\al)=O(1) \qquad \textrm{ and } \qquad \Tr_{\btau}(B)\sim N^{\frac 1 D \sum_c \#(\tau_c) +  \frac 1 D\sum_{c_1<c_2} \#(\tau_{c_1}\tau_{c_2}^{-1})}\tr_{\btau}(\bl) \;.
\]

\begin{figure}[!ht]
\centering
\includegraphics[scale=0.7]{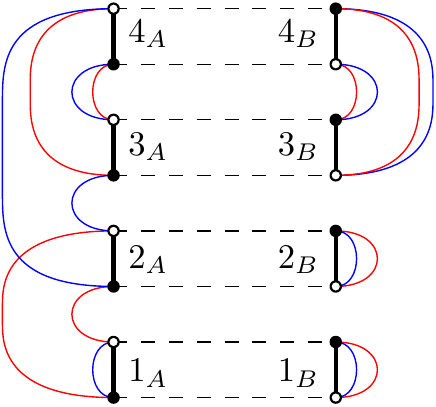}
\hspace{1cm}
\includegraphics[scale=0.7]{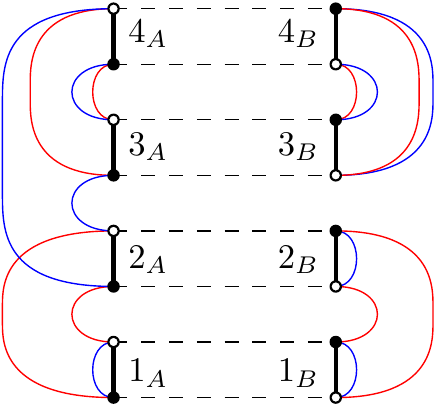}
\hspace{1cm}
\includegraphics[scale=0.7]{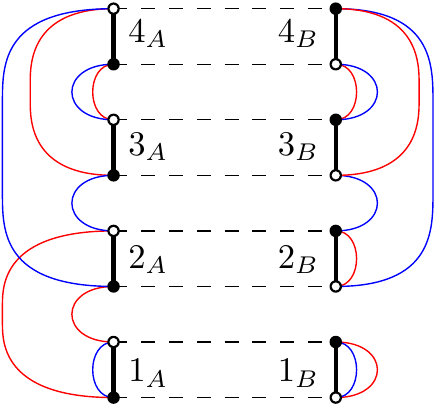}
\caption{Examples of leading order graphs
in the macroscopic boundary regime.}
\label{fig:D2-gen-NC-ent}
\end{figure}

\begin{Lem}\label{lem:rich-entangled}
For $A$  microscopic  ($\epsilon_A = \beta_A = 0$) and $\beta_B = \epsilon_B = 1/D$, we have (see Fig.~\ref{fig:D2-gen-NC-ent}):
\[
\lim_{N\rightarrow +\infty} 
\frac 1 {N} 
C_n\bigl( N^{\frac {D+1}2}\Tr (AUBU^* )\bigr)  =   \sum_{  \substack{{\bsig\, \in \bS_n}\\{\textrm{connected}} }} \;\; \sum_{ \substack{{\btau \preceq \bsig }\\{\Box_{\btau}(\bsig,\btau) = 0 }\\{\omega(\btau)=0} } }\, 
    \tr_{\bsig}(\al) \,  \tr_{\btau^{-1}} (\bl) \, \prod_{c=1}^D \M(\sigma_c \tau_c^{-1}) \;.
\]
The leading order graphs include $(\bsig,\bsig)$ for any $\bsig$ connected melonic, as well as the set $(\bsig,\btau)$ with $\bsig$ connected $(D+1)$-melonic and $\btau\preceq \bsig$.
\end{Lem}

\proof This follows by writing the scaling exponent in \eqref{eq:anfang} in terms of the degree of $\btau$ \eqref{eq:degree-bsig} and of $\Box_{\btau}$ \eqref{eq:boxdef}:
$$
1+ n\bigl[ \ggam - \frac{D+1}2\bigr]  - \frac 1 D\omega(\btau) - \Box_{\btau}(\bsig, \btau) - \big(\lvert \Pi(\bsig, \btau)\rvert - 1 \big)- \sum_c\Bigl( 2g(\sigma_c, \tau_c) +  \#(\sigma_c) - \lvert \Pi(\sigma_c, \tau_c)\rvert \Bigr)  \;,
$$
which is a sum of non-positive terms once $\ggam$ is set to $(D+1)/2$.

From Prop.~\ref{lem:supports-included}, the leading order graphs $(\bsig, \btau)$ are such that $\bsig$ is connected and $\btau\preceq \bsig$.
As $(\bsig, \btau)$ is connected, $f[\bsig, \btau]$ reduces to a product of Moebius functions (Thm.~\ref{th:previous-results}).  

The last assertion follows by noting that (see  \eqref{eq:boxdef}):
\begin{itemize}
 \item for any $\bsig$, we have $\Box_{\bsig}(\bsig,\bsig) = 0 $.
 \item if $\btau \preceq \bsig$, then $\Box_{\bsig}(\bsig,\btau) = 0 $ and from Prop.~\ref{lem:positive-box-2}, $\DeltaC(\bsig, \btau)=\Omega_{D+1}(\bsig)$.
 If moreover $\Box_{\btau}(\bsig, \btau)=0$, then $\Omega_{D+1}(\bsig) = \Omega_{D+1}(\btau)$ and $ \Omega_{D+1}(\btau) =0\Rightarrow  \omega(\btau)=0$. 
\end{itemize}
 
 \qed

While similar, this regime is richer than the $A$-microscopic $B$-macroscopic separable regime of Lemma~\ref{thm:MacroSep}. There exist graphs $(\bsig,\btau)$ with $\Pi(\bsig)=1$, $\btau\preceq \bsig$, $\Box_{\btau}(\bsig,\btau) = \omega(\btau) =0 $ but $\Omega_{D+1}(\bsig)>0$ (the examples in Fig.~\ref{fig:D2-gen-NC-ent} have $\Omega_{D+1}(\bsig)>0$). Also, note that not all connected $\bsig$ are included at leading order as there exist $\bsig $ with $\Pi(\bsig)=1$ such that there exist \emph{no} $\btau\preceq \bsig$ with $\Box_{\btau}(\bsig,\btau) = \omega(\btau) =0$ (right in Fig.~\ref{fig:examples-sigma-box}).

\begin{figure}[!ht]
\centering
\includegraphics[scale=0.7]{K33.pdf}
\caption{A graph $\bsig$ with $\Pi(\bsig)=1$, such that there exist \emph{no} $\btau\preceq \bsig$, $\Box_{\btau}(\bsig,\btau) = \omega(\btau) =0$. }
\label{fig:examples-sigma-box}
\end{figure}

We believe that any connected graph $(\bsig,\btau)$ satisfying the conditions $\Pi(\bsig)=1$, $\btau\preceq \bsig$, $\Box_{\btau}(\bsig,\btau) = \omega(\btau) =0 $ \emph{must also be melonic} ($\omega(\bsig)=0$), but we have not been able to prove it. If this is indeed true, the symmetric macroscopic boundary regime \ref{symAentitemI} of Thm.~\ref{th:other-sym-regimes} below would be richer than the present  $A$-microscopic $B$-macroscopic boundary regime \ref{micAentitemI}. 

\paragraph{All regimes for $A$ microscopic.}
We now give the exhaustive list of asymptotic regimes with $A$ microscopic (Fig.~\ref{fig:summary-results-0-micro}):
\[
\Tr_{\bsig}(A)\sim  \tr_{\bsig}(\al)=O(1) \qquad \textrm{ and } \qquad \Tr_{\btau}(B)\sim N^{\beta \sum_c \#(\tau_c) +  \epsilon\sum_{c_1<c_2} \#(\tau_{c_1}\tau_{c_2}^{-1})}\tr_{\btau}(\bl) \; ,
\]
with $\beta, \epsilon\ge 0$. In the theorem below, the regimes \ref{micAentitemII}, \ref{micAentitemIV}, \ref{micAentitemVI},  and \ref{micAentitemVIII}    
are identical with the regimes \normalfont{II}, \normalfont{IV}, \normalfont{VI}, and \normalfont{VIII} in Thm.~\ref{th:other-sep-microscopic-regimes}, and extend them to non-zero $\epsilon$. The regimes \ref{micAentitemI}, \ref{micAentitemIII}, \ref{micAentitemV}, 
 and \ref{micAentitemVII}
appear only for $\epsilon>0$.

\begin{Th}
\label{th:other-ent-microscopic-regimes}
For $A$ microscopic ($\epsilon_A = \beta_A = 0$) and any $B$ ($\epsilon_B=\epsilon$, $\beta_B=\beta$), the leading order graphs $(\bsig,\btau)$ are such that (see also Fig.~\ref{fig:summary-results-0-micro}):
\begin{enumerate}[label=\normalfont{\Roman*}]
\item\label{micAentitemI} - \underline{The macroscopic boundary regime} (the square red point in Fig.~\ref{fig:summary-results-0-micro}). For $\beta=\epsilon=1/D$, see Lemma~\ref{lem:rich-entangled},
$\bsig$ is arbitrary connected, $\btau \preceq\bsig$, $\omega(\btau)=0$  and 
$\Box_{\btau}(\bsig,\btau) = 0$ (or equivalently $\Omega_{D+1}(\bsig)=\Omega_{D+1}(\btau)$):
\[
\lim_{N\rightarrow +\infty} 
\frac 1 {N} 
C_n\bigl( N^{\frac {D+1}2}\Tr (AUBU^* )\bigr)  =   \sum_{  \substack{{\bsig\, \in \bS_n}\\{\textrm{connected}} }} \;\; \sum_{ \substack{{\btau \preceq \bsig }\\{\Box_{\btau}(\bsig,\btau) = 0 }\\{\omega(\btau)=0} } }\, 
    \tr_{\bsig}(\al) \,  \tr_{\btau^{-1}} (\bl) \, \prod_{c=1}^D \M(\sigma_c \tau_c^{-1}) \;.
\]
This is a {\bf combinatorially  prolific regime}
richer than the regime \ref{micAentitemII}.
\item\label{micAentitemII} -  \underline{The macroscopic separable regime} (the blue line in Fig.~\ref{fig:summary-results-0-micro} including the endpoint $\beta=1$).
For $\beta=1-\epsilon(D-1)> 1 /D$ and $\epsilon \ge 0$, $\bsig$ is an arbitrary connected $(D+1)$-melonic graph and $\btau \preceq\bsig$:
$$
\lim_{N\rightarrow +\infty} 
\frac 1 {N } 
C_n\Bigl( N^{1 + \epsilon \frac{D(D-1)}2}\Tr (AUBU^* )\Bigr)=  \hspace{-0.1cm} \sum_{  \substack{{\bsig\, \in \bS_n \textrm{ conn.}
}\\{ (D+1)\textrm{-melonic}}}} \, \sum_{ \btau, \btau \preceq \bsig }\, 
    \tr_{\bsig}(\al) \,  \tr_{\btau^{-1}} (\bl) \, \prod_{c=1}^D \M(\sigma_c \tau_c^{-1}) \;.
$$
This is a {\bf  combinatorially prolific regime}, the same as in  Lemma.~\ref{thm:MacroSep}.
\item\label{micAentitemIII} - \underline{The mesoscopic boundary regime} (the yellow line in Fig.~\ref{fig:summary-results-0-micro}).
For $0<\beta=\epsilon < 1/D$, $\bsig$ is an arbitrary connected melonic graph and $\btau=\bsig$:
\[
\lim_{N\rightarrow +\infty} 
\frac 1 {N^{\epsilon D} } 
C_n\Bigl( N^{D- \epsilon \frac{D(D-1)}2 }\Tr (AUBU^* )\Bigr)  =   \sum_{  \substack{{\bsig\, \in\, \bS_n
\textrm{connected} }\\{\textrm{melonic}}}} 
    \tr_{\bsig}(\al) \,  \tr_{\bsig^{-1}} (\bl) \; .
\]
\item\label{micAentitemIV} - \underline{The mesoscopic separable regime} (the orange region in Fig.~\ref{fig:summary-results-0-micro}). For $0 \le \epsilon<\beta<1-\epsilon(D-1)$, $\bsig$ is an arbitrary connected $(D+1)$-melonic graph and $\btau = \bsig$:
 \[
\lim_{N\rightarrow +\infty} 
\frac 1 {N^{\beta + \epsilon (D-1)} } 
C_n\Biggl( \frac {N^D}{N^{(D-1)(\beta-\epsilon + \frac  {\epsilon D} 2 )}}\Tr (AUBU^* )\Biggr)  =   \sum_{  \substack{{\bsig\,\in \bS_n \textrm{ connected}
}\\{ (D+1)\textrm{-melonic}}}} 
    \tr_{\bsig}(\al) \,  \tr_{\bsig^{-1}} (\bl)  \;.
\]
\item\label{micAentitemV} - \underline{The entangled regime} (the green region in Fig.~\ref{fig:summary-results-0-micro}).
For $0\le \beta < \min \{1/D , \epsilon \}$
we find that all the $\sigma_c$ and all the $\tau_c$ are the same cycle of length $n$:
\[
\lim_{N\rightarrow +\infty} 
\frac 1 {N^{\beta D} } 
C_n\Bigl( N^{D- \epsilon \frac{D(D-1)}2 }\Tr (AUBU^* )\Bigr)  =   (n-1)! \    \tr(\al^n) \,  \tr (\bl^n) \; ,
\]
reproducing the regime in Lemma~\ref{thm:all-purely-entangled}.
\item\label{micAentitemVI} - \underline{The microscopic regime} (the origin in Fig.~\ref{fig:summary-results-0-micro}).
For $\beta = \epsilon = 0$, see Lemma~\ref{lem:micromicro}, $\bsig$ is an arbitrary connected graph and $\btau=\bsig$:
\[
\lim_{N\rightarrow +\infty} 
C_n\bigl( {N^{D}}\Tr (AUBU^* )\bigr)  =   \sum_{  \bsig\,\in\, \bS_n \ \mathrm{ connected}}
    \tr_{\bsig}(\al) \,  \tr_{\bsig^{-1}} (\bl)  \; .
\]
Although non-prolific, in this regime $\bsig$ is only constrained to be connected (so in particular there is a super-exponential growth of the number of leading order graphs up to relabeling of $1, \ldots, n$).
\item\label{micAentitemVII} - \underline{The hyper-macroscopic boundary regime}  (magenta vertical line in Fig.~\ref{fig:summary-results-0-micro}). For $\epsilon > \beta = 1/D$,  the leading order graphs are $\bsig$  connected, $\btau= (\tau, \dots, \tau)$ for some $\tau\in S_n$, $\btau \preceq \bsig$, and
$\Box_{\btau}(\bsig,\btau)=0$:
\[
\begin{split}
& \lim_{N\rightarrow +\infty} 
\frac 1 N C_n\Bigl( N^{D- \epsilon \frac{D(D-1)}2 }\Tr (AUBU^* )\Bigr) \crcr
& \qquad\qquad = \sum_{  \substack{{\bsig\, \in \bS_n}\\{\textrm{conn.}} }} \; \sum_{ 
\substack{ { \btau = (\tau,\dots ,\tau), \,  \tau \in S_n, }\\
{\btau \preceq \bsig , \; \Box_{\btau}(\bsig,\btau)=0 } }
}  \tr_{\bsig}(\al) \,  \tr_{\btau^{-1}} (\bl) \, \prod_{c=1}^D \M(\sigma_c \tau_c^{-1}) \;.
\end{split}
\]
The leading order graph include $(\bsig,
\mathbf{id} )$ with $\bsig$ connected $(D+1)$-melonic and 
$\mathbf{id}=(\id, \dots, \id ) \in \bS_n$. The regime \ref{micAentitemI} is richer than this regime, as $\omega(\btau)=0$ for $\btau = (\tau, \dots, \tau)$ for any $\tau$. The example displayed on the left in Fig.~\ref{fig:D2-gen-NC-ent} is a leading order graph in this regime, but not the other two. 
While this regime is combinatorially prolific, we expect it to correspond to tensors that exceed the maximal number of degrees of freedom (see the discussion in Sec.~\ref{sub:maxrank}) and which cannot be realized in practice. 
\item\label{micAentitemVIII} - \underline{The hyper-macroscopic regime} (the violet region in Fig.~\ref{fig:summary-results-0-micro}).
For $\beta>\max\big\{1/D , 1-\epsilon(D-1)\big\}$ and $\epsilon \ge 0$, $\bsig$ is an arbitrary connected $(D+1)$-melonic graph and  $\tau_c = \mathrm{id}$ for every $c$:
\[
\lim_{N\rightarrow +\infty} 
\frac 1 {N } 
C_n\Biggl( \frac {N^{D+1}}{N^{\epsilon \frac{D(D-1)}2 + \beta D}}\Tr (AUBU^* )\Biggr)  = \tr (\bl)^n \sum_{  \substack{{\bsig\, \in \,\bS_n\textrm{ connected}
}\\{ (D+1)\textrm{-melonic}}}} 
    \tr_{\bsig}(\al) \,   \prod_{c=1}^D \M(\sigma_c) \;.
\]
\end{enumerate}

The two physical combinatorially prolific regimes are \ref{micAentitemII} and \ref{micAentitemI}. 
\end{Th}

\proof
Items \ref{micAentitemVI} and \ref{micAentitemI} are proven in Lemmata~\ref{lem:micromicro} and \ref{lem:rich-entangled} and the rest in Appendix~\ref{sec:microregproofs}.

\qed

\subsection{Symmetric regimes: Second main theorem}

We now assume a symmetric asymptotic scaling $\epsilon_A= \epsilon_B$ and $\beta_A = \beta_B$.

\paragraph{The symmetric macroscopic separable regime.}
Let $\beta_A = \beta_B=1$,  and $\epsilon_A= \epsilon_B =0 $:
\[
\Tr_{\bsig}(A)\sim N^{\sum_{c=1}^D \#(\sigma_c)}\,\tr_{\bsig}(\al)
 \qquad \textrm{ and } \qquad \Tr_{\btau}(B)\sim N^{\sum_{c=1}^D \#(\tau_c)}\,\tr_{\btau}(\bl) \; .
\]
This generalizes the regime \ref{D=1item1} in Thm.~\ref{th:RegimesD1}, corresponding to the original HCIZ integral \cite{HarishChandra, Itzyk-Zub}, in which all planar non-necessarily connected $(\sigma , \tau)$ contribute at large $N$. 
\begin{Lem}
\label{Th:first-sym-macro-sep}
For a symmetric macroscopic separable scaling ansatz 
$\epsilon_A= \epsilon_B =0 $ and $\beta_A = \beta_B=1$, we have:
\[ 
\lim_{N\rightarrow +\infty} 
\frac 1 {N^2 } 
C_n\bigl( N^{2-D}\Tr (AUBU^* )\bigr)  =   \sum_{  \substack{{\bsig, \btau \, \in\, \bS_n
}\\{
\DeltaC(\bsig, \btau) =0 }\\{\forall c,\, g(\sigma_c, \tau_c)=0
}}} \,  
    \tr_{\bsig}(\al) \,  \tr_{\btau^{-1}} (\bl) \, f \bigl[\mathrm{\bsig, \btau}\bigr] \; ,
\] 
and we stress that the graphs 
$(\bsig, \btau)$ are not necessarily connected. The leading order graphs include all $(\bsig,\bsig)$ with $\bsig$ $(D+1)$-melonic.
\end{Lem}
\proof This follows by writing the scaling \eqref{eq:anfang} as:
\be
\nonumber
2 +  n( \ggam 
+ D-2) -2\sum_c g(\sigma_c, \tau_c) - 2 \DeltaC(\bsig, \btau) \; .
\ee
Note that $g(\sigma_c,\tau_c) =0$ and $\DeltaC(\bsig,\btau)=0$ \emph{do not imply} $\btau \preceq \bsig$: an example is displayed in Fig.~\ref{fig:D2-min-conn}.

\qed

\paragraph{The symmetric macroscopic boundary regime.} This corresponds to the symmetric scaling ansatz:
\[
\begin{split}
& \Tr_{\bsig}(A)\sim N^{\frac 1 D \sum_c \# (\sigma_c)+ \frac 1 D \sum_{c_1<c_2} \#(\sigma_{c_1}\sigma_{c_2}^{-1})} \tr_{\bsig}(\al) \;,\crcr 
& \Tr_{\bsig}(B)\sim N^{\frac 1 D \sum_c \# (\tau_c)+ \frac 1 D \sum_{c_1<c_2} \#(\tau_{c_1}\tau_{c_2}^{-1})} \tr_{\bsig}(\bl) \; .
\end{split}
\]
\begin{Lem}\label{lem:smbr}
For $\epsilon_A=\epsilon_B = \beta_A = \beta_B = 1/D$, the leading order graphs $(\bsig,\btau)$ are the non-necessarily connected graphs with  melonic $\bsig$, $\btau$, and such that $\Box(\bsig, \btau) = 0$ and  $g(\sigma_c, \tau_c)=0, \; \forall c$:
\[
\lim_{N\rightarrow +\infty} \frac 1 {N^{2}}
C_n\Bigl( N \Tr (AUBU^* )\Bigr)  =  \sum_{ \substack{{\bsig, \btau \, \in \bS_n}\\{\omega(\bsig)=\omega(\btau)=0}\\{ \Box(\bsig , \btau)= 0}\\{\forall c,\, g(\sigma_c, \tau_c)=0}}}  \tr_{\bsig}(\al)\, \tr_{\btau^{-1}}(\bl)  \; \pC \;.
\]
The leading order graphs include all $(\bsig,\bsig)$ with $\bsig$ connected melonic.
\end{Lem}

\proof This follows from writing \eqref{eq:anfang} as: 
\[
   2 + n\bigl[ \ggam -1 \bigr]  -2\sum_c g(\sigma_c, \tau_c) - \frac 1 D \bigl( \omega(\bsig) + \omega(\btau)\bigr)  - 2\,\Box(\bsig, \btau) \;.
\]

The last assertion follows by observing that from \eqref{eq:boxdef} we have $\Box_{\bsig}(\bsig,\bsig) = 0$.

\qed

\

From Prop.~\ref{lem:positive-box-2} we have
that if $\Box(\bsig, \btau)= 0$ and $\bsig$ is $(D+1)$-melonic, then $\DeltaC(\bsig, \btau)=0$ and $\btau$ is $(D+1)$-melonic too.
On the other hand, if both $\bsig$ and $\btau$ are melonic but not $(D+1)$-melonic, we do have non-trivial contributions with vanishing $\Box$ but non-vanishing $\DeltaC$. The graph on the left of Fig.~\ref{fig:vanish-box} is such an example. 

\paragraph{All regimes for symmetric scalings.}
The regimes listed below are illustrated on the $\beta - \epsilon$ diagram in Fig.~\ref{fig:summary-results-0-sym}

\begin{Th}
\label{th:other-sym-regimes}
For symmetric scalings $(\epsilon_A = \epsilon_B = \epsilon, \beta_A = \beta_B = \beta)$, the various regimes are located in the exact same regions as the $\beta - \epsilon$ diagram. The leading order graphs in the regimes  \ref{symAentitemIII}, \ref{symAentitemIV},  \ref{symAentitemV}, and \ref{symAentitemVI} of the $\beta - \epsilon$ diagram are \emph{precisely the same} as for $A$ microscopic. They are different for the other regimes. In detail the leading order graphs $(\bsig,\btau)$ are:

\begin{enumerate}[label=\normalfont{S-\Roman*}]
\item\label{symAentitemI} - \underline{The symmetric macroscopic boundary regime}
(square dark red point in Fig.~\ref{fig:summary-results-0-sym}).
For $\epsilon = \beta = 1/D$, see 
Lemma~\ref{lem:smbr}, the leading order $(\bsig,\btau)$ are the non-necessarily connected graphs with $\bsig$ and $\btau$ melonic, $\Box(\bsig, \btau) = 0$ and  $g(\sigma_c, \tau_c)=0, \; \forall c$:
\[
\lim_{N\rightarrow +\infty} \frac 1 {N^{2}}
C_n\Bigl( N \Tr (AUBU^* )\Bigr)  =  \sum_{ \substack{{\bsig, \btau \, \in \bS_n}\\{\omega(\bsig)=\omega(\btau)=0}\\{ \Box(\bsig , \btau)= 0}\\{\forall c,\, g(\sigma_c, \tau_c)=0}}}  \tr_{\bsig}(\al)\, \tr_{\btau^{-1}}(\bl)  \; \pC \;.
\]
The leading order graphs include $(\bsig,\bsig)$ with $\bsig$ connected melonic. 
\item\label{symAentitemII} - \underline{The symmetric macroscopic separable regime:}  (the dark blue line in Fig.~\ref{fig:summary-results-0-sym}, including the endpoint $\beta=1$).
For $\beta=1-\epsilon(D-1)> 1/D$ and $\epsilon \ge 0$, $\DeltaC(\bsig, \btau)=0$ and $g(\sigma_c,\tau_c)=0$:
\[
\lim_{N\rightarrow +\infty} 
\frac 1 {N^2 } 
C_n\bigl( N^{2-D + \epsilon D (D-1)}\Tr (AUBU^* )\bigr)  =   \sum_{  \substack{{\bsig, \btau \, \in\, \bS_n
}\\{
\DeltaC(\bsig, \btau) =0 }\\{\forall c,\, g(\sigma_c, \tau_c)=0
}}} \,  
    \tr_{\bsig}(\al) \,  \tr_{\btau^{-1}} (\bl) \, f \bigl[\mathrm{\bsig, \btau}\bigr] \; ,
\]
where we stress that $(\bsig,\btau)$ is not necessarily connected.
This coincides with the regime at the end point $\beta=1$ in
Lemma~\ref{Th:first-sym-macro-sep}. This regime is richer that the regime \ref{micAentitemII} as $\btau\preceq\bsig$ implies 
$\DeltaC(\bsig, \btau) = \Omega_{D+1}(\bsig)$.
\item\label{symAentitemIII} - \underline{The symmetric mesoscopic boundary regime} (the yellow line in Fig.~\ref{fig:summary-results-0-sym}). For $0<\beta = \epsilon < 1/D $, the leading order graphs are the $(\bsig, \bsig)$, with $\bsig$ a connected melonic graph:
\[
\begin{split}
\lim_{N\rightarrow +\infty} 
\frac 1 {N^{2\epsilon D} } 
C_n\Bigl( N^{D(1- \epsilon (D-1) }\Tr (AUBU^* )\Bigr)  =   \sum_{  \substack{{\bsig\, \in\, \bS_n
\textrm{ connected}} \\{ \textrm{melonic} } }} 
    \tr_{\bsig}(\al) \,  \tr_{\bsig^{-1}} (\bl) \; .
    \end{split}
\]
\item\label{symAentitemIV} - \underline{The symmetric mesoscopic separable regime} (the orange region in Fig.~\ref{fig:summary-results-0-sym}). This coincides with the regime \ref{micAentitemIV}: for $0\le \epsilon<\beta<1-\epsilon(D-1) 
$, the leading order graphs are the $(\bsig, \bsig)$ with $\bsig$ connected $(D+1)$-melonic graph:
\[
\lim_{N\rightarrow +\infty} 
\frac 1 {N ^{2( \beta + \epsilon (D-1))}} 
C_n\Bigl( \frac{N^{D}}{N^{(D-1)(\epsilon(D-2) + 2 \beta)}}\Tr (AUBU^* )\Bigr)  =  \sum_{  \substack{{\bsig\,\in \bS_n \textrm{ connected}
}\\{ (D+1)\textrm{-melonic}}}} 
    \tr_{\bsig}(\al) \,  \tr_{\bsig^{-1}} (\bl) \;.
\]
\item\label{symAentitemV} - \underline{The symmetric entangled regime} (the green region in Fig.~\ref{fig:summary-results-0-sym}). For $0\le \beta < \min \{1/D , \epsilon \}$ the leading order graphs are such that all the $\tau_c,\sigma_c$ are the same cycle:
\[
\lim_{N\rightarrow +\infty} 
\frac 1 {N^{2\beta D} } 
C_n\Bigl( N^{D(1- \epsilon (D-1)) }\Tr (AUBU^* )\Bigr)  =   (n-1)! \    \tr(\al^n) \,  \tr (\bl^n) \; ,
\]
reproducing the regime in Lemma~\ref{thm:all-purely-entangled}.
\item\label{symAentitemVI} - \underline{The microscopic regime} (the origin in Fig.~\ref{fig:summary-results-0-sym}).
For $\beta = \epsilon = 0$, see Lemma~\ref{lem:micromicro}, at leading order $\bsig$ is an arbitrary connected graph, and $\btau=\bsig$:
\[
\lim_{N\rightarrow +\infty} 
C_n\bigl( {N^{D}}\Tr (AUBU^* )\bigr)  =   \sum_{  \bsig\,\in\, \bS_n \ \mathrm{ connected}}
    \tr_{\bsig}(\al) \,  \tr_{\bsig^{-1}} (\bl)  \; .
\]
\item\label{symAentitemVII} - \underline{The symmetric hyper-macroscopic boundary regime} (dark green line in Fig.~\ref{fig:summary-results-0-sym}). 
For $\epsilon > \beta = 1/D$, the leading order graphs are such that for all $c$, $\tau_c = \sigma_c=\sigma$:
\[
\lim_{N\rightarrow +\infty} 
\frac 1 {N^2} C_n\Bigl( N^{D(1- \epsilon (D-1) }\Tr (AUBU^* )\Bigr)  = \sum_{  \sigma\in S_n}  \    \prod_{{\eta \textrm{ cycle }}{\textrm{of $\sigma$}}} \tr\bigl(A^{l(\eta)}\bigr)\tr\bigl(B^{l(\eta)}\bigr) f[\bsig, \bsig] \;,
\]
with $l(\eta)$ the length of the cycle $\eta$.
\item\label{symAentitemVIII} - \underline{The symmetric hyper-macroscopic  regime:} (dark violet region in Fig.~\ref{fig:summary-results-0-sym}).
For $\epsilon \ge 0$ and $\beta>\max\big\{1/D , 1-\epsilon(D-1)\big\}$, at leading order, $\sigma_c=\tau_c = \mathrm{id}$ for all $c$: 
\[
\lim_{N\rightarrow +\infty} 
\frac 1 {N^2 } 
C_n\Biggl( \frac {N^{D+2}}{N^{\epsilon D(D-1) + 2\beta D}} \Tr (AUBU^* )\Biggr)  = \bigl(
\tr (\al)
\cdot
\tr (\bl)\bigr)^n \;.
\]
which incidentally coincides with the moments of the HCIZ integral (Sec.~\ref{sub:moments}), albeit for different $\ggam$ and $\ddel$.
\end{enumerate}

\end{Th}

The theorem is proven in Appendix~\ref{sec:symregproofs}.

\newpage

\appendix

\newpage

\section{Asymptotic regimes for $D=1$, proof of Theorem \ref{th:RegimesD1}}
\label{sec:D1}

The asymptotic expansion of the cumulants of the tensor HCIZ integral \eqref{eq:WC-norm} for $D=1$ reads: 
$$
C_n\bigl( {N^{\ggam}}\Tr (AUBU^* )\bigr)=  N^{n( \ggam - 2 ) }\sum_{  \sigma,\tau \in S_n}
  N^{s(\sigma, \tau) +s_A(\sigma) + s_B(\tau)}   \tr_{\sigma}(\al) \, \tr_{\tau}(\bl) f[\sigma, \tau](1+O(1)) \;, 
$$
where \eqref{eq:scaling-of-weingarten} and the asymptotic scaling ansatz are: 
\[
s(\sigma, \tau) = \#(\sigma\tau^{-1})- 2 \big[ \lvert \Pi(\sigma, \tau)\rvert-1 \big] \;, \qquad \textrm{and} \qquad s_A(\sigma)=\beta_A \#(\sigma) \; ,
\]
and similarly for $B$. Using  \eqref{eq:EulerCharGammac}, we rewrite the exponent of $N$ in a term as:
\be\label{eq:exponentD1}
2 + n(\ggam - 1) - 2g(\sigma, \tau)  - (1 - \beta_A ) \#(\sigma) - (1 - \beta_B) \#(\tau) \;,
\ee
where $g(\sigma, \tau)$ is the genus of the embedded map $(\sigma , \tau)$ discussed in Sec.~\ref{subsec:Gammac}. 
The scaling always selects planar graphs $(\sigma,\tau)$ and favors for $\beta < 1 $ cyclic permutations, for $1<\beta$ the identity permutation, and for $\beta=1$ it is insensitive to the number of cycles of the permutations:
\begin{enumerate}
\item If $\beta_A=\beta_B=1$, \eqref{eq:exponentD1} becomes $2 + n(\ggam - 1) - 2g(\sigma, \tau)$, so that $\ggam=1$, $\ddel = 2$, and the leading order graphs are $(\sigma,\tau)$ planar.
\item If $\beta_A<\beta_B=1$, \eqref{eq:exponentD1} is
$\beta_A + 1 + n(\ggam - 1) - 2g(\sigma, \tau) - (1-\beta_A) (\#(\sigma)-1)
$, so that $\ggam = 1$, $\ddel = \beta_A+1$ and at leading order $g(\sigma, \tau)=0$ and $\#(\sigma)=1$. From
Prop.~\ref{prop:non-cross-cc}, $\tau$ is non-crossing on  $\sigma$ and from Thm.~\ref{th:previous-results}, $f[\sigma, \tau] = \M(\sigma\tau^{-1})$. 
\item If $\beta_A\le \beta_B<1$, we write \eqref{eq:exponentD1}  as: 
$$
\beta_A + \beta_B + n(\ggam -1 ) - 2g(\sigma, \tau)  - ( 1-\beta_A) (\#(\sigma)-1) - (1-\beta_B) (\#(\tau)-1) \; ,
$$
hence $\ddel = \beta_A + \beta_B$, $\ggam = 1$ and at leading order $g(\sigma, \tau) =0$ and $\#(\sigma)=\#(\tau)=1$. From Prop. \ref{lem:supports-included} $\sigma=\tau$ and from Thm.~\ref{th:previous-results} $f \bigl[\sigma, \tau\bigr]=1$.
\item  If $\beta_A < 1 <\beta_B$, we write \eqref{eq:exponentD1} as: 
$$
\beta_A + 1 + n(\ggam + \beta_B - 2 ) - 2g(\sigma, \tau)  - ( 1-\beta_A) (\#(\sigma)-1) - (\beta_B-1) (n- \#(\tau)) \; ,
$$
hence $\ddel = \beta_A+1$, $\ggam = 2-\beta_B$, and at leading order $\#(\tau)=n$, i.e.~$\tau= \mathrm{id}$, and $\#(\sigma)=1$. From Thm.~\ref{th:previous-results}, $f[\sigma, \id] = \M(\sigma) = \frac{(-1)^{n-1}}{n}\binom{2n-2}{n-1}$ since $\#(\sigma)=1$. As in the regime \ref{D=1item2}, the contribution is the same for all cycles.

\item If $\beta_A =1 <\beta_B$, we write \eqref{eq:exponentD1} as
$2 + n(\ggam + \beta_B - 2 ) - 2g(\sigma, \tau)   - (\beta_B-1) (n- \#(\tau))
$,
hence $\ddel = 2$, $\ggam = 2-\beta_B$, and at leading order $\#(\tau)=n$, i.e.~$\tau= \mathrm{id}$, and $\sigma$ is arbitrary.

\item If  $1<\beta_A \le \beta_B$, we write \eqref{eq:exponentD1} as: 
$$
2 + n(\ggam + \beta_A + \beta_B - 3 ) - 2g(\sigma, \tau)  -  (\beta_A -1)(n -  \#(\sigma)) - (\beta_B -1)(n  - \#(\tau) ) \;,
$$
hence $\ddel =2$, $\ggam=3-\beta_A-\beta_B$. At leading order, $\sigma=\tau=\id$ and $f[\id, \id]$ follows from Thm.~\ref{th:previous-results}.
\end{enumerate}
This concludes the proof of Thm. \ref{th:RegimesD1}.

\newpage
\section{Proof of 
Theorem~\ref{th:other-ent-microscopic-regimes}: regimes with $A$ microscopic}
\label{sec:microregproofs}
\addtocontents{toc}{\protect\setcounter{tocdepth}{1}}

We check the rest of the regimes in 
Theorem~\ref{th:other-ent-microscopic-regimes}.

\subsection{Regime \ref{micAentitemII}}

\begin{Lem}
For $\beta=1-\epsilon(D-1)>\frac 1 D$ and $\epsilon \ge 0$, the leading order graphs are the $(\bsig, \btau)$ with $\bsig$ connected  $(D+1)$-melonic and $\btau \preceq \bsig$:
\[
\lim_{N\rightarrow +\infty} 
\frac 1 {N } 
C_n\Bigl( N^{1 + \epsilon \frac{D(D-1)}2}\Tr (AUBU^* )\Bigr)  =   \sum_{  \substack{{\bsig\, \in \bS_n \textrm{ connected}
}\\{ (D+1)\textrm{-melonic}}}} \, \sum_{ \substack{{\btau ,\; \btau \preceq \bsig}}}\, 
    \tr_{\bsig}(\al) \,  \tr_{\btau^{-1}} (\bl) \, \prod_{c=1}^D \M(\sigma_c \tau_c^{-1}).
\]
\end{Lem}

\proof Using $\omega(\btau)$ \eqref{eq:degree-bsig}, $\Box_{\btau}$ \eqref{eq:boxdef}, and $\DeltaC$ \eqref{eq:def-DeltaK}, the scaling in \eqref{eq:anfang} reads:
\begin{align*}
& 1 + n\bigl[ \ggam 
-D + \epsilon \frac{D(D-1)}2 + (1-\epsilon D)(D-1)\bigr]  - (\lvert \Pi(\bsig, \btau)\rvert - 1)  \\ 
&\qquad \quad -\sum_c\bigl( 2g(\sigma_c, \tau_c) +\#(\sigma_c) - \lvert \Pi(\sigma_c, \tau_c)\rvert\bigr) - \epsilon \omega(\btau) - \epsilon D \Box_{\btau}(\bsig, \btau) -(1-\epsilon D)\DeltaC(\bsig, \btau) \;, 
\end{align*}
thus $\Pi(\bsig,\btau)=1$ and from 
Prop.~\ref{lem:supports-included} we get
$\btau\preceq \bsig$ and $\Pi(\bsig) = \Pi(\bsig,\btau) =1$. As $\btau\preceq \bsig$, from Prop.~\ref{lem:positive-box-2} we get that $ \DeltaC(\bsig, \btau)=0$ if and only if $\bsig$ is $(D+1)$-melonic.
On the other hand, $  \DeltaC(\bsig, \btau) = 0$ ensures that   
$\Omega_{D+1}(\btau)=0$, hence $\omega(\btau)=0$.

\qed

\subsection{Regime \ref{micAentitemIII}} 
\begin{Lem} 
For $0<\beta=\epsilon<1/D $, the leading order graphs are $(\bsig, \bsig)$ with $\bsig$ connected melonic:
\[
\lim_{N\rightarrow +\infty} 
\frac 1 {N^{\epsilon D} } 
C_n\Bigl( N^{D- \epsilon \frac{D(D-1)}2 }\Tr (AUBU^* )\Bigr)  =   \sum_{  \substack{{\bsig\, \in\, \bS_n
\textrm{connected} }\\{\textrm{melonic}}}} 
    \tr_{\bsig}(\al) \,  \tr_{\bsig^{-1}} (\bl).
\]
\end{Lem}
\proof Using \eqref{eq:degree-bsig}, \eqref{eq:boxdef} and \eqref{eq:def-DeltaK}, we rewrite \eqref{eq:anfang} as:
\begin{align*}
& \epsilon D + n\bigl[ \ggam -D + \epsilon \frac{D(D-1)}2\bigr]   -\sum_c\Bigl( 2g(\sigma_c, \tau_c) +\#(\sigma_c) - \lvert \Pi(\sigma_c, \tau_c)\rvert \Bigr) - \epsilon \omega(\btau) \\ &\hspace{2cm} - (2- \epsilon D) (\lvert \Pi(\bsig, \btau)\rvert - 1) - \epsilon D\, \Box_{\btau}(\bsig, \btau) -(1-\epsilon D)\sum_c\bigl[\#(\tau_c) - \lvert \Pi(\sigma_c, \tau_c)\rvert\bigr] \; . 
\end{align*}
From Prop.~\ref{lem:supports-included} we conclude that $\btau = \bsig$ and the lemma follows.

\qed

\subsection{Regime \ref{micAentitemIV}}
\begin{Lem}
For $0\le \epsilon<\beta<1-\epsilon(D-1)$, the leading order graphs are the $(\bsig, \bsig)$ with $\bsig$ connected $(D+1)$-melonic, and:
\[
\lim_{N\rightarrow +\infty} 
\frac 1 {N^{\beta + \epsilon (D-1)} } 
C_n\Biggl( \frac {N^D}{N^{(D-1)(\beta-\epsilon + \frac  {\epsilon D} 2 )}}\Tr (AUBU^* )\Biggr)  =   \sum_{  \substack{{\bsig\,\in \bS_n \textrm{ connected}
}\\{ (D+1)\textrm{-melonic}}}} 
    \tr_{\bsig}(\al) \,  \tr_{\bsig^{-1}} (\bl)  \;.
\]
\end{Lem}

\proof Using the expressions for $\omega(\btau)$ \eqref{eq:degree-bsig}, $\Omega_{D+1}(\btau)$ \eqref{eq:c-degree-bsig} and $\Box_{\btau}$ \eqref{eq:boxdef}, the scaling in $N$ in \eqref{eq:anfang} becomes:
\begin{align*}
&   \beta + \epsilon (D-1) + n\bigl[ \ggam -D + \epsilon \frac{D(D-1)}2 + (\beta-\epsilon )(D-1)\bigr]    \\ & 
\qquad - (\lvert \Pi(\bsig, \btau)\rvert - 1)- \epsilon \omega(\btau) -\sum_c\bigl( 2g(\sigma_c, \tau_c) + \#(\sigma_c) - \lvert \Pi(\sigma_c, \tau_c)\rvert \bigr)
\\ & \qquad\qquad  -  \Box_{\btau}(\bsig, \btau) -(\beta-\epsilon )\Omega_{D+1}(\btau) -(1+\epsilon - \beta - \epsilon D)(\lvert \Pi(\btau)\rvert-1) .\nonumber
\end{align*}
The leading order graphs are connected. Moreover, since $1+\epsilon - \beta - \epsilon D>0$, they have $\lvert \Pi(\btau)\rvert=1$, so that $ \Box_{\btau}(\bsig, \btau) = \sum_c\bigl[\#(\tau_c) - \lvert \Pi(\sigma_c, \tau_c)\rvert\bigr] =0$.
As $g(\sigma_c, \tau_c)=0$ and $\#(\sigma_c) - \lvert \Pi(\sigma_c, \tau_c)\rvert=0$, from Prop.~\ref{lem:supports-included} we conclude that $\btau=\bsig$. Taking into account that
$\Omega_{D+1}(\btau)=0 \Rightarrow \omega(\btau)=0$, we conclude that $\bsig=\btau$ and $\bsig$ is a connected $(D+1)$-melonic graph. 

\qed

\subsection{Regime \ref{micAentitemV}} 
\begin{Lem}
For $\epsilon > \beta$, $\epsilon > 0$ and $\beta < 1/D $, the leading order graphs are such that all the $\sigma_c,\tau_c$ are the same cycle:
\[
\lim_{N\rightarrow +\infty} 
\frac 1 {N^{\beta D} } 
C_n\Bigl( N^{D- \epsilon \frac{D(D-1)}2 }\Tr (AUBU^* )\Bigr)  =   (n-1)! \    \tr(\al^n) \,  \tr (\bl^n).
\]
\end{Lem}
\proof For  $\epsilon > \max(\beta, 0)$ and $\beta < 1/D$, we rewrite \eqref{eq:anfang} as:
\begin{align*}
&\beta D + n\bigl[ \ggam -D + \epsilon \frac{D(D-1)}2\bigr]   -\sum_c\bigl( 2g(\sigma_c, \tau_c) + \#(\sigma_c) - \lvert \Pi(\sigma_c, \tau_c)\rvert \bigr) - (\lvert \Pi(\bsig, \btau)\rvert - 1) \\ 
&\qquad - \epsilon \omega(\btau) - \Box_{\btau}(\bsig, \btau) -(\epsilon - \beta)\bigl[\sum_c\#(\tau_c) - D\lvert \Pi(\btau)\rvert\bigr] - (1 - \beta D) (\lvert \Pi(\btau)\rvert-1) \; .
\end{align*}
At leading order, $\lvert \Pi(\btau)\rvert=1$, so that $\sum_c\#(\tau_c) - D\lvert \Pi(\btau)\rvert=\sum_c\#(\tau_c) - D $, which vanishes if and only if for all $c$, $\tau_c$ is a cycle of length $n$. Since $\tau_c\preceq \sigma_c$, this imposes that $\sigma_c=\tau_c$, which implies $\Box_{\btau}(\bsig, \btau)=0$. Finally, from \eqref{eq:degree-bsig}, $\omega(\btau)=0$ and each $\tau_c$ is a cycle if and only if all the $\tau_c$ are equal. 

\qed

\subsection{Regime \ref{micAentitemVII}} 

\begin{Lem}
For $\epsilon > \beta = 1/D$,  the leading order graphs are such that 
$\btau= (\tau, \dots, \tau)$ for some $\tau\in S_n$, $\bsig$ is connected, $\btau \preceq \bsig$, and
$\Box_{\btau}(\bsig,\btau)=0$:
\[
\lim_{N\rightarrow +\infty} 
\frac 1 N C_n\Bigl( N^{D- \epsilon \frac{D(D-1)}2 }\Tr (AUBU^* )\Bigr)  = \sum_{  \substack{{\bsig\, \in \bS_n}\\{\textrm{connected}} }} \; \sum_{ 
\substack{ { \btau = (\tau,\dots ,\tau), \,  \tau \in S_n, }\\
{\btau \preceq \bsig , \; \Box_{\btau}(\bsig,\btau)=0 }  }
} \hspace{-0.3cm}  \tr_{\bsig}(\al) \,  \tr_{\btau^{-1}} (\bl) \, \prod_{c=1}^D \M(\sigma_c \tau_c^{-1}).
\]
The leading order graph include $(\bsig,
\mathbf{id} )$ with $\bsig$ connected $(D+1)$-melonic and 
$\mathbf{id}=(\id, \dots, \id ) \in \bS_n$. 
\end{Lem}

\proof For  $\epsilon > \beta$ and $\beta =1/D$, we rewrite \eqref{eq:anfang} as:
\begin{align*}
&1 + n\bigl[ \ggam -D + \epsilon \frac{D(D-1)}2\bigr]   -\sum_c\bigl( 2g(\sigma_c, \tau_c) + \#(\sigma_c) - \lvert \Pi(\sigma_c, \tau_c)\rvert \bigr) - (\lvert \Pi(\bsig, \btau)\rvert - 1) \\ 
&\qquad  - \epsilon \omega(\btau) - \Box_{\btau}(\bsig, \btau) -(\epsilon - \frac 1 D)\bigl[\sum_c\#(\tau_c) - D\lvert \Pi(\btau)\rvert\bigr]  \; .
\end{align*}
The leading order graphs $(\bsig,\btau)$ are connected, and from Prop.~\ref{lem:supports-included} they satisfy $\btau \preceq \bsig$ and $\Pi(\bsig)=1$. From \eqref{eq:degree-bsig}, we note that $\btau$ satisfies $\omega(\btau)=0$ and $\sum_c\#(\tau_c) = D\lvert \Pi(\btau)\rvert$ if and only if all 
$\tau_c =\tau$ from some $\tau\in S_n$.
 
The last assertion follows by noting that $\mathbf{id} \preceq \bsig $; \eqref{eq:c-degree-bsig} and \eqref{eq:def-DeltaK} imply that $\DeltaC(\bsig,\mathbf{id} ) = \Omega_{D+1}(\bsig) $,
and from Prop.~\ref{lem:positive-box-2}, 
$\Box_{\btau}(\bsig, \mathbf{id} )=0$ if and only if 
$\DeltaC(\bsig,\mathbf{id}) = 0 $.

\qed

\subsection{Regime \ref{micAentitemVIII}} 
\begin{Lem}
For $\epsilon \ge \frac 1 D$,  $\beta > \frac 1 D$, and $0\le\epsilon < \frac 1 D$, $\beta > 1- \epsilon(D-1)$, the leading order graphs are $(\bsig, \btau)$ with $\bsig$ connected  $(D+1)$-melonic, and
$\tau_c = \id$ for every $c$:
\[
\lim_{N\rightarrow +\infty} 
\frac 1 {N } 
C_n\Biggl( \frac {N^{D+1}}{N^{\epsilon \frac{D(D-1)}2 + \beta D}}\Tr (AUBU^* )\Biggr)  = \tr (\bl)^n \sum_{  \substack{{\bsig\, \in \,\bS_n\textrm{ connected}
}\\{ (D+1)\textrm{-melonic}}}} 
    \tr_{\bsig}(\al) \,   \prod_{c=1}^D \M(\sigma_c).
\]
\end{Lem}

\proof We divide the proof for this regime in two regions: $\epsilon \ge 1/D $ and $\beta > 1/D$ on one hand, and $0\le\epsilon < 1/D $ and $\beta > 1- \epsilon(D-1)$ on the other hand. 

For $\epsilon \ge 1/D $ and $\beta > 1/D $, we rewrite the scaling with $N$ in \eqref{eq:anfang} as:
\begin{align*}
&   1 + n\bigl[ \ggam -D + \epsilon \frac{D(D-1)}2 + \beta D- 1\bigr]  -\sum_c\bigl( 2g(\sigma_c, \tau_c) +\#(\sigma_c) - \lvert \Pi(\sigma_c, \tau_c)\rvert\bigr) - \epsilon \omega(\btau) \\&\hspace{0.6cm} - (\lvert \Pi(\bsig, \btau)\rvert - 1)  -  \Box_{\btau}(\bsig, \btau) - (\beta- \frac 1 D )\sum_c(n- \#(\tau_c)) -(\epsilon - \frac 1 D)(\sum_c  \#(\tau_c) - D\lvert \Pi(\btau)\rvert)  \; .
\end{align*}
The leading order graphs are connected and such that for all $c$, $\#(\tau_c)=n$, that is, $\tau_c=\id$. From \eqref{eq:boxdef}, we have $\Box_{\bsig}(\bsig,\btau)=0$, that is, 
$\DeltaC(\bsig,\btau) = \Omega_{D+1}(\bsig)$, while from Prop.~\ref{lem:positive-box-2}  we get $\Box_{\btau}(\bsig,\btau) = \DeltaC(\bsig,\btau)$. It follows that at leading order, $\bsig$ is $(D+1)$-melonic and connected (as $\Pi(\bsig,\btau)=1$).
For $0\le\epsilon < 1/D$ and $\beta > 1- \epsilon(D-1)$, we follow the same reasoning starting from rewriting \eqref{eq:anfang} as:
\begin{align*}
&   1+ n\Bigl[ \ggam -D + \epsilon \frac{D(D-1)}2 + (1-\epsilon D)(D- 1) + (\beta - \epsilon + \epsilon D - 1)D\Bigr]   - (\lvert \Pi(\bsig, \btau)\rvert - 1)  \\
&\qquad  - \sum_c \bigl( 2g(\sigma_c, \tau_c)  + \#(\sigma_c) - \lvert \Pi(\sigma_c, \tau_c)\rvert\bigr)- (\beta - \epsilon + \epsilon D - 1)\sum_c\bigl(n -   \#(\tau_c)\bigr)\\&\hspace{5cm}  - \epsilon \omega(\btau) -  \epsilon D\, \Box_{\btau}(\bsig, \btau) - (1 - \epsilon D )\DeltaC(\bsig, \btau)   \; .
\end{align*}
\vspace{-0.2cm}
\qed

\vspace{-0.2cm}

\section{Proof of Theorem~\ref{th:other-sym-regimes}: symmetric regimes}
\label{sec:symregproofs}

We now check the rest of the regimes in 
Thm.~\ref{th:other-sym-regimes}.

\subsection{Regime \ref{symAentitemII}}

\begin{Lem}
For $\beta=1-\epsilon(D-1)> 1/D$ and $\epsilon \ge 0$, we have:
\[
\lim_{N\rightarrow +\infty} 
\frac 1 {N^2 } 
C_n\bigl( N^{2-D + \epsilon D (D-1)}\Tr (AUBU^* )\bigr)  =   \sum_{  \substack{{\bsig, \btau \, \in\, \bS_n
}\\{
\DeltaC(\bsig, \btau) =0 }\\{\forall c,\, g(\sigma_c, \tau_c)=0
}}} \,  
    \tr_{\bsig}(\al) \,  \tr_{\btau^{-1}} (\bl) \, f \bigl[\mathrm{\bsig, \btau}\bigr] \;.
\]
\end{Lem}

\proof  Using $\omega(\btau)$ from  \eqref{eq:degree-bsig}, $\Box$ from Prop.~\ref{lem:positive-box-2}, $\Box_{\btau}$ from \eqref{eq:boxdef}, and $\DeltaC$ from \eqref{eq:def-DeltaK}, we rewrite \eqref{eq:anfang} as: 
\begin{align*}
& 2 + n\bigl[ \ggam 
-D + \epsilon D(D-1) + 2(1-\epsilon D)(D-1)\bigr]   \\
&\qquad \quad - 2 \sum_c g(\sigma_c, \tau_c)  - \epsilon (\omega(\bsig) + \omega(\btau)) - 2\epsilon D\, \Box(\bsig, \btau)- 2 (1-\epsilon D)\DeltaC(\bsig, \btau) \; .
\end{align*}
From Prop.~\ref{lem:positive-box-2}, $\DeltaC(\bsig, \btau)=0$ implies that both $\Box(\bsig, \btau)=0$ and $\Omega_{D+1}(\bsig)=\Omega_{D+1}(\btau) =0$, which from Prop.~\ref{thm:coloredgr} implies that $\omega(\bsig)=\omega(\btau)=0$. 

\qed

\subsection{Regime \ref{symAentitemIII}}

\begin{Lem}
For $\beta_A=\beta_B=\epsilon_A=\epsilon_B=\epsilon$, with $0<\epsilon< 1/D $, the leading order graphs are the $(\bsig, \bsig)$, with $\bsig$ a connected melonic graph:
\[
\begin{split}
\lim_{N\rightarrow +\infty} 
\frac 1 {N^{2\epsilon D} } 
C_n\Bigl( N^{D(1- \epsilon (D-1) }\Tr (AUBU^* )\Bigr)  =   \sum_{  \substack{{\bsig\, \in\, \bS_n
\textrm{ connected}} \\{\omega(\bsig)=0} }} 
    \tr_{\bsig}(\al) \,  \tr_{\bsig^{-1}} (\bl) \; .
    \end{split}
\]
\end{Lem}
\proof 
In this regime, the scaling in \eqref{eq:anfang} reads: 
\begin{align*}
& 2\epsilon D + n\bigl[ \ggam -D + \epsilon D(D-1)\bigr]   - 2 \sum_c g(\sigma_c, \tau_c) - \epsilon \bigl(\omega(\bsig) + \omega(\btau)\bigr) - 2(1- \epsilon D) \bigl(\lvert \Pi(\bsig, \btau)\rvert - 1\bigr)  \\ 
&\qquad  - 2\epsilon D\, \Box(\bsig, \btau)  -(1-\epsilon D)\sum_c\bigl[\#(\sigma_c) + \#(\tau_c) - 2 \lvert \Pi(\sigma_c, \tau_c)\rvert\bigr] \; .
\end{align*}
The leading order graphs have $\omega(\bsig)=0$. They also have $g(\sigma_c, \tau_c)=0$, $\#(\sigma_c) - \lvert \Pi(\sigma_c, \tau_c)\rvert=0$, and $\#(\tau_c) - \lvert \Pi(\sigma_c, \tau_c)\rvert =0 $, and  hence have $\bsig = \btau$ (Prop.~\ref{lem:supports-included}) and 
 are connected. But then $\lvert \Pi(\btau)\rvert= |\Pi(\bsig)|=1$, so that $\Box_{\btau}(\bsig, \btau) = \sum_c\bigl[\#(\tau_c) - \lvert \Pi(\sigma_c, \tau_c)\rvert\bigr]$ and $\Box_{\bsig}(\bsig, \btau) = \sum_c\bigl[\#(\sigma_c) - \lvert \Pi(\sigma_c, \tau_c)\rvert\bigr]$. 

\qed

\subsection{Regime \ref{symAentitemIV}}

%
\begin{Lem}
For $\beta_A=\beta_B=\beta$ and $\epsilon_A=\epsilon_B=\epsilon$, with  $0\le \epsilon<\beta<1-\epsilon(D-1) 
$, the leading order graphs are the $(\bsig, \bsig)$ with $\bsig$ a connected $(D+1)$-melonic graph:
\[
\lim_{N\rightarrow +\infty} 
\frac 1 {N ^{2( \beta + \epsilon (D-1))}} 
C_n\Bigl( \frac{N^{D}}{N^{(D-1)(\epsilon(D-2) + 2 \beta)}}\Tr (AUBU^* )\Bigr)  =  \sum_{  \substack{{\bsig\,\in \bS_n \textrm{ connected}
}\\{ (D+1)\textrm{-melonic}}}} 
    \tr_{\bsig}(\al) \,  \tr_{\bsig^{-1}} (\bl) \;.
\]
\end{Lem}

\proof Using $\omega(\btau)$ from  \eqref{eq:degree-bsig}, $\Box$ from Prop.~\ref{lem:positive-box-2}, $\Box_{\btau}$ from \eqref{eq:boxdef}, and $\Omega_{D+1}(\btau)$ from \eqref{eq:c-degree-bsig}, we rewrite \eqref{eq:anfang} as: 
\begin{align*}
& 2\beta + 2\epsilon(D-1)+  n\bigl(\ggam - D + \epsilon D (D-1) + 2(\beta - \epsilon)(D-1) \bigr) -2\sum_c g(\sigma_c, \tau_c)    \\ & \qquad
- \epsilon(\omega(\btau) + \omega(\bsig)) - (\beta - \epsilon)(\Omega_{D+1}(\bsig) + \Omega_{D+1}(\btau) ) -  2\, \Box(\bsig, \btau) \\ 
& \qquad\qquad 
 -(1+\epsilon - \beta - \epsilon D)(\lvert \Pi(\btau)\rvert + \lvert \Pi(\bsig)\rvert- 2) \; . 
\end{align*}
Since $1+\epsilon - \beta - \epsilon D>0$, at leading order we have $\lvert \Pi(\btau)\rvert=\lvert \Pi(\bsig)\rvert=1$, so that:
\[
\Box_{\btau}(\bsig, \btau) = \sum_c\bigl[\#(\tau_c) - \lvert \Pi(\sigma_c, \tau_c)\rvert\bigr] =0 , \qquad \mathrm{and} \qquad \Box_{\bsig}(\bsig, \btau) = \sum_c\bigl[\#(\sigma_c) - \lvert \Pi(\sigma_c, \tau_c)\rvert\bigr] =0 \; ,
\]
which together with $g(\sigma_c,\tau_c)=0$
imposes that $\bsig= \btau$
(Prop.~\ref{lem:supports-included}). 
As (see Prop.~\ref{thm:coloredgr}) $\Omega_{D+1}(\btau)=0$ implies $\omega(\btau)=0$, we find that at leading order $\bsig$ is connected $(D+1)$-melonic, and $\btau = \bsig$. 

\qed

\subsection{Regime \ref{symAentitemV}} 
\begin{Lem}For $\epsilon > \beta$, $\epsilon>0$ and $\beta < 1/D$, the leading order graphs are such that
all the $\tau_c,\sigma_c$ are the same cycle, and:
\[
\lim_{N\rightarrow +\infty} 
\frac 1 {N^{2\beta D} } 
C_n\Bigl( N^{D(1- \epsilon (D-1)) }\Tr (AUBU^* )\Bigr)  =   (n-1)! \    \tr(\al^n) \,  \tr (\bl^n) \; .
\]
\end{Lem}

\proof For  $\epsilon > \beta$ and $\beta < 1/D $ we rewrite \eqref{eq:anfang} as:
\begin{align*}
& 2\beta D + n\bigl[ \ggam -D + \epsilon D(D-1)\bigr]   -(\epsilon - \beta)\bigl[\sum_c \bigl(\#(\sigma_c) + \#(\tau_c)\bigr) - D\bigl(\lvert \Pi(\bsig)\rvert + \lvert \Pi(\btau)\rvert\bigr)\bigr]   \\ 
&\qquad   -\sum_c  2g(\sigma_c, \tau_c)- \epsilon (\omega(\bsig) + \omega(\btau)) - 2\,\Box(\bsig, \btau)  - (1 - \beta D) (\lvert \Pi(\bsig)\rvert+\lvert \Pi(\btau)\rvert-2) \; . 
\end{align*}
The leading order graphs are such that $\lvert \Pi(\btau)\rvert=1$, which implies $\sum_c\#(\tau_c) - D\lvert \Pi(\btau)\rvert=\sum_c\#(\tau_c) - D\ge0$, which in turn vanishes if and only if $\#(\tau_c)=1$. They also have $\omega(\btau)=0$, which imposes that all the $\tau_c$ are the same cycle. The same goes for $\bsig$. For the genera $g(\sigma_c, \tau_c)$ to vanish, $\sigma_c$ and $\tau_c$ must be the same cycle.  $\Box(\bsig, \btau)$ indeed vanishes for such contributions. 

\qed

\subsection{Regime \ref{symAentitemVII}} 

\begin{Lem}
For $\epsilon > \beta = 1/D$, the leading order graphs are such that for all $c$, $\tau_c = \sigma_c=\sigma$, not necessarily connected, and:
\[
\lim_{N\rightarrow +\infty} 
\frac 1 {N^2} C_n\Bigl( N^{D(1- \epsilon (D-1) }\Tr (AUBU^* )\Bigr)  = \sum_{  \sigma\in S_n}  \    \prod_{{\eta \textrm{ cycle }}{\textrm{of $\sigma$}}} \tr\bigl(A^{l(\eta)}\bigr)\tr\bigl(B^{l(\eta)}\bigr) f[\bsig, \bsig] \;,
\]
with $l(\eta)$ the length of the cycle $\eta$.
\end{Lem}
We can give a close expression for $f[\bsig, \bsig]$ using \eqref{eq:comb-expr-leading-cum-Weing}:
\[
f[\bsig, \bsig] = 2^{nD} 
\sum_{\substack{{\pi_1 ,\ \ldots\ ,\  \pi_D}\\
{|\Pi(\sigma)\vee\pi_1\vee\ldots\vee\pi_D | = 1 }\\{\sum_c  \lvert \Pi(\pi_c)\rvert= 1 + nD - \lvert \Pi(\sigma)\rvert }}} \   \prod_{c=1}^D \prod_{B\in \pi_c} \frac{(3\lvert B \rvert  - 3)!}{(2\lvert B \rvert )!} \;.
\]

\proof For  $\epsilon > \beta = 1/D$, \eqref{eq:anfang} becomes:
\begin{align*}
& 2 + n\bigl[ \ggam -D + \epsilon D(D-1)\bigr]   -\sum_c 2g(\sigma_c, \tau_c) - \epsilon ( \omega(\bsig)  + \omega(\btau)) - 2\,\Box(\bsig, \btau) \\
&\qquad   -(\epsilon - \frac 1 D)\bigg[\sum_c \bigl(\#(\sigma_c) + \#(\tau_c)\bigr) - D\bigl(\lvert \Pi(\bsig)\rvert + \lvert \Pi(\btau)\rvert\bigr)\bigg] \; .
\end{align*}
At leading order, $\bsig$ and $\btau$ must be melonic,  and since $\sum_c \#(\sigma_c) =D\lvert \Pi(\bsig)\rvert $ and $\sum_c \#(\tau_c) = D \lvert \Pi(\btau)\rvert$, from  \eqref{eq:degree-bsig} it follows that $\sigma_c =\sigma$ and $\tau_c = \tau$ for all $c$. With these constraints, $\Box(\bsig, \btau)=0$ is equivalent to $\#(\sigma)=\#(\tau) = \lvert \Pi(\sigma, \tau)\rvert$ and since $g(\sigma, \tau)=0$, $\sigma=\tau$ (Prop.~\ref{lem:supports-included}). 

\qed

\subsection{Regime \ref{symAentitemVIII}}

\begin{Lem}
In the regions $\epsilon \ge 1/D$, $\beta > 1/D$ and $0\le\epsilon < 1/D$,  $\beta > 1- \epsilon(D-1)$, the leading order graphs are such that $\sigma_c=\tau_c = \id$ for all $ c$:
\[
\lim_{N\rightarrow +\infty} 
\frac 1 {N^2 } 
C_n\Biggl( \frac {N^{D+2}}{N^{\epsilon D(D-1) + 2\beta D}} \Tr (AUBU^* )\Biggr)  = \bigl(
\tr (\al)
\cdot
\tr (\bl)\bigr)^n.
\]
\end{Lem}

\proof We divide the proof for this regime in two regions: $\epsilon \ge 1/D$ and $\beta > 1/D$ on one hand, and $0\le\epsilon < 1/D $ and $\beta > 1- \epsilon(D-1)$ on the other. 

For $\epsilon \ge  1/D $ and $\beta > 1/D $, we rewrite \eqref{eq:anfang} as:
\begin{align*}
&   2 + n\bigl[ \ggam -D + \epsilon D(D-1) + 2\beta D- 2\bigr]  -2\sum_c g(\sigma_c, \tau_c)  - \epsilon (\omega(\bsig) + \omega(\btau)) - 2\,  \Box(\bsig, \btau) 
\\& - (\beta- \frac 1 D )\sum_c \bigl[n-\#(\sigma_c) +n- \#(\tau_c)\bigr] -(\epsilon - \frac 1 D)\bigg[\sum_c \bigl(\#(\sigma_c) + \#(\tau_c)\bigr) - D\bigl(\lvert \Pi(\bsig)\rvert + \lvert \Pi(\btau)\rvert\bigr)\bigg]  \; ,
\end{align*}
which at leading order, due to the first term in the second line, restricts to $\sigma_c=\tau_c=\id$ for all $c$. 
For $0\le\epsilon < 1/D $ and $\beta > 1- \epsilon(D-1)$, we reach the same conclusion by writing \eqref{eq:anfang} as:
\begin{align*}
&   2+ n\Bigl[ \ggam -D + \epsilon D(D-1) + 2(1-\epsilon D)(D- 1) + 2(\beta - \epsilon + \epsilon D - 1)D\Bigr]    \\
& \qquad  - \sum_c 2g(\sigma_c, \tau_c)  - (\beta - \epsilon + \epsilon D - 1)\sum_c \bigl[n-\#(\sigma_c) +n- \#(\tau_c)\bigr]\\
&\qquad\qquad  - \epsilon (\omega(\bsig) + \omega(\btau)) -  2\,\epsilon D\, \Box(\bsig, \btau) - 2(1 - \epsilon D )\DeltaC(\bsig, \btau)  \;  .
\end{align*}
\qed


\vspace{-0.2cm}

\begin{thebibliography}{99}



\bibitem{HarishChandra} 
Harish-Chandra, ``Differential Operators on a Semisimple Lie Algebra.'' American Journal of Mathematics 79, no. 1 (1957): 87-120.

\bibitem{Itzyk-Zub} 
C.~Itzykson and J.-B.~Zuber, ``The planar approximation. II'', Journal of Mathematical Physics 21:3, 411-421.

\bibitem{CGL} 
B.~Collins, R.~Gurau, L.~Lionni, 
``The tensor Harish-Chandra--Itzykson--Zuber integral I: Weingarten calculus and a generalization of monotone Hurwitz numbers'', accepted for publication in J. Eur. Math. Soc. (JEMS),
[arXiv:2010.13661].


\bibitem{CHS} 
B.~Collins, T.~Hasebe, N.~Sakuma, 
``Free probability for purely discrete eigenvalues of random matrices.'', 
J. Math. Soc. Japan 70 (2018), no. 3, 1111–1150.


\bibitem{LU-1}
R.~Horodecki, P.~Horodecki, M.~Horodecki, and K.~Horodecki,
``Quantum entanglement'',
Rev. Mod. Phys. 81, 865,
[arXiv:quant-ph/0702225]. 

\bibitem{LU-2}
B.~Kraus,
Local Unitary Equivalence of Multipartite Pure States
Phys. Rev. Lett.  {\bf104}, 020504,
[arXiv:0909.5152].  

\bibitem{LU-3}
B.~Kraus.
Local unitary equivalence and entanglement of multipartite pure states
Phys. Rev. A {\bf 82}, 032121,
[arXiv:1005.5295]. 

\bibitem{LU-4}
M.~Walter, D.~Gross, and J.~Eisert, 
``Multipartite Entanglement'',
 In Quantum Information (eds D. Bruss and G. Leuchs), 2016,
 [arXiv:1612.02437].
 
   \bibitem{DNL}
S.~Dartois, L.~Lionni, I.~Nechita, ``On the joint distribution of the marginals of multipartite random quantum states'', Random Matrices: Theory and Applications, {\bf 9}(03):2050010, 2020
[arXiv :1808.08554].


\bibitem{random-meas-1}
M.C.~Tran, B.~Daki\'{c}, W.~Laskowski, and T.~Paterek,
``Correlations between outcomes of random measurements'',
Phys. Rev. A {\bf 94}, 042302,
[arXiv:1605.08529]. 


\bibitem{random-meas-2}
A.~Ketterer, N.~Wyderka, and O.~G\"{u}hne,
``Characterizing Multipartite Entanglement with Moments of Random Correlations'',
Phys. Rev. Lett. {\bf 122}, 120505,
[arXiv:1808.06558].

\bibitem{random-meas-3}
L.~Knips, J.~Dziewior, W.~K{\l}obus, et al.,
``Multipartite entanglement analysis from random correlations'', 
npj Quantum Inf {\bf 6}, 51 (2020),
[arXiv:1910.10732]. 

\bibitem{random-meas-4}
S.~Imai, N.~Wyderka, A.~Ketterer, O.~G\"{u}hne,
``Multiparticle correlations and bound entanglement from randomized measurements'',
[arXiv:2010.08372].

\bibitem{random-meas-5}
A.~Ketterer, N.~Wyderka, O.~G\"{u}hne, 
``Entanglement characterization using quantum designs'',
 Quantum 4, 325 (2020), 
 [arXiv:2004.08402].
 
   \bibitem{NicaSpeicher}
  A.~Nica and R.~Speicher, 
  ``Lectures on the combinatorics of free probability," 
  volume 13. Cambridge University Press, 2006.
  
  \bibitem{Spei03}
  R.~Speicher, ``Free Probability Theory and Random Matrices'', In: Vershik A.M., Yakubovich Y. (eds) Asymptotic Combinatorics with Applications to Mathematical Physics. Lecture Notes in Mathematics, vol 1815. Springer, Berlin, Heidelberg (2003).
  


 
     \bibitem{uniform1} A.~J.~Scott, ``Multipartite entanglement, quantum-error-correcting codes, and entangling power of quantum evolutions'',  Phys. Rev. A {\bf 69}, 052330 (2004), [arXiv:0310137].  
     
    \bibitem{uniform2} P.~Facchi, G.~Florio, G.~Parisi, and S.~Pascazio,  ``Maximally multipartite entangled states'', Physical review A {\bf 77}, 060304(R) (2008), [arXiv:0710.2868].   


    \bibitem{uniform3} D.~Goyeneche, K.~Zyczkowski, ``Genuinely multipartite entangled states and orthogonal arrays'', Physical review A {\bf 90}, 022316 (2014), [arXiv:1404.3586].     
    
    

  
    \bibitem{maximally-entangled-pure}
    D.~Cavalcanti, F.G.S.L.~Brand{\~a}o, and M.O.~Terra Cunha,
``Are all maximally entangled states pure?''
Phys. Rev. A 72, 040303(R) , 2005 [arXiv:quant-ph/0505121].
  
  
    \bibitem{Gurau:2011xp} 
  R.~Gurau and J.~P.~Ryan,
  ``Colored Tensor Models - a review,''
  SIGMA {\bf 8}, 020 (2012)
  doi:10.3842/SIGMA.2012.020
  [arXiv:1109.4812].

\bibitem{Gurau:2019qag} 
  R.~Gurau,
  ``Notes on Tensor Models and Tensor Field Theories,''
  [arXiv:1907.03531].
  
  \bibitem{FLT} 
  E.~Fusy, L.~Lionni, and A.~Tanasa, ``Combinatorial study of graphs arising from the
Sachdev-Se-kitaev Model'', European Journal of Combinatorics, 86:103066 (2020), [arXiv:1810.02146].

  
  \bibitem{ZJ} P.~Zinn-Justin, ``Adding and multiplying random matrices: A generalization of Voiculescu's formulas'', 
Phys. Rev. E {\bf 59}, 4884.

\bibitem{Collins03} 
B.~Collins, 
``Moments and Cumulants of Polynomial random variables on unitary groups, the Itzykson Zuber integral and free probability'', 
Int. Math. Res. Not. {\bf 2003}(17), 953-982 (2003).

    \bibitem{HCIZ-Hurwitz-1}
I.~Goulden, M.~Guay-Paquet, and J.~Novak, ``Monotone Hurwitz Numbers and the HCIZ Integral II,'' (2011) [arXiv:1107.1001].



\end{thebibliography}
\end{document}